\documentclass[11pt,prd,groupedaddress,showpacs,nofootinbib]{revtex4-1}
\usepackage{graphicx}
\usepackage{amsmath}
\usepackage{amssymb}
\usepackage[colorlinks=true, urlcolor=blue, linkcolor=blue, citecolor=blue]{hyperref}
\usepackage{cleveref}
\usepackage{slashed}
\usepackage{eufrak}

\usepackage{bbm}  
\usepackage{amsmath}                                                

\newcommand{\nc}{\newcommand}
\nc{\beq}{\begin{equation}}  \nc{\eeq}{\end{equation}}
\nc{\bea}{\begin{eqnarray}}  \nc{\eea}{\end{eqnarray}}
\nc{\baa}{\begin{array}}     \nc{\eaa}{\end{array}}
\nc{\bit}{\begin{itemize}}   \nc{\eit}{\end{itemize}}
\nc{\ben}{\begin{enumerate}} \nc{\een}{\end{enumerate}}
\nc{\bce}{\begin{center}}    \nc{\ece}{\end{center}}
\nc{\bpm}{\begin{pmatrix}}   \nc{\epm}{\end{pmatrix}}
\nc{\bvt}{\begin{verbatim}}  \nc{\evt}{\end{verbatim}}
\nc{\bal}{\begin{align}}
\def\mcr{\nonumber\\[6pt]}

\def\half{\frac12}	

\def\to{\rightarrow}

\def\boldoverdot{\,{\raise6pt\hbox{\bf.}\!\!\!\!\>}}

\def\then{{\quad\Rightarrow\quad}}
\def\acal{{\cal A}}

\def\dcal{{\cal D}}

\def\fcal{{\cal F}}
\def\gcal{{\cal G}}

\def\jcal{{\cal J}}

\def\lcal{{\cal L}}
\def\mcal{{\cal M}}

\def\ocal{{\cal O}}
\def\pcal{{\cal P}}

\def\wcal{{\cal W}}

\def\ee{{\bf e}}

\def\BB{{\bf B}}

\def\EE{{\bf E}}

\def\WW{{\bf W}}

\def\gBB{{\mathbbm G}}

\def\zBB{{\mathbbm Z}}
\def\mati{{\mathbbm1}}

\usepackage{bm}
\def\pibf{{\bm\pi}}		
\def\sigbf{{\bm\sigma}}		
\def\ssb{spontaneous symmetry breaking}
\def\vev{vacuum expectation value}

\def\tr{ \hbox{tr}}
\def\det{\hbox{det}}

\def\diag{\hbox{\diag}}

\def\gev{\hbox{GeV}}
\def\tev{\hbox{TeV}}

\def\vevof#1{\left\langle #1 \right\rangle}

\def\doubleundertext#1{
{\undertext{\vphantom{y}#1}}\par\nobreak\vskip-\the\baselineskip\vskip4pt%
\undertext{\hbox to 2in{}}}
\def\inbox#1{\vbox{\hrule\hbox{\vrule\kern5pt
     \vbox{\kern5pt#1\kern5pt}\kern5pt\vrule}\hrule}}
\def\sqr#1#2{{\vcenter{\hrule height.#2pt
      \hbox{\vrule width.#2pt height#1pt \kern#1pt
         \vrule width.#2pt}
      \hrule height.#2pt}}}

\def\today{\ifcase\month\or
  January\or February\or March\or April\or May\or June\or
  July\or August\or September\or October\or November\or December\fi
  \space\number\day, \number\year}
\def\pmb#1{\setbox0=\hbox{#1}%
  \kern-.025em\copy0\kern-\wd0
  \kern.05em\copy0\kern-\wd0
  \kern-.025em\raise.0433em\box0 }

\def\pmbb#1{\setbox0=\hbox{#1}%
  \kern-.02em\copy0\kern-\wd0
  \kern.04em\copy0\kern-\wd0
  \kern-.02em\raise.03464em\box0 }
\def\up#1{^{\left( #1 \right) }}
\def\inv#1{\frac1{#1}}
\def\su#1{{SU(#1)}}
\def\ui{U(1)}
\def\sumprime_#1{\setbox0=\hbox{$\scriptstyle{#1}$}
  \setbox2=\hbox{$\displaystyle{\sum}$}
  \setbox4=\hbox{${}'\mathsurround=0pt$}
  \dimen0=.5\wd0 \advance\dimen0 by-.5\wd2
  \ifdim\dimen0>0pt
  \ifdim\dimen0>\wd4 \kern\wd4 \else\kern\dimen0\fi\fi
\mathop{{\sum}'}_{\kern-\wd4 #1}}
\def\wc{w} 
\def\wcc{c} 
\def\leff{\lcal_{\tt eff}}
\def\seff{S_{\tt eff}}
\def\me{m_{\tt e}} 
\def\smvev{{\tt v_{\tt SM}}}

\def\be{\begin{equation}}
\def\ee{\end{equation}}

\def\lcal{\mathcal{L}}

\def\nsm{$\nu$SM}

\def\leff{\lcal_{\rm eff}}
\def\rh{right-handed}
\def\lh{left-handed}
\def\vp{\varphi}

\def\mn{{\mu\nu}}
\def\eps{\epsilon}

\def\phit{\widetilde\phi}
\def\ptd{\phit^\dagger}
\def\ncb{\overline{\nu^c}}
\def\ecb{\overline{e^c}}
\def\lcb{\overline{\ell^c}}
\def\Ncb{\overline{N^c}}
\def\mw{m_{\rm w}}
\def\mz{m_{\rm z}}
\def\mh{m_{\rm H}}
\def\cw{c_{\rm w}}

\def\mdm{m_{\tt DM}}

\def\li{{\mathfrak{l}}} 
\def\he{{\mathfrak{h}}} 

\def\qr {{\mathfrak{q}}} 

\def\eps{\varepsilon}
\def\wt{\widetilde}

\def\phij {\mbox{${\phi^\dag i\,\raisebox{2mm}{\boldmath ${}^\leftrightarrow$}\hspace{-4mm} D_\mu\,\phi}$}}
\def\phijt{\mbox{${\phi^\dag i\,\raisebox{2mm}{\boldmath ${}^\leftrightarrow$}\hspace{-4mm} D_\mu^{\,I}\,\phi}$}}

\def\fm{F} 

\begin{document}

\markboth{S. Bhattacharya, J. Wudka}
{Effective theories with DM applications}

\title{EFFECTIVE THEORIES WITH DARK MATTER APPLICATIONS}

\author{SUBHADITYA BHATTACHARYA}

\address{Department of Physics, Indian Institute of Technology Guwahati\\
North Guwahati, Assam-781039, India\\
 email: {\tt subhab@iitg.ac.in}}

\author{JOS\'E WUDKA}

\address{Department of Physics {\it\&} Astronomy, University of California Riverside\\
Riverside, CA 92521-0413, USA\\
email: {\tt jose.wudka@ucr.edu}}

\begin{abstract}

Standard Model (SM) of particle physics has achieved enormous success in describing  the interactions among the known fundamental constituents of nature, yet it fails to describe phenomena for which there is very strong experimental evidence, such as the existence of dark matter, and which point to the existence of new physics not included in that model; beyond its existence, experimental data, however, have not provided clear indications as to the nature of that new physics. The effective field theory (EFT) approach, the subject of this review, is designed for this type of situations; it provides a consistent and unbiased framework within which to study new physics effects whose existence is expected but whose detailed nature is known very imperfectly. We will provide a description of this approach together with a discussion of some of its basic theoretical aspects. We then consider applications to high-energy phenomenology and conclude with a discussion of the application of EFT techniques to the study of dark matter physics and its possible interactions with the SM. In several of the applications we also briefly discuss specific models that are ultraviolet complete and may realize the effects described by the EFT. 

\end{abstract}

\maketitle

\tableofcontents

 \newpage

\section{Introduction}
\label{sec:introduction}
Effective field theories (EFT) \cite{Polchinski:1992ed,Brivio:2017vri,Dobado:1997jx}~\footnote{There is a large number of review articles and several books that discuss effective theories, the ones selected complement and deepen the discussion presented below; the reference list is not intended to be comprehensive.}, are an efficient and consistent way of dealing with degrees of freedom that are either difficult to include exactly (e.g. calculating QCD effects at low energy, where {\it ab-initio} calculations are impossible), or whose existence is hypothesized and constraints on their properties are desired. For this last case a frequent application has been the parametrization of the possible effects of physics beyond the Standard Model (SM) in collider and high-precision measurements, where one looks for deviations from the SM, and, absent those, to extract meaningful constraints on classes of new physics; especially the scale at which they become manifest. A related application, which has been gaining attention recently, is the parameterization of hypothesized couplings between dark matter (DM) and the SM. Many of the concepts presented in this review have appeared elsewhere,  nonetheless we consider that there is a paucity of review papers that discuss effective theories for both the SM and for DM physics, and so we believe this contribution will be a useful addition to the literature.

There are two basic ingredients needed to construct an EFT: the canonical degrees of freedom and their symmetries. Given these the process is straightforward: one simply writes all local operators $ \ocal $ involving the fields corresponding to these degrees of freedom, and which obey the given symmetries. The EFT Lagrangian $ \leff$ is then a linear combination of these operators with arbitrary coefficients $\wc$, 
\beq
\leff = \sum_a \wc_a \ocal_a\,;
\label{eq:leff}
\eeq
the $\wc$ are commonly referred to as Wilson coefficients.

At this point, however, $\leff$ is not useful or manageable, since there is an infinite number of $ \wc $. The formalism then needs a third and final ingredient \cite{Preskill:1990fr}: a hierarchy among the $ \ocal $ so that the sum over these operators can be split,
\beq
\leff = \leff\up0 + \leff\up1 + \leff\up2 + \cdots\,, \qquad \leff\up n  =  \sum_{a \in A_n } \wc_a \ocal_a\,;
\label{eq:hierarchy}
\eeq
where under the hierarchy the set of indices $a$ is segregated into subsets $A_n$ whose union cover all possible indices, and such that the effects of $ \leff\up{n+1} $ are subdominant to those of $ \leff\up n $. 

The specific hierarchy imposed on the EFT depends on the situation and requires additional assumptions. For example, if we are interested in describing the effective interactions generated by particles much heavier than the available energies,  and we assume that the effects of such particles decouple, then the EFT hierarchy is obtained by an expansion on the (inverse) scale of these heavy modes. If we are interested in low-energy phenomena, then an expansion in powers of the momenta is appropriate, and the hierarchy is defined by the number of derivatives in each $ \ocal $.

Once a hierarchy is selected the formalism becomes useful since the effects of $ \leff\up n$ can be neglected for $n$ sufficiently large because their effects will be smaller than the experimental error.

There is another issue relevant for a consistent choice of hierarchy, and that concerns radiative corrections. The Lagrangian in \cref{eq:leff} is renormalizable: any divergence generated by it will correspond to a local operator obeying the symmetries of the theory, and since (by definition) $ \leff$ contains all such operators, such divergences can be absorbed in a redefinition of one (or more) of the $\wc$; finite results can be obtained by replacing $ \wc_a \to \wc_a \up 0 = \wc_a + \delta \wc_a $ for appropriately chosen counterterms $ \delta \wc_a $.   However, if the hierarchy is to be useful, it must be the case that  $ \delta\wc_a $ for $ a\in A_n $ (see \cref{eq:hierarchy}) should only depend on the $ \wc_b $ for $ b \in A_l,\, l\le n$. Otherwise the running of the $ \wc$ will depend on couplings we are supposed to be able to ignore. In practical applications this third conditions is obeyed.

\subsection{Illustration}

 Consider QED with at energies well below the electron mass $ \me $; in these circumstances the electrons cannot be observed directly, but their virtual effects generate photon ($A_\mu$) self-interactions. The action of the full theory is~\footnote{Throughout this review we use the conventions of Itzykson {\it\&} Zuber \cite{Itzykson:1980rh}.}
\beq
S_{\tt QED} = \int d^4x \left[ - \inv4 F_{\mu\nu} F^{\mu\nu} + \bar \psi \left( i \slashed D - \me \right) \psi \right]\,; \qquad D_\mu = \partial_\mu + i e A_\mu \,,
\eeq
where $F_{\mu\nu}=\partial_\mu A_\nu-\partial_\nu A_\mu$ is the field strength tensor, and 
$e$ is the electron charge. The generating function of connected Green's function, $W[j]$ is then given in terms of a functional integral (here $j$ denotes an external source) 
\bal
e^{i W[j]} 
&= \int [dA] \int [d\psi\,d\bar\psi] e^{i S_{\tt QED} + i \int d^4x j_\mu A^\mu } \,,\mcr
&= \int [dA]  e^{i \seff + i \int d^4x j_\mu A^\mu } \,,
\end{align}
where the effective action $ \seff $ can be expressed  in terms of a fermionic determinant:
\beq
\seff = - \inv4 \int d^4 x F_{\mu\nu} F^{\mu\nu} -i \ln \text{Det} \left( i \slashed D - \me \right)\,,
\eeq
and the second term is a formal representation of an infinite series,
\bal
-i \ln \text{Det} \left( i \slashed D - \me \right) &=
-i \text{Tr} \ln \left[S^{-1}(\mati  - e S \slashed A )\right]\,; \qquad S = (i \slashed \partial - \me)^{-1} \,,\mcr
&= \text{const.} + i \frac{e^2}2 \text{Tr}\{(S \slashed A)^2 \} +i  \frac{e^4}4 \text{Tr}\{(S \slashed A)^4 \} + \cdots \,,
\end{align}
where  the terms odd in $A$ are absent because of charge-conjugation symmetry and the first (constant) term is independent of $A$.

The second order calculation is straightforward. Using dimensional regularization one readily obtains~\footnote{If $ 4a>1$, $ \gcal_{\tt R} $ must be analytically continued using $ \me \to \me - i \epsilon$.}:
\bal
i \frac{e^2}2 \text{Tr}\{(S \slashed A)^2 \} &=- \frac\alpha{2 \pi} \int d^4 x \, d^4 y F^{\mu\nu}(x) G(x-y) F_{\mu\nu}(y) \,; \quad G(x) = \inv6 C_{\tt UV} \delta(x) + \int \frac{d^4k}{(2\pi)^4} e^{i k\cdot x} \gcal_{\tt R}(\me^2/k^2) \,,\mcr
\gcal_{\tt R}(a) &=  - \int_0^1 du\, u (1-u) \ln \left[ 1 - \frac{u(u-1)}a \right] = \frac{5+12a}{18}+ \left( \frac{8a^2+2a-1}3 \right) \frac{\text{arccoth}(\sqrt{1-4a})}{\sqrt{1-4a}} \,,\mcr
&=   \inv{30\,a}   + \inv{280\, a^2}  + \cdots \,, 
\end{align}
where $ \alpha = e^2/(4\pi) $ and $ C_{\tt UV} = 2/(n-4) + \gamma_{\tt Euler} + \ln (4\pi\mu/\me) $ ($n$ is the dimension of space-time, $\gamma_{\tt Euler}$ the Euler constant, and $ \mu$ the renormalization scale). The power series expansion is accurate only as long as $ |k^2| \ll \me^2$.

The effective action then becomes,
\beq
\seff = \int d^4 x \left[ - \frac{1 + \alpha C_{\tt UV}/(3 \pi)}4F_{\mu\nu} F^{\mu\nu} + \frac\alpha{60\pi\me^2} F_{\mu\nu} \Box F^{\mu\nu} - \frac\alpha{560\pi\me^4} F_{\mu\nu} \Box^2 F^{\mu\nu} + O\left(\me^{-6},\,\alpha^2 \right) \right]\,.
\label{eq:qed.leff.2}
\eeq
The divergent term can be absorbed in a wave-function renormalization of the photon field. After this is performed the dispersion relation for the photon field becomes 
\beq
\left[ 1- \frac\alpha{2\pi} \gcal_{\tt R}(\me^2/k^2) \right] k^2 =0 \,,
\label{eq:tach.pole}
\eeq
so that the photon propagator seems to have, aside from the usual $ k^2=0 $ pole, another where $ \alpha \gcal_{\tt R} = 2\pi $; the (tachyonic) solution to the second relation, $ k^2 \sim - \me^2 \, \exp(12\pi/\alpha) $, is unphysical since $ |k^2| \gg \me^2 $, in violation of the basic assumption behind the expansion used.

Following this procedure, obtaining the higher-order terms (in the number of fields) is cumbersome, but alternative methods are available, at least in some interesting cases. For example, when $A_\mu $ is slowly varying it is much simpler to use the Fock-Schwinger proper-time method \cite{Schwartz:2013pla,Itzykson:1980rh}. The idea is the following: let $ S_A = 1/(i\slashed D - \me) $ and $ H = (i\slashed D - \me)(i\slashed D + \me) $ then it is straightforward to show that
\beq
i \frac{\delta \seff}{\delta A_\mu(x)} = e \, \text{tr} \vevof{ x \left| \gamma^\mu S_A \right|x }= e \, \text{tr} \vevof{ x \left| \gamma^\mu (i \slashed D + \me) \inv{H+ i \epsilon} \right|x }\,,
\eeq
where we replaced $ \me^2 \to \me^2 - i \epsilon$ in $H$ and the trace ($\text{tr}$) is over the Dirac matrix indices. Noting then that $H$ has an even number of $\gamma $ matrices, that it commutes with $ \slashed D$, and that $ \delta H = e \{ i \slashed D, \delta\slashed{ A} \}$ when $ A_\mu $ is varied, we find
\beq
\frac{\delta \seff}{\delta A_\mu(x)} = \frac{\delta }{\delta A_\mu(x)} \int_{-\infty}^0 \frac{d\tau}{2 i \tau} e^{\epsilon \tau} \, \text{Tr}~ e^{- i \tau H}\,,
\eeq
where Tr denotes the trace over both Dirac matrix indices and coordinates. For the case where $ F_{\mu\nu} $ is a constant Tr$\,\exp(-i \tau H ) $ is known~\cite{Itzykson:1980rh}:
\beq
\text{Tr} ~e^{- i \tau H} = - \frac i{(4\pi \tau)^2}e^{i \tau \me^2 }\int d^4 x \left[ \det \left( \frac{\sinh\fcal}\fcal \right) \right]^{-1/2} \text{tr}\left[\exp\left( \frac i2 \sigma_{\mu\nu} \fcal^{\mu\nu} \right)\right]\,,
\eeq
where $ \fcal_{\mu\nu} = e\tau F_{\mu\nu}$, the determinant is over the Lorentz indices and the trace over the Dirac matrix ones. This expression can then be expanded in powers of the field (or, equivalently, power of $e$):
\beq
\text{Tr} ~e^{- i \tau H} = - \frac i{4\pi^2 \tau^2}e^{i \tau \me^2 }\int d^4 x \left[1 + \frac{(e\tau)^2}3 a - \frac{(e\tau)^4}{45} (a^2 + 7 b^2) + \cdots\right]\,,
\eeq
where
\beq
a = \EE^2 - \BB^2\,,\qquad b = \EE\cdot \BB\,.
\eeq
The first term inside the bracket gives a divergent but $A_\mu$-independent contribution to $\seff $. The second term is also divergent and proportional to the free action $ \int d^4 x F_{\mu\nu} F^{\mu\nu} $; it can be absorbed in a wave-function renormalization of the $A_\mu $ field, as in the previous calculation. Note that this $O(e^2) $ term does not reproduce all of \cref{eq:qed.leff.2} because we are assuming here that $ F_{\mu\nu} $ is a constant. Finally, the $O(e^4) $ term gives
\beq
\seff\up4 = \frac{e^4}{360 \pi^2 \me^4} \int d^4 x (a^2 + 7 b^2)\,.
\label{eq:EH}
\eeq
This is the Euler-Heisenberg Lagrangian \cite{Heisenberg:1935qt}~\footnote{For a pedagogical derivation see \cite{Schwartz:2013pla,Itzykson:1980rh}.} that summarizes the leading photon-photon interactions (for slowly varying fields).

The low-energy Lagrangian is then composed of terms which are local and gauge invariant, they are suppressed by powers of $ 1/\me $ and by numerical factors $ \sim 1/(16\pi^2) $. The first two characteristics follow from the large $ \me $ expansion and from the fact that the charged fermion decouples \cite{Appelquist:1974tg,Witten:1975bh,Weinberg:1980wa}~\footnote{For a clear pedagogical discussion see \cite{Collins:2008pla}.} in the $ \me \to \infty $ limit. The numerical suppression occurs because the interactions induced by the fermions are generated at 1 loop. We will see that all these features are reproduced in general.

\subsection{Lessons learned}
\label{sec:lessons-learned}
In the simple model of the previous section the effective theory at low energies exhibits several features that are worth emphasizing. 

The EFT is obtained by an expansion in inverse powers of $\me$, and becomes unreliable when applied to energies close to that scale, the scale of `new' physics. Effects obtained while ignoring this limitation are unphysical (such as the spurious tachyonic pole briefly discussed below \cref{eq:tach.pole}).

Not all terms are suppressed by inverse powers of $ \me$. For example, this is the case of the divergent term $ \propto C_{\tt UV}$ in \cref{eq:qed.leff.2}. But, as noted in the example above, such contributions are unobservable, and can be absorbed in a renormalization of the low-energy parameters and fields. This is a consequence of the decoupling theorem, discussed briefly in section \ref{sec:dec}.

Lastly, in addition to the inverse powers of $ \me $, the coefficients of the effective Lagrangian are suppressed by powers of the coupling constant $e$ and by loop suppression factors $\sim 1/(16\pi^2) $. These terms would not be present in the coefficients of operators if generated at tree level.

As we will see in the following sections, these are generic features, common to all EFT. The first two are important in ensuring the EFT predictions are reliable and physical. The last is useful in gauging the importance of the low-energy effects of heavy new physics, and in identifying observables that can be most sensitive to them.

\subsection{Effective field theory approach}
The discussion above took advantage of our foreknowledge of both the light theory, containing only photons, and the heavy physics, containing only the electron; because of this (and since the models is perturbative) we were able to calculate all low-energy effects in terms of the electron mass and charge. 

The EFT approach is most useful in a related situation, where only the low-energy theory (with its particle content and symmetries) is known. The effective Lagrangian can then be used to estimate the possible low-energy (virtual) effects that could be generated by unknown new physics, without specifying nature of that new physics -- assuming only that the heavy and low-energy theories are separated by an energy gap. In this scenario the low energy observable effects of the heavy dynamics are parameterized by the coefficients (the $ \wc_a $ in \cref{eq:leff}) of higher dimensional operators that contain only light fields and respect the local symmetries of the light theory. 

For the specific example considered above, the parameterization takes the form
\beq
\leff = - \inv4  F_{\mu\nu} F^{\mu\nu} + \frac{c_1}{\Lambda^4} \underbrace{ \left( \half F_{\mu\nu} F^{\mu\nu} \right)^2}_{\ocal_1} + \frac{c_2}{\Lambda^4} \underbrace{ \left( \inv8 \epsilon_{\mu\nu\rho\sigma} F^{\rho\sigma} F^{\mu\nu} \right)^2}_{\ocal_2}   +  \cdots \,,
\eeq
where $ \Lambda $  denotes the scale of the new physics generating these low-energy effects and $ c_1,\,c_2,\,\ldots $ depend on the nature of that physics. The effects of higher-dimensional terms (corresponding to the ellipsis) are subdominant because of the increased number of inverse powers of $ \Lambda $. In addition, we will see (Sect. \ref{sec:tree.loop}) that, independent of the type of heavy physics (as long as it is weakly coupled and decoupling), both operators listed are  generated by loop diagrams; accordingly the coefficients are suppressed by an additional numerical factor of $ 1/(4\pi)^2 $ (in this respect the example above is generic). Finally, gauge fields are universally coupled, so they appear together with the corresponding gauge coupling: $ c_{1,2} \propto e^4$. For the specific type of heavy physics considered above $ \Lambda = \me $ and $ c_1 = (2/45) \times [e^4/(4\pi)^2],\,c_2 = (14/45) \times [e^4/(4\pi)^2] $; other types of new physics will generate similar coefficients but with different numerical factors multiplying $e^4/(4\pi)^2$ (and a different scale).

These results and observations can be readily applied to more realistic cases. We summarize the application to SM physics in  sections \ref{sec:SMEFT} and \ref{sec:SMEFT-obs}, and to some aspects of dark matter physics in section \ref{sec:DM}.

\bigskip

\section{The effective Lagrangian and the decoupling theorem}
\label{sec:dec}
Having provided a simple example and a general overview of the ideas behind EFT we now turn to some of the theoretical underpinnings of this approach. In this section we look at two important such aspects. We first discuss briefly the decoupling theorem, which provides the formal justification of the expansion in \cref{eq:leff}, and the types of  physics for which it is useful. We then turn to the role of gauge invariance in effective theories. We continue these formal considerations in Section \ref{sec:formal} with a discussion of the equivalence theorem, which allows for the elimination of a significant number of terms in the effective Lagrangian (when low-energy applications are considered); we conclude our formal considerations with a discussion of several aspects of the estimation of the effective Lagrangian coefficients (the $ \wc_a$ in \cref{eq:leff}) that are useful in providing reliable and natural estimates of the effects of new physics.

\subsection{Decoupling}

In the example above all observable effects of the heavy physics were suppressed by inverse power of a heavy-physics scale. This is not fortuitous, but follows from a general result known as the {\em decoupling theorem} \cite{Appelquist:1974tg,Witten:1975bh,Weinberg:1980wa,Collins:1978wz}. In its simplest terms this theorem shows that if a theory has light and heavy fields (denoted by $ \phi$ and $ \Phi $, respectively), and if the action $ S[\phi,\,\Phi=0] $ describes a renormalizable theory for the light fields invariant under all low-energy local symmetries, then all observable effects of the $ \Phi $ at low energies are suppressed by inverse powers of the heavy $ \Phi $ masses; where `low energy' refers to energies well below those same heavy scales. Any low-energy effects of the $ \Phi $ that grow with the heavy scales can be absorbed in a renormalization of the parameters of the low-energy theory.

The decoupling theorem then provides the conditions under which the low-energy observable effects of some unspecified dynamics take the EFT form \cref{eq:leff} whose Wilson coefficients $ \wc_a $ vanish as the heavy scale goes to infinity:
\beq
\leff = \sum \frac{ \wcc_a} {\Lambda_a^{ n_a}} \ocal_a\,,
\eeq
where the $n_a$ are integers~\footnote{Strictly speaking, $n_a+4$ is equal to the dimension of $ \ocal_a $ which may differ from an integer in certain cases where the effects of some strong interactions become important.} and the $ \Lambda_a$ are heavy scales associated with the operator $ \ocal_a $. These operators are local,  involve only light fields, and respect all the local symmetries of the light theory. The proof of this theorem is perturbative, so it is also assumed that the heavy modes are weakly coupled.

There being excellent technical \cite{Weinberg:1980wa,Collins:1978wz} and pedagogical \cite{Collins:2008pla} reviews of the decoupling theorem, we will provide no proof of this result, but will only comment on a basic though important point. The basic premise of the theorem, that there are light and heavy fields, presupposes that the kinetic and mass terms in the Lagrangian are diagonal; any mixing ({\it e.g.} a mass term of the form $ \phi \Phi $) must be eliminated through appropriate field rotations before light and heavy fields are identified.

It is important to note that the EFT methodology is not restricted to situations where the decoupling theorem hold. For example, one can use this approach to investigate the low energy effects of a heavy Higgs particle in the SM \cite{Longhitano:1980tm,Kuti:1987ns,Nyffeler:1996nh}.

Though any effect that grows with the $ \Lambda_a $ can be absorbed in a renormalization of the low-energy parameters, and is therefore unobservable, it has been argued \cite{Cheng:1980cx,Susskind:1982mw,Weinberg:1984jb,Altarelli:1999gu,Masina:2013wja,Bar-Shalom:2014taa,Grzadkowski:2009mj} that such contributions are unnatural when too large. The best known example of this argument pertains to the Higgs mass, which gets a  contribution $ \propto \Lambda^2 $ from a variety of new physics~\footnote{There are similar contributions $ \propto \smvev^2$ from SM particles, where $ \smvev$ is the SM \vev.}. Arguing that such contributions should be at most of order $ m_{\tt Higgs}^2 $ leads to an upper (naturality) bound on the scale of heavy physics. This argument is an independent assumption and does not follow from the decoupling theorem.

\subsection{The role of gauge invariance}

At various points in the above discussion we have insisted that the effective operators must respect the local symmetries of the light theory.  In this section we wish to examine this constraint. We begin by reviewing a construction, originally due to St\"uckelberg \cite{Stueckelberg:1900zz} that allows to describe {\em any} Lagrangian as that of a gauge theory when a specific gauge is chosen. We then comment on the implications of this result.

\subsubsection{The St\"uckelberg construction}

We begin with a theory that contains a number of vector bosons $W^n_\mu,\,n=1,2,\ldots, N $ and other fields, that we denote collectively by $ \chi $; their interactions is described by an action $ S[W,\chi]$. What we like to do is to construct a gauge theory whose action reduces to this same $ S $ in the unitary gauge.

To this end we choose a gauge group $\gBB$ of dimension $N$ (which may or may not be simple) with generators $\{T^n\}$ that we choose Hermitian and normalized according to
\beq
\text{tr} \left\{ T^n T^m \right\} = \delta_{nm}\,.
\eeq
Using these we {\em define} a derivative operator
\beq
D_\mu = \partial_\mu + i \sum_{n=1}^N T^n W^n_\mu \,.
\eeq
Next, we introduce a field $U$ which is unitary, with $ U(x) \in U(N)$ for all $x$; for reasons that will become clear we refer to $U$ as an auxiliary field. We assume that the infinitesimal transformation of $U$ under $\gBB$ is given by (note that this is not the  transformation for a field in the adjoint representation) 
\beq
\delta U = i \sum_{n=1}^N \epsilon_n T^n U \,,
\label{eq:U.transf}
\eeq
and use them to define the composite fields
\beq
\wcal^n_\mu = \tr \left\{ T^n U^\dagger D_\mu U \right\}\,.
\eeq
These fields are invariant under $\gBB$ gauge transformations assuming that {\it(i)} the $W$ transform as gauge fields, and {\it(ii)} \cref{eq:U.transf} gives the transformation rule for $U$. In addition there is a gauge choice for which $U=1$, which we call the unitary gauge, also
\beq
\wcal|_{\tt unit.\, gauge} = W \,.
\eeq
It follows that if we replace $ W \to \wcal $ in $S$ we obtain a gauge invariant Lagrangian $S[\wcal,\chi] $ which reverts to $S[W,\chi ]$ in the unitary gauge.

Since any Lagrangian can be taken as a gauge theory, one may question whether the requirement of gauge invariance has any content \cite{Burgess:1992va}. In fact, it does:
\bit
\item Note that in the above construction all matter fields are gauge singlets. So in this case there are no predictions such as lepton universality, which are accurately verified by observations. Within the above formalism such results would be accidental.
\item The action $S[\wcal,\chi] $ involves a unitary field, and so it will become strongly coupled at sufficiently high energies. For the case where the low-energy theory is the SM this scale is $ \sim 4 \pi \smvev \sim 3\, \tev $. The model is not useful at higher energies.
\item Gauge transformations do not mix terms with different canonical dimensions, so that all terms in $S[\wcal,\chi]$ with a given dimension are separately gauge invariant. For the case where the SM is the low-energy theory, all available experimental data is described to high accuracy by a  $\su3\times\su2\times\ui $ gauge theory, which strongly suggests that  $ \gBB$ is identical to this group, or at least contains it.
\eit
Because of these points, the fact that the St\"uckelberg procedure can be used to ``gauge invariantize'' any theory does not diminish the importance and relevance of requiring the effective operators to be gauge invariant under the SM gauge group, with non-trivial transformation properties of the matter fields ($ \chi $).

\bigskip

Before closing this section we also remark on the classical results of Cornwall {\it et al.} \cite{Cornwall:1973tb,Cornwall:1974km} (see also. \cite{Gunion:1990kf}), who showed that any model containing scalars, vectors and fermions, and which is unitary at tree level, is a spontaneously broken gauge theory, assuming all terms in the Lagrangian have at most 2 fermions and at most 4 vectors.

\section{Formal developments}
\label{sec:formal}
 
Before proceeding with the construction of effective theories for the SM and for dark-matter physics  we take a detour to discuss a few formal aspects of the EFT approach. 

\subsection{The equivalence theorem -- independent operators and the equation of motion}
\label{sec:equiv.thm}

The {\em equivalence theorem}~\footnote{This name is shared with a quite unconnected result that relates the high-energy effects of gauge bosons in spontaneously broken gauge theories to those of a pure-scalar theory.}\cite{Chisholm:1961tha,Kamefuchi:1961sb,Kallosh:1972ap,Tyutin:2000ht,Arzt:1993gz,Einhorn:2013kja,Passarino:2016saj} is a consequence of the re-parameterization freedom in quantum field theories. Its main application in the context of effective field theories is the following. Suppose that, to lowest order in $ 1/\Lambda $, two operators $ \ocal $ and $\ocal'$ obey 
\beq
a \ocal - \ocal' = \acal \frac{\delta S_{\tt light}}{\delta\phi }\,,
\label{eq:oo-identity}
\eeq
where $a$ is a constant, $ \acal $ is a local operator composed of light fields and their derivatives, and $ S_{\tt light} $ the action of the light fields (denoted collectively by $ \phi $). Then the {\sl S} matrix will depend on $ \wc + a \wc' $ (where $ w,\,w'$ are the Wilson coefficients of $ \ocal,\,\ocal' $, respectively), and not on these Wilson coefficients separately.

Several remarks are in order. First, though this is obtained to lowest order in $ 1/\Lambda $ it can be extended to $ O(1/\Lambda^n) $ provided $ S_{\tt light} $ is replaced with the EFT up to order $ 1/\Lambda^{n-1} $. Second, this is a low-energy result, it implies that at such scales the effects of $ \ocal$ and $ \ocal' $ cannot be differentiated, the heavy physics may generate either, or both operators through quite independent processes.

\subsubsection{The equivalence theorem in quantum mechanics }

Before considering the situation for field theory we present a simple (1-dimensional) quantum mechanical example that illustrates the results, as well as some of the technical issues involved.

Consider a classical non-relativistic Lagrangian
\beq
L = \half m \dot x^2 - V(x) \,,
\eeq
and modify it by adding a term that vanishes when the classical equations of motion hold
\bal
L \to L_\epsilon &= L - \epsilon A(x) ( m \ddot x +V') \,,\mcr
&= L + \epsilon( m A' \dot x^2 -  A V') + \dot B \,,
\end{align}
where $ B = - \epsilon m \dot x A $ and $ A' = dA/dx$. Dropping the last term (since it does not affect the classical dynamics), and following the standard canonical procedure we find the Hamiltonian
\beq
H_\epsilon = \frac{p^2}{2m} + V + \epsilon \left( A V' - \inv m A' p^2 \right)\,.
\eeq

Next we consider the quantum theory obtained from $ H_\epsilon$. This operator has an ordering ambiguity that we resolve by replacing
\beq
A' p^2 \to \inv 4 \{ \{ p,A'\},p\}\,.
\eeq
It then only requires a straightforward calculation to show that
\beq
H_\epsilon = U \left( \frac{p^2}{2m} + V \right) U + O(\epsilon^2) \,; \quad U = \exp \left( - \frac i2 \epsilon \{ p,A\} \right)\,,
\eeq
thus, to first order in $ \epsilon $, $H_{\epsilon} $ and $ H_{\epsilon=0} $ are unitarily equivalent and, to this order, the physics derived from either is the same. This is the form the equivalence theorem takes in this simple example: if we add to the Lagrangian a term that vanishes when the classical equations of motion are obeyed, the perturbative dynamics is unchanged.

\subsubsection{The equivalence theorem in field theory}

Now, for the case of field theory we provide only a sketch of the proof, that ignores for the most part renormalization issues. We will also restrict ourselves to the case where the light fields reduce to a single spinless boson $ \phi $, and we denote the light-field Lagrsangian by $ \lcal_{\tt light} $. Suppose then that two local operators $ \ocal,\,\ocal' $, generated by some heavy physics of scale $ \Lambda $, are related by \cref{eq:oo-identity}. We assume for definiteness that the leading observable contributions from the new physics are described by operators of dimension 6, and that $ \ocal,\,\ocal' $ belong to this set, so their Wilson coefficients are of the form $\text{const.}/\Lambda^2$. Then
\beq
\leff =  \lcal_{\tt light} + \inv{\Lambda^2} \left( c\ocal + c' \ocal' + \cdots \right) + O(\Lambda^{-3})\,.
\eeq
If we replace in this expression $ \phi \to \phi + c \acal/\Lambda^2 $ we obtain
\beq
\leff  \to \leff'=  \lcal_{\tt light} + \inv{\Lambda^2} \left[ (c + a c') \ocal + \cdots \right] + O(\Lambda^{-3})\,;
\eeq
so, with this change of variables, the effective action does not depend on $c,\,c'$ independently but only on the combination $ c + a c' $. Next we consider the Jacobian {\sf J} of the above transformation:
\bal
{\sf J} &= \text{Det} \left(\mati + \frac c{\Lambda^2} \frac{\delta\acal}{\delta\phi} \right) \,,\mcr
 &= 1  + \frac c{\Lambda^2}\text{Tr} \left(\frac{\delta\acal}{\delta\phi} \right)  + \cdots. \,,\mcr
 &= \exp \left\{  \frac c{\Lambda^2}\text{Tr} \left(\frac{\delta\acal}{\delta\phi} \right)  + \cdots \right\}\,.
 \end{align}
But, since $ \acal $ is a local operator, $ \text{Tr} ( \delta\acal/\delta\phi) $ will consist of terms proportional to $ \delta\up4(x=0)$ and its derivatives, all of which vanish if we use dimensional regularization \cite{Collins:2008pla}~\footnote{If one uses another regularization scheme one can show that these terms can be absorbed in a renormalization of the light Lagrangian $ \lcal_{\tt light} $.}.

Finally, we consider the generating function for the effective theory
\beq
e^{W[j]} = \int [d\phi] \exp \left\{ i \int d^4x (\leff + j\phi) \right\}\,,
\eeq
where $j$ is an external source. Under the above replacements
\beq
e^{W[j]} = \int [d\phi] \exp \left\{ i \int d^4x \left( \leff' + j\phi  + \frac c{\Lambda^2} j \acal \right) \right\}\,.
\eeq
The last term, containing $ j \acal $ does not contribute to the {\sl S} matrix, unless $ \acal $ has a term linear in $ \phi $ (other terms in $ \acal $ do not generate single-particle poles, and the corresponding contributions will vanish under the `amputation' process); if $ \acal $ does have a contribution linear in $ \phi $, $ \acal = b \phi + \cdots $, it can be absorbed by a wave-function renormalization. Thus, the generating function $W'$,
\beq
e^{W'[j]} = \int [d\phi] \exp \left\{ i \int d^4x (\leff' + j\phi) \right\}
\eeq
will produce the same {\sl S} matrix as $W$, whence the {\sl S} matrix depends on $c + a c' $ only.

As noted above, this is a statement about the low-energy effects of the heavy physics: to order $ 1/\Lambda^2 $ the effects of $ \ocal$ and $ \ocal' $ cannot be differentiated; this is a  property of the light theory in the sense that the relation in \cref{eq:oo-identity} depends on $S_{\tt light} $. It is also easy to see that the equivalence of operators of dimension $n$ is determined by the terms of $ \leff$ up to and including dimension $ n-1$ (which replace as $ \lcal_{\tt light} $ in the calculation).

It is important to note that the equivalence theorem refers to {\sl S} matrix elements, not to Green's functions. The former only depend on $ c + a c' $, while the latter will in general depend on $c$ and $ c' $ each multiplied by different momentum-dependent functions.

Since the effects of $ \ocal' $ cannot be distinguished from those of $ \ocal $ (to this order in $ 1/\Lambda$), one can drop the former operator from $ \leff$. This is the practical use of this equivalence theorem: it reduces the number of operators, and therefore the number of unknown Wilson coefficients, from the effective theory. Elimination of operators can be done in a systematic way \cite{Grzadkowski:2010es}; using this procedure for the case where the SM is the light theory. The reduced basis of dimension 6 operators is presented in section \ref{sec:SMEFT}.

\subsubsection{Field theory example}

Consider the case where $ S_{\tt light} $ corresponds to a real scalar, and is symmetric under $ \phi \to - \phi $, which is also respected by the heavy physics. In this case 
\beq
\lcal_{\tt light} = \half (\partial\phi)^2 - \half m^2 \phi^2 - \frac\lambda{24} \phi^4\,,
\eeq
and the dimension 6 operators~\footnote{The parity symmetry forbids dimension 5 operators.} are
\beq
\phi^6\,, \quad \phi^3 \Box\phi\,, \quad \left( \Box\phi \right)^2\,.
\eeq
The equation of motion for the light field is
\beq
\Box\phi + m^2 \phi + \frac\lambda6 \phi^3 =0 \,;
\eeq
then, refering to \cref{eq:oo-identity},
\beq
\underbrace{\frac\lambda6}_a \underbrace{\left( \frac{6m^2}\lambda \phi^4 + \phi^6 \right)}_{\ocal} + \underbrace{\phi^3\Box\phi}_{-\ocal'} =  \underbrace{\phi^3}_{\acal} \underbrace{ \left(\Box\phi + m^2 \phi + \frac\lambda6 \phi^3 \right)} _ {\delta S_{\tt light}/\delta\phi}\,.
\eeq
This means that to $O(1/\Lambda^2)$ the low-energy physics cannot distinguish between the effects of $ \phi^3 \Box\phi $ and the operator $ ( 6 m^2/\lambda) \phi^4 + \phi^6 $; moreover, the effects of the first (quartic) term can be absorbed in a finite renormalization of $ \lambda$, namely $ \lambda - \wc( 6 m^2)/(\Lambda^2 \lambda) \to \lambda $, where $\wc$ is the operator coefficient for $ \ocal $. With this we see that  $ \phi^3 \Box \phi $ is equivalent to $ \phi^6 $.  One can follow a similar procedure to show that, modulo finite renormalizations, $ ( \Box\phi)^2 $ is also equivalent to $ \phi^6 $. This example shows that, for this toy model, all the $O(1/\Lambda^2)$ effects of {\em any} type of heavy physics are parameterized by a single Wilson coefficient, that of $ \phi^6$.

It is an instructive exercise to show this equivalence perturbatively using Feynman diagrams. As an example we consider the operator
\beq
\ocal' = \phi^3 \left( \Box\phi + m^2 \phi\right)\,,
\eeq
which, according to the above theorem, should be equivalent to $ \ocal=\phi^6 $ with $ a = - \lambda/6 $ in \cref{eq:oo-identity}. The vertices generated by $ \ocal,\,\ocal' $ are
\bal
\ocal:& \quad i  \frac{6!\,c}{\Lambda^2} \,;\mcr
\ocal':& \quad i \frac{3!\,c'}{\Lambda^2} \sum_{i=1}^4 (m^2 - k_i^2)\,;
\end{align}
where $k_i $ is the incoming momentum in the $i$-th leg. Consider next the contribution of these operators to the amplitude for the $2 \phi \to 4 \phi $ process. The contributions from $ \ocal,\, \ocal' $ are generated by diagrams of the form
\beq
\includegraphics[width=3in]{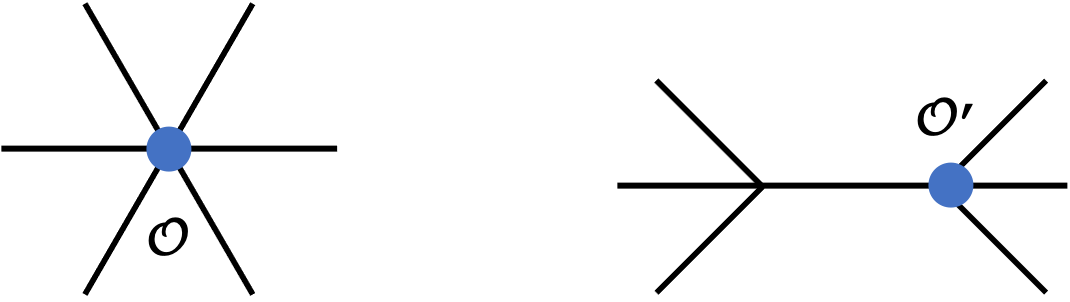}
\eeq
The graph involving $ \ocal $ gives the amplitude
\beq
\acal_\ocal = i \frac{720 c} {\Lambda^2} \,.
\eeq
For $ \ocal' $ they are of two kinds of graphs: those where the factor $ m^2 - k^2  $ corresponds to an external leg, and those where it corresponds to an internal line. The first group of diagrams do not contribute to the cross section since they will vanish when the external propagators are amputated and the (external) momenta are on-shell. The second class of diagrams do contribute to the cross section; moreover, the factor $ m^2 - k^2 $ identically cancels the internal propagator, leaving only a factor of $-i$; finally, there are $\inv{3!}{6\choose4}=20$ ways of attaching 3 of the the `standard' $ \phi^4 $ vertex legs to the external lines. The amplitude is then
\beq
\acal_{\ocal'} = 20 \times (- i \lambda) \times (-i) \times i \frac{6\, c'}{\Lambda^2} + \cdots = - \frac\lambda6 \times i \frac{720 c} {\Lambda^2} + \cdots \,,
\eeq
where the ellipses denote terms that do not contribute to the cross section. Thus
\beq
\acal_{\ocal'} + \frac\lambda6 \acal_\ocal + \cdots =0 \,,
\eeq
which is the same result obtained using the equivalence theorem.

\subsection{Loop and tree generated operators}
\label{sec:tree.loop}

When the new physics is weakly coupled and decoupling the Wilson coefficient for an operator $ \ocal $ is of the form $ c /\Lambda^n$, where $n-4$ is the canonical dimension of $ \ocal $. Depending on the details of the new physics, the coefficient $c$ can be generated at tree level or at one or higher loops (by graphs involving only heavy-particle internal lines), so we have
\beq
c \sim \frac{\prod \text{(coupling constants)}}{(4\pi)^{2L}}\,,
\eeq
where $L$ is the lowest number of loops at which $ \ocal $ is generated. The precise value of the factor generated by the momentum integration can differ from $ \sim 1/(16\pi^2) $ by a numerical constant of $O(1) $, so the above expression should be considered only as an order of magnitude estimate.

The above observation regarding the structure of $c$ is useful in EFT phenomenological applications. As noted previously (Sect. \ref{sec:introduction}) this approach is valid and useful in cases where the new physics is not directly observed, so that its effects are noticed by deviations from the established low-energy theory ({\it e.g.} the Standard Model). The largest deviations can then be expected in processes involving operators with the largest (or to be more accurate, the least suppressed) Wilson coefficients, and these correspond to operators that are generated at tree level.

Unfortunately which operators are generated at tree level depends on the details of the new physics. In contrast, there are operators that are {\em always} generated by loops \cite{Arzt:1994gp,Einhorn:2013kja}~\footnote{At least as long as the underlying theory is assumed to be a local quantum field theory containing scalars, fermions and vectors.}, independently of the details of the new physics. We refer to this last set as ``loop generated (LG) operators''. Operators that are not LG may or may not be generated at tree level, depending on the details of the (unknown) new physics, and we refer to them as ``potentially tree generated (PTG) operators''. Therefore, the set of operators of a given dimension can be separated in two subsets: LG and PTG operators. 

We illustrate these results by considering the dimension 6 operators in the effective theory where the SM describe the low-energy dynamics (abbreviated as SMEFT). To this end assume that the underlying physics is described by a spontaneously broken gauge theory with gauge vectors $V$, scalars $ \vartheta $ and fermions $ \zeta $, that can be separated into light and heavy fields:
\beq
V = \{ A\, \text{(light)};\,\, X\, \text{(heavy)} \} \,, \quad
\vartheta = \{ \phi\, \text{(light)};\,\, \Phi\, \text{(heavy)} \} \,, \quad
\zeta = \{ \psi\, \text{(light)};\,\, \Psi\, \text{(heavy)} \} \,.
\eeq
We denote the group structure constants by $f$ (that we can assume fully antisymmetric), and the generators by $T$; we also denote light gauge indices (associated with the $A$) by $ \li$, and heavy gauge indices (associated with $X$) by $ \he$. We will also assume that the underlying gauge symmetry is broken spontaneously (in one or more steps) leaving no physical Goldstone bosons. Then we make the following observations
\bit
\item The light generators close into an algebra -- symbolically $ [T_\li,\,T_\li] = T_\li $, which implies $ f_{\li\li\he} =0 $. Therefore there are no $AAX$ and $AAAX$ vertices.
\item The $\zeta$-$\zeta$-$V$ couplings are generated by the fermion kinetic terms in the underlying theory, then, since the light (unbroken) generators $ T_\li$ do not mix light and heavy modes, there are no $ \psi$-$\Psi$-$A$ vertices.
\item Similarly, The $\vartheta$-$\vartheta$-$V$ couplings are generated by the boson kinetic terms in the underlying theory, so there are no $\phi$-$\Phi$-$A$ vertices.
\item Only the heavy scalars get a large \vev: $ \vevof\Phi \sim \Lambda $, the vectors $ T_\he \vevof\Phi $ point along the Goldstone directions that are orthogonal to the physical scalars.
\item Finally, we note that the SM quantum number assignments forbid the cubic vertex $ \phi^3 $.
\eit
Summarizing, the forbidden vertices are \cite{Arzt:1994gp}:
\bal
\text{cubic}:&\quad \phi\phi\phi,~A\phi\Phi,~ A\psi\Psi,~AA\phi,~A\phi X,~\phi XX,~AA \Phi,~AX\Phi,~AAX \,,\mcr
\text{quartic}:& \quad AA\phi\Phi,~AAAX\,.
\label{eq:forbidden-vertices}
\end{align}

Next we consider all dimension 6 operators involving SM fields and ask {\it(i)} what type of tree-level graphs can generate them; and {\it(ii)} whether all these graphs involve vertices of the type \cref{eq:forbidden-vertices}. If the answer to the second question is `yes' then the operator in question is of the LG type, otherwise it is of the PTG type. It is then a straightforward exercise \cite{Arzt:1994gp} to show that the following sets are loop generated
\beq
\text{LG}:\quad A^\mu{}_\nu A^\nu{}_\rho A^\rho{}_\mu\,, \quad \phi \phi A_{\mu\nu}A^{\mu\nu}\,, \quad (\bar\psi \sigma_{\mu\nu} \psi) A^{\mu\nu}\,;
\label{eq:LG-ops}
\eeq
where we ignored all gauge indices, and $ A_{\mu\nu} $ denotes the field strength for the $A_\mu $. This results implies that, should the underlying theory be weakly coupled and decoupling, the Wilson coefficients of these three classes of operators will be suppressed by a factor $ 1/(16\pi^2) \sim 0.006 $ and their observable effects will be correspondingly weaker: LG operators may provide striking signals, but the corresponding event rates will be negligible.

As an example, we consider the NP effects on the so-called triple vector-boson vertices (such as NP contributions to the $WW\gamma$ coupling). These correspond to the first class of operators in \cref{eq:LG-ops}, so the deviations from the SM will be of order $ \mw^2/( 4 \pi \Lambda)^2 $. A limit of $0.02 $ \cite{PhysRevD.98.030001} on this Wilson coefficient then implies $ \Lambda > 45 \gev $, which is not useful. In order to probe physics at scales $ \sim 3 \tev $ these effects must be measured to precision $ \sim 5 \times 10^{-6} $.

Operators {\em not} of the types \cref{eq:LG-ops} may or may not be generated at tree-level by the underlying theory. For the SMEFT these include 4-fermion operators of the form $ (\bar\psi_R\gamma_\mu \psi_R)(\bar\psi_R\gamma^\mu \psi_R) $ that are generated at tree-level only if the underlying theory contains a heavy vector boson with appropriate couplings to the $ \psi_R$. Absent such a particle this operator may still be generated, but only at 1 or higher loops. 

We provide below (Sect. \ref{sec:SMEFT}) the full list of dimension-6 SMEFT operators, indicating their PTG or LG character.

\subsection{Naturality considerations}

Radiative corrections can be calculated in effective field theories as in any other quantum field theory. In a loop expansion the observed physical quantities are calculated as a sum of the contributions from tree-level, 1 loop, 2 loops, etc.  processes~\footnote{We assume that there is a tree-level term but the argument is readily expanded to cases where the first non-vanishing contribution occurs at $n$ loops. We will not discuss the summability of this series.}. Symbolically
\beq
\wp_{\tt physical} = \wp_{\tt tree} + \wp_{\tt 1-loop} + \wp_{\tt 2-loops} + \cdots\,,
\eeq
which is useful provided the higher orders do not dominate, in this case the theory is called {\em natural}. It is important to note that though this naturality condition may seem quite reasonable, it does not follow that unnatural theories are mathematically inconsistent, but only that the usual perturbative approach is not useful in such cases. It is important to note that not only theories with small parameters are natural; the standard example is the theory of strong interactions at low energies as described by a chiral Lagrangian \cite{Gasser:1983yg,Manohar:1983md,Georgi:1992dw,Georgi:1985kw}. Here we summarize some of the consequences of imposing the naturality requirement on generic effective field theories

As a first step we consider an effective operator containing $b$ bosonic fields (denoted symbolically by $B$), $f$ fermionic fields $F$ and $d$ derivatives $D$, so that its generic form is
\beq
\ocal \sim D^d B^b F^f\,.
\label{eq:generic.op}
\eeq
The Wilson coefficient for $ \ocal $ will be of the form
\beq
\wc_\ocal \sim \frac{\lambda(b,f)}{\Lambda^{\Delta_\ocal}}\,, \qquad \Delta_\ocal = \text{dim}(\ocal) - 4 = d + b + \frac32f - 4 \,;
\label{eq:wc.for.O}
\eeq
where we allow in $ \lambda $ an explicit dependence on the number of boson and fermion fields (the dependence on $d$ is determined by dimensionality). We will only be interested in obtaining order of magnitude estimates on the coefficients $ \lambda $ for which purpose no further differentiation among operators with the same $b$ and $f$ is needed.

Now, consider an $L$-loop graph that renormalizes $ \ocal $, and which has $ I_b$ internal bosonic lines, $I_f$ internal fermionic lines, and whose vertices are generated by a series of operators $ \{\ocal_v\}$. The naive degree of divergence of such a graph is
\beq 
N_{\tt div} = 4 L - 2 I_b - I_f + \sum_v d_v - d = \sum_v \Delta_{\ocal_v} - \Delta_\ocal\,,
\label{eq:N-div}
\eeq
where the last equality follows from the standard expressions \cite{Itzykson:1980rh} relating the number of internal lines to the number of vertices. 

If we now assume that the effective theory is regularized by an ultraviolet cutoff of order $ \Lambda $ (the limit of applicability of the theory), then the above graph will get $ N_{\tt div} $ powers of $ \Lambda $ from the loop integrations and $ - \Delta_{\ocal_v} $ from the Wilson coefficient for each of the vertices. As a result the total power of $ \Lambda $ is $ -\Delta_\ocal $, the same as in the Wilson coefficient for $ \ocal $ in \cref{eq:wc.for.O}, as required for consistency.

Next we derive the constraints on the couplings $ \lambda(b,f) $ derived from naturality as defined above. The graph under consideration generates a radiative contribution 
\beq
\delta\lambda(b,f) \sim \inv{(16\pi^2)^L} \prod_v \lambda(b_v,f_v)\,,
\eeq
and naturality requires $ \delta \lambda \lesssim \lambda $, with the naturality `limit' being $ \delta\lambda \sim \lambda $. Imagine now that we replace any of the vertex operators $ \ocal_v $ with another one with two additional identical boson fields: $ \ocal_v \to B^2 \ocal_v $. Correspondingly we have a graph contributing to $ \wc_\ocal $ derived from the previous one where these two additional bosons fields are contracted. In the new graph the $ \ocal_v $ coupling is replaced by $ \lambda(b_v+2,f_v) $, and this graph also has an additional loop (from the contraction of the new boson fields), so there is an additional factor of $ 1/(16\pi^2) $. In the naturality limit both this and the old graph give similar contributions to $\wc_\ocal$ whence
\beq
\inv{16\pi^2} \lambda(b_v+2,f_v) \sim \lambda(b_v,f_v) \then \lambda(b,f) = ( 4\pi)^{b-2} \lambda(2,f)\,.
\eeq
Similarly for fermions we find $ \lambda(b,f) \sim ( 4\pi)^{f-2} \lambda(b,2) $. Combining these results, and including the case of purely bosonic and purely fermionic operators we find
\beq
\lambda(b,f) \sim (4\pi)^{N_\ocal} \bar\lambda\,, \qquad N_\ocal = b + f -2\,,
\eeq
so that $ N_\ocal + 2$ is the number of fields (bosonic + fermionic) in $ \ocal $. Using this we obtain the following naturality constraint on the Wilson coefficients
\beq
\wc_\ocal \lesssim \frac{(4\pi)^{N_\ocal} }{\Lambda^{\Delta_\ocal }}\,.
\eeq

Now, if $ N_{\tt div}=0 $ the graph is logarithmically divergent, while if $ N_{\tt div} > 0 $ the graph has a logarithmic {\em sub}divergence, which is of interest because it is the logarithmic divergences that generate the renormalization group (RG) running of the Wilson coefficients. Writing $ \wc_\ocal = c_\ocal/\Lambda^{\Delta_\ocal} $ this logarithmic term, denoted by $\delta_{\tt log} c_\ocal$, is given by
\bal
N_{\tt div} = 0: & \quad \delta_{\tt log} c_\ocal = (4\pi)^{N_\ocal} \times \text{power of} \ln\Lambda \,.\mcr
N_{\tt div} > 0: & \quad \delta_{\tt log} c_\ocal = (4\pi)^{N_\ocal} \times \left( \frac m\Lambda \right)^{N_{\tt div}} \times \text{power of} \ln\Lambda\,;
\end{align}
where $m$ represents a mass in the light theory. Thus, graphs with power divergences ($N_{\tt div} >0$) generate sub-leading contributions to the RG running of the Wilson coefficients; leading effects stem from $ N_{\tt div} =0 $ graphs for which $ \sum_v \Delta_{\ocal_v} = \Delta_\ocal $ ({\it cf.} \cref{eq:N-div}).

All operators have $ \Delta_\ocal \ge0 $ {\em except} those generating super-renormalizable vertices ($b=3,\,d=f=0$) for which $ \Delta_\ocal=-1 $; these  vertices have then dimensional coefficients. If these are of $O(m)$, their effect on the RG running is subdominant, but not if they are $O(\Lambda ) $. In this last case, however, this term also generates a tree-level mass of $ O(\Lambda) $ for at least one of the scalars involved in this cubic interaction (unless the theory is fine tuned); but then such particle would not belong to the low-energy theory: we conclude that, barring fine-tuning, $ O(\Lambda) $ coefficients of super-renormalizable vertices are absent.

The above arguments also imply that the leading RG effects are generated by operators with $ \Delta_\ocal \ge 0 $, and correspond to graphs satisfying $ \sum_v \Delta_{\ocal_v} = \Delta_\ocal $. In particular, the leading RG effects for $ \ocal $ are generated by operators with $ \Delta_{\ocal_v} \le \Delta_{\ocal} $ for each $v$. This is of interest because, for weakly coupled and decoupling NP, we have imposed a hierarchy on the effective Lagrangian determined by the canonical dimension of the operators, and we have argued that experimental precision allows us to ignore all operators with a sufficiently large dimension. This argument then implies that the leading RG running of the Wilson coefficients of the operators we retained is generated also by this subset of operators; there is no RG `feeding down' from operators we ignored.

We will not discuss here the renormalization group running of the Wilson coefficients nor their anomalous dimensions. For the case of the EFT constructed `above' the SM these issues are well studied in the literature \cite{Jenkins:2013zja,Jenkins:2013wua,Alonso:2013hga,Elias-Miro:2014eia,Elias-Miro:2013mua,Elias-Miro:2013gya}.

\bigskip

The above results can be generalized in a convenient way by defining the `index' $s_\ocal $ of an operator \cite{Wudka:1994ny} $ \ocal $ by
\beq
s_\ocal(u) = \Delta_\ocal + \frac{u-4}2 N_\ocal\,; \quad 0 \le u \le 4\,,
\eeq
where we recall that $ \Delta_\ocal +4 $ is the canonical dimension of $ \ocal $, and $ N_\ocal + 2 $ the total number of fields in $ \ocal $. The real parameter $u$ will be discussed below. In terms of this index the naive degree of divergence of an $L$-loop graph \cref{eq:N-div} can be written
\beq
N_{\tt div} = \sum_v s_{\ocal_v} - s_\ocal + (4-u)L\,.
\eeq
As shown above, the leading RG effects on $c_\ocal $ are characterized by $ N_{\tt div} =0,\, \Delta_\ocal \ge 0 $ in which case
\beq
s_\ocal = \sum_v s_{\ocal_v} + ( 4-u)L \ge \sum_v s_{\ocal_v} \ge  s_{\ocal_v}\,,
\eeq
which implies that the leading RG running of $c_\ocal $ is generated by operators of lower or equal index. If we then write
\beq
\leff = \sum_{\text{index}=s} \lcal_s\,,
\eeq
where $ \lcal_s $ denotes all terms with index $s$, then the RG evolution of $ \lcal_s $ is generated by the $ \lcal_{s'}$ with $ s' \le s $. 

We consider three examples; the parameters $b,\,f,\,d$ are defined in \cref{eq:generic.op}.
\ben
\item Assume $u=1$ in a theory with only fermion fields ($b=0$) and where all operators have at least one derivative ($d\ge 1$). In this case $ s = d-1 $ and the generic form of the naturality bound on the Wilson coefficient is
\beq
\wc_\ocal \lesssim\frac{16\pi^2}{ (4\pi)^{2s/3} \Lambda_\psi^{\Delta_\ocal}} \,, \quad \lambda_\psi = \frac\Lambda{(4\pi)^{2/3}}\,;
\eeq
where $ \Delta = d + 3f/2-4$. An example of when this situation is of interest is in a nonlinear realization of supersymmetry \cite{Deser:1977uq,Volkov:1972jx} where the Lagrangian is
\beq
\lcal_{\tt NL\, SUSY} = - \inv{2 \kappa^2} \text{det}A\,, \quad A^a_\mu = \delta^a_\mu + i \kappa^2 \psi \sigma^a \stackrel\leftrightarrow
\partial_\mu \bar\psi\,.
\eeq
When expanded, $\lcal_{\tt NL\, SUSY}$ generates a series of operators of the form \cref{eq:generic.op} with $ b=0,\,f=2d$: $\ocal ~\sim \kappa^{2d-2} \psi^{2d} \partial^d $; the above bound then gives
\beq
\kappa \lesssim \frac{4\pi}{\Lambda^2}\,,
\eeq
which relates the parameter $ \kappa $ to the scale of the physics underlying this model. Alternatively the above implies that one can perform perturbative calculations with this model up to a scale $ \sim \sqrt{4\pi/\kappa} $.

\item Assume $u=2$. Then the index is independent of the number of boson fields in the operator, $ s = d + f/2 -2 $, and
\beq
\wc_\ocal \lesssim \inv{ (4\pi)^{s + 2} \Lambda_\phi^{\Delta_\ocal}} \,, \quad \Lambda_\phi = \frac\Lambda{4\pi}\,;
\eeq
where $ \Delta = d + b+ 3f/2-4$. An interesting case where this applies is that of chiral theories (here considered without fermions, $f=0$, for simplicity). Defining the unitary field $ U = \exp ( i \sigbf \cdot \pibf/f_\pi) $ where $ f_\pi $ is a constant with dimensions of mass, the Lagrangian \cite{Georgi:1985kw} takes the form
\beq
\lcal_{\tt chiral} = - f_\pi^2 \text{tr} \left[ \partial_\mu U^\dagger\, \partial^\mu U \right]  + \bar c_4\up1 \left[ \partial_\mu U^\dagger\, \partial^\mu U \right]^2 + \cdots + \frac{\bar c_{2n}}{f_\pi^{2n-4}} \times \partial^{2n} U^k  + \cdots\,.
\eeq
This then gives $ f_\pi = \Lambda_\phi $ and $ \bar c_{2n}\lesssim (4\pi)^{2-2n} $. It is worth noting that when this formalism is applied to low-energy meson physics \cite{Gasser:1983yg,Georgi:1985kw,Gasser:1982ap}, the experimentally derived constraints on the $ \bar c_{2n} $ coefficients are consistent with this bound.

\item Assume $ u = 4 $, then $s = d+b +(3/2)f - 4$, so $ s+4$ is the canonical dimension of the operator and we find
\beq
\wc_\ocal \lesssim \frac{(4\pi)^{N_\ocal}}{\Lambda^{\Delta_\ocal}}\,.
\eeq
If we use the same arguments for a dimensionless couplings of a renormalizable theory we find that Yukawa couplings are bounded by $ (4\pi)$ while scalar self-interaction couplings are bound by $(4\pi)^2$. These are the  standard perturbativity bounds \cite{Lee:1977eg,Lee:1977yc,Dawson:1988va,Durand:1992wb,Durand:1989zs,Albrecht:2010wf,Argyres:1992nqz,Grzadkowski:2009mj}: larger values of these couplings indicate a breakdown in perturbation theory in the sense that radiative corrections are comparable to tree-level contributions. This, as noted earlier, does not necessarily imply that the theory is then inconsistent.

\een

\section{The SM effective theory up-to operators of dimension 7}
\label{sec:SMEFT}
In this section, we will discuss the effective operators that can be constructed up to mass dimension 7 using Standard Model as known physics, i.e. 
\be
\leff = \lcal_{\tt SM} + \sum_{i,\,n\ge5}  \frac{c_i\up n}{\Lambda^{n-4}} \ocal_i\up n\,,
\label{eq:leff-gen}
\ee
where the $ \ocal_i\up n $ are gauge-invariant local operators with mass dimension $n$ constructed using SM fields and their derivatives, and the $ c_i\up n $ are unknown coefficients. 

The field content therefore consists of $ \phi $, the SM scalar isodoublet;  $l$ and $q$,  \lh\ lepton and quark isodoublets, respectively; $e$ the \rh\ charged leptons, and $u,\,d$  \rh\ up and down-type quarks respectively (we drop the subindices $R,\,L$ to indicate right or left-handed fields in order to simplify the notation). We use $D$ for the covariant derivatives, and denote the $\su3_c,\, \su2_L$ and $ \ui_Y$ gauge fields by $G,\,W$ and $B$, respectively. We shall occasionally consider the possibility that the low-energy spectrum contains right-handed neutrinos, which we then denote by $ \nu $.

\subsection{SM-EFT Operators of dimension five}

In this section we construct all possible dimension five effective operators using SM fields; our goal is purely pedagogical as the result is well known.

A generic operator ({\it cf.} \cref{eq:generic.op}) with $f$ fermions, $b$ bosons\footnote{Here we will use the fact that vector bosons appear only inside covariant derivatives, and field strengths are expressed of commutators of such derivatives; hence, a boson refers to the SM scalar isodoublet.}, and $d$ derivatives has dimension $ (3/2)f+b + d $, where $f$ must be even; then $ f \le 2 $ for a dimension $5$ operator. If $f=0$, then $ b + d = 5 $ which, given the SM field content, one can readily verify correspond to operators that violate gauge and/or Lorentz invariance; it follows that $ f = 2 = b+d $. When $ d=2 $ Lorentz invariance demands an operator of the form $ \bar\psi \gamma_\mu \gamma_\nu D^\mu D^\nu  \psi' $ (up to integration by parts),  the fermions must then have opposite chiralities, whence such operators violate $\su2_L$ gauge invariance. When $ d=1 $ the fermions must have the same chirality; sine such operators contain one scalar isodoublet, then, assuming again Lorentz invariance, they also violate $\su2_L$ gauge invariance. Finally, when $d=0$ the operator contains two scalar isodoublets and two fermions with opposite chiralities; going through all the possible fermion choices one can readily check that the only gauge-invariant operator of this kind is~\footnote{The notation is motivated by the fact that in the unitary gauge $ N = (\smvev/\sqrt{2}) \nu_L + \cdots $.} 
\beq
\ocal\up5_{l\phi} = \bar N N^c\,, \qquad N = \phi^\dagger \varepsilon l\,;
\label{eq:weinberg-op}
\eeq
where the superscript $c$ denotes charge conjugation, $ \varepsilon = - i \tau_2$ (with $ \tau_2 $ the usual Pauli matrix) acting on gauge indices, and we have suppressed family indices. This is the single dimension 5 operator that can be constructed using SM fields, as first noted by Weinberg \cite{Weinberg:1979sa}; it violates lepton number $L$ by two units, so its scale $ \Lambda $ is naturally associated with that of $L$ violation. In the unitary gauge, and after \ssb,
\beq
\ocal\up5_{l\phi} \to  \half \frac{(\smvev + h )^2}{\Lambda} (\bar\nu_L \nu^c_L)\,, \qquad  \phi \to \frac{\smvev + h}{\sqrt{2}} \begin{pmatrix}0\cr1\end{pmatrix}\,;
\eeq
where $ \smvev$ denotes the SM \vev\ and $h$ the Higgs field. The $h$-independent term corresponds to a Majorana mass term for the neutrinos. There is a vast body of work that concerns the origins and possible effects of this effective operator, ranging from neutrino oscillations to the baryon asymmetry of the universe. These considerations, however, fall outside the scope of this review so we limit ourselves to referring the reader to the literature: ~\cite{GonzalezGarcia:2007ib,McKeown:2004yq,Dydak:2004qb,Vergados:2002pv,Altarelli:1999gu,Zuber:1998xe,Avignone:2007fu,Bilenky:1987ty}; we merely remark that the smallness of the neutrino mass is naturally associated with a very large $ \Lambda $.

If light right-handed neutrinos are present, two additional 5 dimensional operators can be constructed,
\beq
\ocal\up5_{\nu \phi} = (\bar\nu \nu^c) (\phi^\dagger\phi) \,, \qquad \ocal\up5_{\nu B} = \bar\nu \sigma^{\mu\nu} \nu^c B_{\mu\nu}
\eeq
The effects and possible origins of these operators are discussed in \cite{Aparici:2009fh,Aparici:2009oua}.

\subsection{SM-EFT Operators of dimension six}

A number of publications have provided lists of dimension six operators without right-handed neutrinos ($\nu$) \cite{Buchmuller:1985jz,Leung:1984ni,Grzadkowski:2010es}. A careful discussion of the equivalence theorem as applied to the SM effective theory is provided in \cite{Grzadkowski:2010es}; this reference also gives a list of operators that cannot be further reduce using the equations of motion. This list is not unique (see {\it e.g.} \cite{Einhorn:2013kja}), but it is irreducible. In the following we adopt this operator basis (commonly called the Warsaw basis), which contains 59 operators (for a single family) that conserve lepton and baryon number, denoted by $L$ and $B$, respectively. For completeness these operators are listed in Tables~\ref{tab:dim-6-1} and ~\ref{tab:dim-6-2}.

\begin{table}[h]
{\footnotesize
\renewcommand{\arraystretch}{1.5}
\begin{tabular}{||c|c||c|c||c|c||}
\hline \hline
\multicolumn{2}{||c||}{$X^3$\ \bf{(LG)} } & 
\multicolumn{2}{|c||}{$\phi^6$~ and~ $\phi^4 D^2$ \bf{(PTG)} } &
\multicolumn{2}{|c||}{$\psi^2\phi^3$\ \bf{(PTG)} }\\
\hline
$\ocal_G$                & $f^{ABC} G_\mu^{A\nu} G_\nu^{B\rho} G_\rho^{C\mu} $ &  
$\ocal_\phi$       & $(\phi^\dag\phi)^3$ &
$\ocal_{e\phi}$           & $(|\phi|^2)(\bar l_p e_r \phi)$\\
$\ocal_{\wt G}$          & $f^{ABC} \wt G_\mu^{A\nu} G_\nu^{B\rho} G_\rho^{C\mu} $ &   
$\ocal_{\phi\Box}$ & $ |\phi|^2   \Box |\phi|^2 $ &
$\ocal_{u\phi}$           & $(|\phi|^2)(\bar q_p u_r \phit)$\\
$\ocal_W$                & $\eps^{IJK} W_\mu^{I\nu} W_\nu^{J\rho} W_\rho^{K\mu}$ &    
$\ocal_{\phi D}$   & $ \left| \phi^\dag D_\mu\phi\right|^2$ &
$\ocal_{d\phi}$           & $(|\phi|^2)(\bar q_p d_r \phi)$\\
$\ocal_{\wt W}$          & $\eps^{IJK} \wt W_\mu^{I\nu} W_\nu^{J\rho} W_\rho^{K\mu}$ &&&&\\    
\hline \hline
\multicolumn{2}{||c||}{$X^2\phi^2$\ \bf{(LG)}} &
\multicolumn{2}{|c||}{$\psi^2 X\phi$\ \bf{(LG)}} &
\multicolumn{2}{|c||}{$\psi^2\phi^2 D$\ \bf{(PTG)} }\\ 
\hline 
$\ocal_{\phi G}$     & $|\phi|^2\, G^A_{\mu\nu} G^{A\mu\nu}$ & 
$\ocal_{eW}$               & $(\bar l_p \sigma^{\mu\nu} e_r) \tau^I \phi W_{\mu\nu}^I$ &
$\ocal_{\phi l}^{(1)}$      & $(\phij)(\bar l_p \gamma^\mu l_r)$\\
$\ocal_{\phi\wt G}$         & $|\phi|^2\, \wt G^A_{\mu\nu} G^{A\mu\nu}$ &  
$\ocal_{eB}$        & $(\bar l_p \sigma^{\mu\nu} e_r) \phi B_{\mu\nu}$ &
$\ocal_{\phi l}^{(3)}$      & $(\phijt)(\bar l_p \tau^I \gamma^\mu l_r)$\\
$\ocal_{\phi W}$     & $|\phi|^2\, W^I_{\mu\nu} W^{I\mu\nu}$ & 
$\ocal_{uG}$        & $(\bar q_p \sigma^{\mu\nu} T^A u_r) \phit\, G_{\mu\nu}^A$ &
$\ocal_{\phi e}$            & $(\phij)(\bar e_p \gamma^\mu e_r)$\\
$\ocal_{\phi\wt W}$         & $|\phi|^2\, \wt W^I_{\mu\nu} W^{I\mu\nu}$ &
$\ocal_{uW}$               & $(\bar q_p \sigma^{\mu\nu} u_r) \tau^I \phit\, W_{\mu\nu}^I$ &
$\ocal_{\phi q}^{(1)}$      & $(\phij)(\bar q_p \gamma^\mu q_r)$\\
$\ocal_{\phi B}$     & $ |\phi|^2\, B_{\mu\nu} B^{\mu\nu}$ &
$\ocal_{uB}$        & $(\bar q_p \sigma^{\mu\nu} u_r) \phit\, B_{\mu\nu}$&
$\ocal_{\phi q}^{(3)}$      & $(\phijt)(\bar q_p \tau^I \gamma^\mu q_r)$\\
$\ocal_{\phi\wt B}$         & $|\phi|^2\, \wt B_{\mu\nu} B^{\mu\nu}$ &
$\ocal_{dG}$        & $(\bar q_p \sigma^{\mu\nu} T^A d_r) \phi\, G_{\mu\nu}^A$ & 
$\ocal_{\phi u}$            & $(\phij)(\bar u_p \gamma^\mu u_r)$\\
$\ocal_{\phi WB}$     & $ \phi^\dag \tau^I \phi\, W^I_{\mu\nu} B^{\mu\nu}$ &
$\ocal_{dW}$               & $(\bar q_p \sigma^{\mu\nu} d_r) \tau^I \phi\, W_{\mu\nu}^I$ &
$\ocal_{\phi d}$            & $(\phij)(\bar d_p \gamma^\mu d_r)$\\
$\ocal_{\phi\wt WB}$ & $\phi^\dag \tau^I \phi\, \wt W^I_{\mu\nu} B^{\mu\nu}$ &
$\ocal_{dB}$        & $(\bar q_p \sigma^{\mu\nu} d_r) \phi\, B_{\mu\nu}$ &
$\ocal_{\phi u d}$   & $i(\phit^\dag D_\mu \phi)(\bar u_p \gamma^\mu d_r)$\\
\hline \hline
\end{tabular}
\caption{Dimension-six operators other than the four-fermion ones ($\tilde{\phi}=i\sigma_2\phi^{*}$, and the subindices $p,\,r$ denote family labels).
\label{tab:dim-6-1}}
} 
\end{table}
\begin{table}[h]
{\footnotesize
\renewcommand{\arraystretch}{1.5}
\begin{tabular}{||c|c||c|c||c|c||}
\hline\hline
\multicolumn{2}{||c||}{$(\bar LL)(\bar LL)$\ \bf{(PTG)}} & 
\multicolumn{2}{|c||}{$(\bar RR)(\bar RR)$\ \bf{(PTG)}} &
\multicolumn{2}{|c||}{$(\bar LL)(\bar RR\ \bf{(PTG)})$}\\
\hline
$\ocal_{ll}$        & $(\bar l_p \gamma_\mu l_r)(\bar l_s \gamma^\mu l_t)$ &
$\ocal_{ee}$		& $(\bar e_p \gamma_\mu e_r)(\bar e_s \gamma^\mu e_t)$ &
$\ocal_{le}$		& $(\bar l_p \gamma_\mu l_r)(\bar e_s \gamma^\mu e_t)$ \\
$\ocal_{qq}^{(1)}$  & $(\bar q_p \gamma_\mu q_r)(\bar q_s \gamma^\mu q_t)$ &
$\ocal_{uu}$        & $(\bar u_p \gamma_\mu u_r)(\bar u_s \gamma^\mu u_t)$ &
$\ocal_{lu}$		& $(\bar l_p \gamma_\mu l_r)(\bar u_s \gamma^\mu u_t)$ \\
$\ocal_{qq}^{(3)}$  & $(\bar q_p \gamma_\mu \tau^I q_r)(\bar q_s \gamma^\mu \tau^I q_t)$ &
$\ocal_{dd}$        & $(\bar d_p \gamma_\mu d_r)(\bar d_s \gamma^\mu d_t)$ &
$\ocal_{ld}$		& $(\bar l_p \gamma_\mu l_r)(\bar d_s \gamma^\mu d_t)$ \\
$\ocal_{lq}^{(1)}$	& $(\bar l_p \gamma_\mu l_r)(\bar q_s \gamma^\mu q_t)$ &
$\ocal_{eu}$		& $(\bar e_p \gamma_\mu e_r)(\bar u_s \gamma^\mu u_t)$ &
$\ocal_{qe}$		& $(\bar q_p \gamma_\mu q_r)(\bar e_s \gamma^\mu e_t)$ \\
$\ocal_{lq}^{(3)}$	& $(\bar l_p \gamma_\mu \tau^I l_r)(\bar q_s \gamma^\mu \tau^I q_t)$ &
$\ocal_{ed}$		& $(\bar e_p \gamma_\mu e_r)(\bar d_s\gamma^\mu d_t)$ &
$\ocal_{qu}^{(1)}$	& $(\bar q_p \gamma_\mu q_r)(\bar u_s \gamma^\mu u_t)$ \\ 
&& 
$\ocal_{ud}^{(1)}$	& $(\bar u_p \gamma_\mu u_r)(\bar d_s \gamma^\mu d_t)$ &
$\ocal_{qu}^{(8)}$	& $(\bar q_p \gamma_\mu T^A q_r)(\bar u_s \gamma^\mu T^A u_t)$ \\ 
&& 
$\ocal_{ud}^{(8)}$	& $(\bar u_p \gamma_\mu T^A u_r)(\bar d_s \gamma^\mu T^A d_t)$ &
$\ocal_{qd}^{(1)}$	& $(\bar q_p \gamma_\mu q_r)(\bar d_s \gamma^\mu d_t)$ \\
&&&&
$\ocal_{qd}^{(8)}$ & $(\bar q_p \gamma_\mu T^A q_r)(\bar d_s \gamma^\mu T^A d_t)$\\
\hline\hline 
\multicolumn{2}{||c||}{$(\bar LR)(\bar RL)$ and $(\bar L R)(\bar L R)$\ \bf{(PTG)}}\\
\cline{1-2}
$\ocal_{ledq}$ & $(\bar l_p^j e_r)(\bar d_s q_t^j)$ \\
$\ocal_{quqd}^{(1)}$ & $(\bar q_p^j u_r) \eps_{jk} (\bar q_s^k d_t)$ \\
$\ocal_{quqd}^{(8)}$ & $(\bar q_p^j T^A u_r) \eps_{jk} (\bar q_s^k T^A d_t)$ \\
$\ocal_{lequ}^{(1)}$ & $(\bar l_p^j e_r) \eps_{jk} (\bar q_s^k u_t)$ \\
$\ocal_{lequ}^{(3)}$ & $(\bar l_p^j \sigma_{\mu\nu} e_r) \eps_{jk} (\bar q_s^k \sigma^{\mu\nu} u_t)$ \\
\cline{1-2}
\cline{1-2}
\end{tabular}
\caption{Four-fermion operators conserving baryon number (the subindices $p,\,r,\,s$ and $t$ denote family labels). 
\label{tab:dim-6-2}}
} 
\end{table}

When right-handed neutrinos, denoted by $ \nu $ are added to the low-energy theory, there are additional $B$ and $L$ conserving operators (table \ref{tab:dim-6-3}\footnote{The basis of operators presented has not been reduced by use of the equivalence theorem ({\it cf.} section \ref{sec:equiv.thm}); for a reduced basis see \cite{Liao:2016qyd}.} ). Operators violating baryon number (with and without $ \nu $) are listed in table \ref{tab:dim-6-4}. 

\begin{table}[h]
{\footnotesize
\renewcommand{\arraystretch}{1.5}
\begin{tabular}{||c|c||c|c||c|c||}
\hline
$\ocal_{d u \nu e} $&$ (\bar d \gamma^\mu u)(\bar \nu \gamma^\mu e)              $&$\ocal_{q u \nu l} $&$ (\bar q u)(\bar \nu l)                    $&$\ocal_{\nu\phi }     $&$ (\phi^\dagger \phi) (\bar l \nu\tilde\phi)                   $\cr
$\ocal_{u \nu}     $&$ (\bar u \gamma_\mu u)(\bar \nu \gamma^\mu \nu)            $&$\ocal_{q \nu}     $&$ |\bar q \nu|^2                            $&$\ocal_{ \phi \nu   } $&$ i (\phi^\dagger D_\mu \phi) (\bar \nu \gamma^\mu \nu)          $\cr
$\ocal_{d \nu}     $&$ (\bar d \gamma_\mu d)(\bar \nu \gamma^\mu \nu)            $&$\ocal_{l \nu}     $&$ |\bar l \nu|^2                            $&$\ocal_{ D\nu       } $&$ (\bar l D_\mu \nu) D^\mu \tilde \phi                         $\cr
$\ocal_{e \nu}     $&$ (\bar e \gamma_\mu e)(\bar \nu \gamma^\mu \nu)            $&$\ocal_{l e \nu}   $&$ (\bar l \nu)\varepsilon (\bar l e)        $&$\ocal_{ \bar D \nu } $&$ (D_\mu \bar l \nu) D^\mu \tilde \phi                         $\cr
$\ocal_{\nu}       $&$ \half (\bar \nu \gamma_\mu \nu) (\bar \nu \gamma^\mu \nu) $&$\ocal_{q d \nu}   $&$ (\bar l \nu)\varepsilon (\bar q d)        $&$\ocal_{ \nu W      } $&$ (\bar l \sigma^{\mu\nu} \tau^I \nu) \tilde \phi W_{\mu\nu}^I $\cr
$\ocal_{q \nu}     $&$ (\bar q \gamma_\mu q)(\bar \nu \gamma^\mu \nu)            $&$\ocal_{l d q \nu} $&$ (\bar q \nu)\varepsilon (\bar l d)        $&$\ocal_{ \nu B      } $&$ (\bar l \sigma^{\mu\nu} \nu) \tilde \phi B_{\mu\nu}          $\cr
$\ocal_{\ell \nu}  $&$ (\bar\ell \gamma_\mu \ell)(\bar \nu \gamma^\mu \nu)       $&$\ocal_{\nu\phi }  $&$ (\phi^\dagger\phi) (\bar l \nu\tilde\phi) $&$\ocal_{\phi\phi\nu } $&$ i (\bar\nu \gamma^\mu e)(\phi^T \varepsilon D_\mu \phi)        $\cr \hline
\end{tabular}
\caption{Dimension 6 four-fermion operators conserving lepton and baryon number containing right-handed neutrinos $ \nu $ (family indices are omitted). \label{tab:dim-6-3}}
} 
\end{table}

\begin{table}[h]
{\footnotesize
\renewcommand{\arraystretch}{1.5}
\begin{tabular}{||c|c||c|c||c|c||}
\hline
\multicolumn{2}{||c||}{$\Delta B = 0,\,\Delta L=4$} & \multicolumn{4}{c||}{$\Delta B = \Delta L = 1$ } \cr \hline
$\ocal_\nu $ & $ (\bar\nu \nu^c)^2 $ & $ \ocal_{q d \nu} $ & $ (\bar q q^c)(\bar d \nu^c) $ & $ \ocal_{q l    } $ & $ (\bar q q^c)(\bar q  l^c) $ \cr
             &                       & $ \ocal_{u \nu d} $ & $ (\bar u \nu^c)(\bar d d^c) $ & $ \ocal_{q u e  } $ & $ (\bar q q^c)(\bar u e^c) $ \cr
             &                       & $ \ocal_{u d \nu} $ & $ (\bar u d^c)(\bar d \nu^c) $ & $ \ocal_{q e u  } $ & $ (\bar q \gamma^\mu u^c)(\bar q \gamma_\mu e^c)  $ \cr
             &                       &                     &                                & $ \ocal_{q u d l} $ & $ (\bar q \gamma^\mu u^c)(\bar d \gamma_\mu  l^c) $ \cr
             &                       &                     &                                & $ \ocal_{q l d u} $ & $ (\bar q l^c)(\bar d u^c) $ \cr
             &                       &                     &                                & $ \ocal_{u d e  } $ & $ (\bar u u^c)(\bar d e^c) $ \cr
             &                       &                     &                                & $ \ocal_{u e d u} $ & $ (\bar u e^c)(\bar d u^c) $ \cr\hline
\end{tabular}
\caption{Dimension 6 four-fermion operators violating lepton and/or baryon number, with and without right-handed neutrinos $ \nu $ (family labels omitted). \label{tab:dim-6-4}}
} 
\end{table}

\subsection{SM-EFT operators of dimension $\ge7$ and other developments}

There are several publications listing the effective operators of dimension 7 \cite{Bhattacharya:2015vja,Liao:2020zyx,Liao:2016hru}, dimension 8 \cite{Li:2020gnx} and dimension 9 \cite{Liao:2020jmn}. There are also studies on the general structure of the EFT bases \cite{Henning:2017fpj} and software is available for generating bases of arbitrary dimension \cite{Henning:2015alf}. A Hilbert-series method for constructing effective operators is discussed in \cite{Anisha:2019nzx}. Practical applications of these results are limited as the effects are suppressed by increasingly higher values of the new physics scale $ \Lambda $. As an example we present the list of dimension 7 operators in appendix \ref{sec:ap-A}, here we only remark that all such operators violate lepton and/or baryon number conservation.

We close this section by mentioning a few other approaches to probe physics beyond the SM within the EFT approach. References \cite{Bakshi:2020eyg} and \cite{deBlas:2017xtg} provide classifications of EFT operators in relation to classes of UV complete models and discuss their relevance to specific experimental observables; the  Mathematica based package {\tt CoDEx} \cite{Bakshi:2018ics} has been developed to compute the  Wilson coefficients of an effective operator generated by a UV complete model. The low energy effective theory below the electroweak symmetry breaking scale has also been studied, see {\it e.g.} \cite{Chakrabortty:2020mbc}.

\section{Observability of SMEFT effects}
\label{sec:SMEFT-obs}

We now turn to the application of EFT developed in the previous sections to specific high-energy processes. There is an enormous number of examples of such applications, and it is not our intent to provide a thorough overview, but limit ourselves to a few illustrative cases.

\subsection{Example 1: Effects of EFT in determining the top-Higgs Yukawa coupling}

With the discovery of the Higgs scalar at the LHC \cite{Chatrchyan:2012ufa,Aad:2012tfa} the full particle content of the Standard Model was confirmed. So far, the measured properties of this scalar particle favor those expected from the SM Higgs (an isodoublet with $Y=1/2$), yet accurate confirmation is still lacking (though progress continues); for a detailed review see Ref. \cite{Zyla:2020zbs}. Effective operators  have played a key role in estimating the possible effects of physics beyond the SM in Higgs related observable(s). The relevant operators ({\it cf.} table \ref{tab:dim-6-1}) are~\footnote{The operator $|\phi|^2|D^\mu \phi|^2$ is redundant \cite{Grzadkowski:2010es} and is therefore not included.}
\beq
\begin{array}{lccccc}
\text{PTG:} \quad & \ocal_{\phi\Box}		& \quad \ocal_\phi 				& \quad \ocal_{u\phi} 			& \quad \ocal_{d\phi} & \quad \ocal_{e\phi}  \cr
\text{LG (CP conserving)}:\quad& \ocal_{\phi B} 			& \quad \ocal_{\phi W} 			& \quad \ocal_{\phi G} 			& & \cr
\text{LG (CP violating)}:\quad& \ocal_{\phi \tilde B} 	& \quad \ocal_{\phi \tilde W} 	& \quad \ocal_{\phi \tilde G} 	& & 
\end{array}
\eeq

\begin{figure}[htb]
$$
\includegraphics[height=5.0cm]{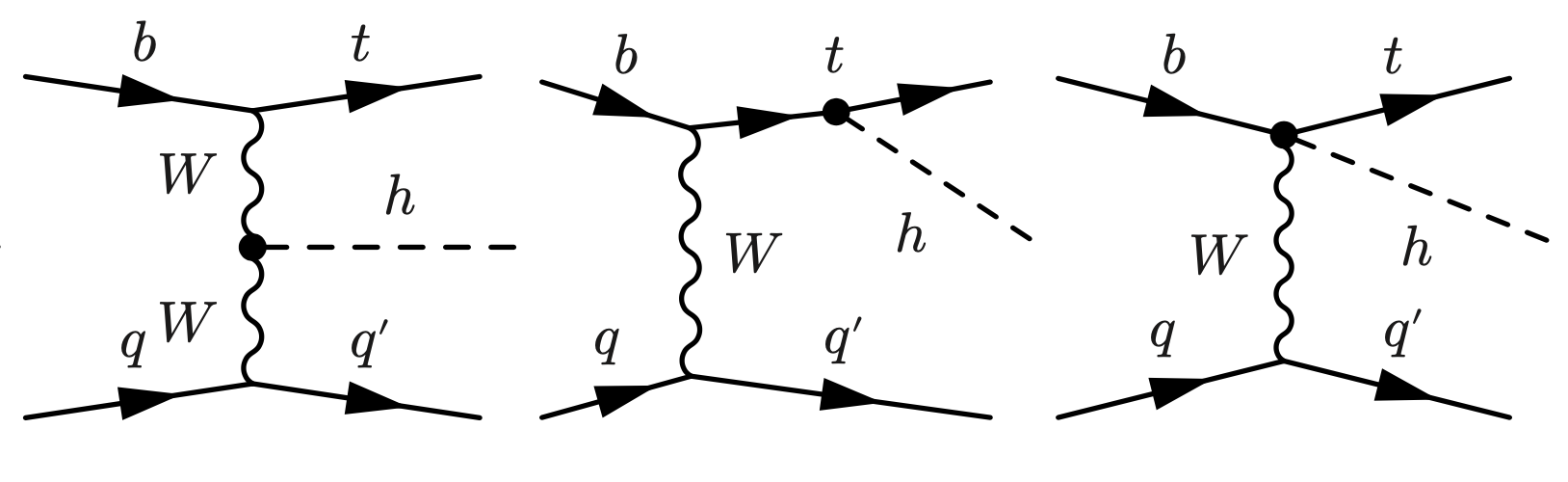}
$$
\caption{Diagrams contributing $tH+$light jet production at the LHC. The two graphs on the left receive SM contributions, the one on the right is an example of a pure EFT  contribution from \cref{eq:eft-thb} that can interfere with the SM amplitude.}
\label{fig:fd-2}
\end {figure}

As an interesting example, Higgs-top Yukawa coupling $ y_t$ is not still well measured, and even its sign is unknown \cite{Zyla:2020zbs} (relative to that of the $WWh$ coupling). The main channels through which the Higgs-top Yukawa coupling can be accessed are $ht\bar{t}$ production, or $ h \to \gamma \gamma $ decay. However, $ht\bar{t}$ production receives its main contributions from terms not involving $ y_t $ and also has a significant NLO and higher order QCD corrections \cite{Sirunyan:2020sum}; while the Higgs to photon decay process is suppressed since it generated only at one loop. Therefore, single top production in association with Higgs, see Fig.~\ref{fig:fd-2}, proves to be an important channel where the Higgs top Yukawa coupling can be also probed. The two contributing diagrams interfere constructively or destructively depending on their relative phase, and since they have similar magnitudes, this process is particularly sensitive to the sign of $ y_t$ \cite{Biswas:2012bd}. Moreover, the SM predicts the interference will be destructive, suppressing SM effects and making this reaction particularly sensitive to new physics effects.

Assuming  $y_t$ is a free parameter to be determined, and assuming the standard $WWh$ coupling, it was shown in ref. \cite{Sirunyan:2018lzm} that LHC data disfavors values $ y_t <-0.9 $ (in our conventions $y_t=1$ in the SM). However, new physics effects may manifest themselves not only in a non-standard value for $ y_t$, and the situation becomes more involved once all other EFT contributions~\footnote{A non-standard value of $ y_t$ corresponds to a contribution from the operator $ \ocal_{u\phi} $ when $ u = t$: $ \delta y_t = - \wc_{t\phi} \smvev^2/\Lambda^2$.} to the process are included. For example, the operator
\beq
\mathcal{O}_{\phi ud}=i (\tilde{\phi}^{\dagger}D_\mu \phi)(\bar{t}\gamma^\mu b)
\label{eq:eft-thb}
\eeq
(see table \ref{tab:dim-6-1}; we replaced $ u_3 = t,\, d_3 = b$ for the \rh\ 3$^{\text{rd}}$ generation quarks) will contribute to the single top $thj$ production ($j$ represents a light-quark jet) and interferes with the SM amplitude, as shown in Fig. \ref{fig:fd-2}. Note that there are other operators that can also contribute ({\it cf.} tables \ref{tab:dim-6-1} ,\ref{tab:dim-6-2}):  $ \ocal_{dW},\,\ocal_{uW} $ and $ \ocal_{\phi W} $ that are LG and whose contributions are suppressed by a factor $\sim 1/16\pi^2$, as well as $ \ocal_{u\phi},\, \ocal_{\phi\Box} $ and $\ocal\up3_{\phi q} $ and several 4-fermion operators, that are PTG. 

A full investigation of the top Yukawa coupling, allowing for non-SM effects must therefore consider the possible effects of these higher-dimensional operators. A partial analysis in this direction, assuming $y_t$ has its SM value was published in \cite{Degrande:2018fog}. The authors are not aware of a complete analysis where no {\it a-priori} assumptions are made.

SMEFT has been used to probe other properties of Higgs boson, such as its coupling to the gauge bosons \cite{Kumar:2015eea}, the CP properties of Higgs \cite{Ananthanarayan:2013cia}, and Higgs self coupling \cite{Gupta:2013zza, Kumar:2019bmk}. The role of modified kinematics \cite{Banerjee:2013apa}, event ratios \cite{Banerjee:2015bla}, and differential SMEFT \cite{Banerjee:2019twi} have been pursued extensively to constrain NP effects in Higgs interactions.

\subsection{Example 2: Same sign dilepton events generated by a dimension 7 operator}

In general, the effects of operators with dimension $ \ge7 $ are  difficult to observe because of their small couplings $ \propto 1/\Lambda^3 $. However, all  dimension 7 operators operators violate $B-L$ (see appendix \ref{sec:ap-A}) and so can generate signatures that have negligible SM backgrounds, improving their observability. In this section we will consider one such case.

In the SM there are a variety of processes that contain two same-sign charged leptons in the final state; however, because of lepton-number conservation, these leptons must be accompanied by neutrinos leading to a significant amount of `missing' energy. In contrast, $L$-violating dimension-7 operators can generate the same pair of same-sign charged leptons, but now without neutrinos or missing energy. Thus the absence of missing energy can be used to significantly suppress the SM contribution, making this a good process in which to look for new physics.

As a specific example of dimension 7 effective operators that can contribute to the same-sign dilepton collider signal at LHC \cite{Bhattacharya:2015vja} we consider the  PTG operators
\beq
\ocal_1=(\lcb \eps D^\mu \phi) (\ell \eps D_\mu \phi)\,; \qquad \ocal_3=( \lcb \eps  D_\mu \ell)( \phi \eps D^\mu \phi)\,,
\eeq 
({\it cf.} \cref{eq:PTG-nonu}). It is readily seen that in unitary gauge they both contain the same lepton-number violating vertex involving two 
$W$ gauge bosons and two left-handed charged leptons:
\beq
( \lcb \eps  D_\mu \ell)( \phi \eps D^\mu \phi) ,~ (\lcb \eps D^\mu \phi) (\ell \eps D_\mu \phi) \supset \mw^2  \left(1 + \frac h\smvev \right)^2  {W^+}^2 (e_L^T Ce_L) \,.
\label{op:ug}
\eeq
Note that $ \ocal_{1,\,3} $ violate lepton number by two units.

We define the linear combination  
\beq
\ocal_{\ell\ell} = f_1 \ocal_1 + f_3 \ocal_3\,,
\label{eq:oll}
\eeq
and consider its effect on  same-sign dilepton production at the LHC, that is, in the process $pp\rightarrow \ell \ell j j$, where $\ell=e, \mu$ have the {\em same} sign, and the $j$ denote light-quark jets. Since this reaction violates lepton number there is no (perturbative) SM contribution and the cross section will scale as $ \Lambda^{-6} $.

It is worth noting that though there are no contributions from dimension 6 operators, there are contributions from diagrams containing two insertions of the dimension 5 operator \cref{eq:weinberg-op}. We do not include these because this same operator also generates Majorana masses for the neutrinos, and so the corresponding graphs are of order $ (m_\nu/\smvev)^2 $, and therefore negligible. We also remark that $ \ocal_{\ell\ell}$ contributes to neutrinoless double beta decay  \cite{delAguila:2012nu} and the experimental limits imply $ \Lambda/f_{1,3} > 7.5 \, \tev $, assuming no cancellations.

\begin{figure}[thb]
$$
\begin{array}{ccc}
\vspace{-.1in}
{\includegraphics[height=3cm]{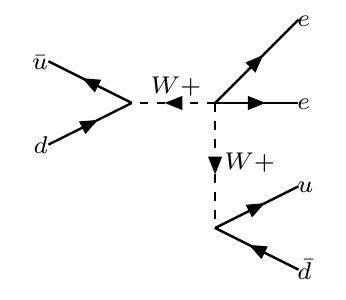}}&
{\includegraphics[height=3cm]{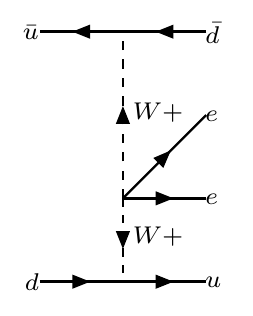}}&
{\includegraphics[height=3cm]{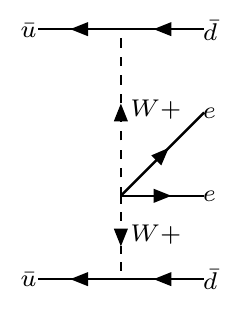}} \cr
(a) & (b) & (c) \cr
&&\cr
\vspace{-.1in}
{\includegraphics[height=3cm]{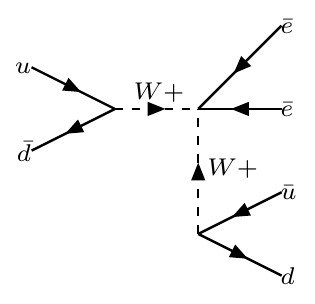}}&
{\includegraphics[height=3cm]{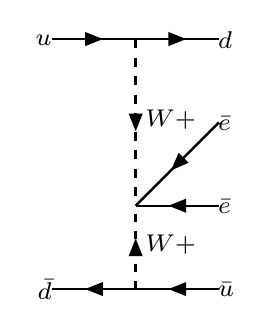}}&
{\includegraphics[height=3cm]{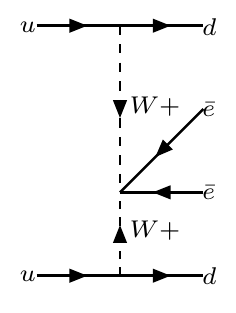}}\cr
(d) & (e) & (f)
\end{array}
$$
\caption{Leading Feynman diagrams that contribute to $pp \to \ell\ell jj$ at the LHC; the $eeWW$ vertex \cref{op:ug} is generated by  $\ocal_{\ell\ell}$ in \cref{eq:oll}.}
\label{fig:fd12}
\end{figure}

In Fig.~(\ref{fig:fd12}) we show the dominant parton level Feynman diagrams that contribute to the  $\ell\ell jj$ final state generated by this operator. The calculation of the cross section $ \sigma(pp\rightarrow \ell \ell j j)$ is straightforward and was carried out using {\tt FeynRules} \cite{Christensen:2008py} and {\tt MadGraph} \cite{Alwall:2011uj}.

A detailed analysis for such a process and the SM background (mainly from $t\bar{t}$ production) is performed in \cite{Bhattacharya:2015vja}; the absence of a signal requires $ \Lambda > 327 \, \gev $, which is the current collider limit for the scale of the new physics that can generate $ \ocal_{\ell\ell} $. One can also obtain the reach of the high-luminosity LHC for a CM energy of $14 \, \tev $ and an estimated integrated luminosity of $ 3000 \, \text{fb}^{-1} $: in this case a signal of 5 standard deviations above the SM (reducible) background would be seen provided $ \Lambda \le 387\,\gev$.

Though the above limits are modest compared to the ones obtained from neutrinoless double beta decay, we highlight this calculation because it illustrates an important consideration in the application of the effective theory. As emphasized in Sect. \ref{sec:lessons-learned}, the formalism is not applicable if the energies involved are comparable or larger than $ \Lambda $, and given the above limits on this scale and the large CM energy of the LHC, it is important to verify that the calculations are in fact reliable. For the case at hand the question is whether the heavy particle that generates $ \ocal_{\ell \ell } $ carries an energy comparable to $ \Lambda $. To investigate this we display in Fig.~(\ref{fig:PTG_graphs}) the types of heavy physics that could generate the operators in question at tree level; we see that there are two possible cases: 
 
\begin{figure}[thb]
\vspace{1cm}
$$
\includegraphics[height=3cm]{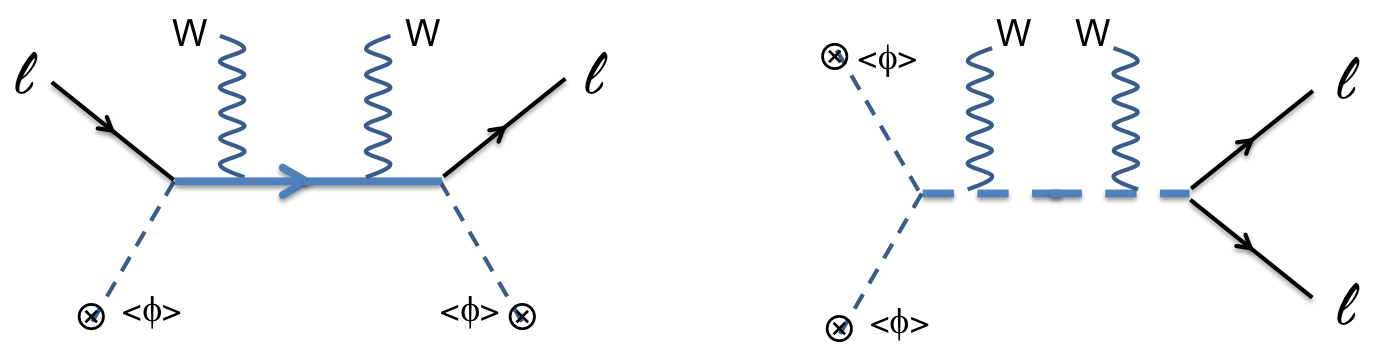}
$$
\caption{Tree-level graphs that can generate the operators $ \ocal_1$ and $ \ocal_3 $ by the exchange of a heavy fermion (left) or scalar (right), 
both isotriplets with unit hypercharge (denoted by the thick internal lines).}
\label{fig:PTG_graphs}
\end{figure}

\ben

\item When the heavy particles carry a momentum equal to the lepton-$W$ invariant mass (Fig.~\ref{fig:PTG_graphs}, left), in which case the heavy particle  will be a heavy fermion isotriplet or isosinglet of zero hypercharge, or,

\item When the heavy particles carry a momentum equal to the invariant dilepton mass (Fig.~\ref{fig:PTG_graphs}, right), in which case the heavy particle is a boson isotriplet of unit hypercharge.
\een

A careful analysis from collider simulation shows that when $ \Lambda \sim 400 ~\gev$ case 2 is problematic, as the momentum of one heavy boson is $ > \Lambda $ in a large region of phase space as revealed from the invariant mass distribution from dilepton final state. Case 1, however, is different: the lepton-$W$ invariant mass peaks at an energy $ \lesssim 200~\gev $, significantly below the limit on $ \Lambda $ (see \cite{Bhattacharya:2015vja}). It follows immediately that the limits quoted above are applicable for the case where the new heavy physics corresponds to the same fermions associated with Type III seesaw mechanisms for neutrino mass generation.

\subsection{Example 3: Discrete symmetries and flavor changing processes}

\begin{figure}[htb]
$$
\includegraphics[height=4cm]{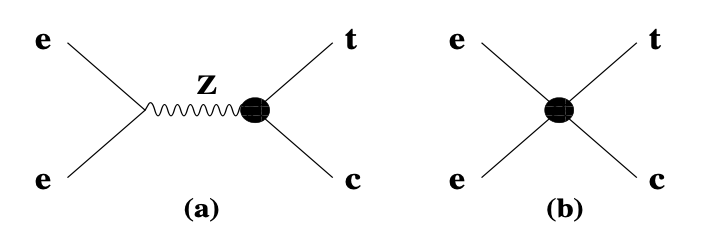}
$$
\caption{Feynman diagrams that give rise to $ e^+ e^- \to t \bar c $ in the presence of effective $Ztc$ and $tcee$ couplings.}
\label{fig:tc-fig1}
\end {figure}

As a last example of the applications of the SMEFT to high energy phenomenology we summarize the results of the studies of $tc$ production at an $e^+ e^-$ collider \cite{BarShalom:1999iy,Han:1998yr}. This reaction has a very small SM contribution that occurs at 1-loop and is GIM~\cite{Glashow:1970gm} suppressed, so it provides a sensitive probe of possible flavor-changing effects in physics beyond the SM. Ignoring the SM contributions, the relevant graphs are those in Fig. \ref{fig:tc-fig1}. There are other related studies that are also of interest such as $ e^+ e^- \to Z t \bar c $ which is sensitive to a flavor-changing $ htc$ vertex, and the $W$-fusion reaction $ e^+ e^- \to \nu_e \bar\nu_e W^+ W^- \to \nu_e\bar\nu_e t \bar c $, sensitive to the $Ztc$, $htc$ and $Wtd$ effective couplings; details can be found in \cite{BarShalom:1999iy}. 

Referring to tables \ref{tab:dim-6-1} and \ref{tab:dim-6-2} the operators that contribute to the process are
\bal
Ztc \text{ vertex}&:\qquad \ocal_{\phi q}\up{1,3}\,, \quad \ocal_{\phi u} \mcr
eetc \text{ vertex}&:\qquad \ocal_{lq}\up{1,3}\,, \quad \ocal_{eu}\,, \quad \ocal_{lu}\,, \quad \ocal_{qe}\,, \quad \ocal\up{1,3}_{lequ}\,.
\end{align}
Using these the relevant Feynman rules are obtained from
\bal
\lcal_{\tt Z t c} &= g { v^2 \over \Lambda^2 } \bar t \gamma_\mu  \left(a_L L +a_R R \right) c + \text{H.c.} \mcr
& a_L = \frac{1}{4 \cw} \left( c_{\phi q}^{(1)} - c_{\phi q}^{(3)} \right) ~,\qquad a_R = \frac{1}{4 \cw} c_{\phi u} ~,
\end{align} 
where $\cw= \cos\theta_{\tt w}$ and $\theta_{\tt w}$ is the weak mixing angle, and
\bal
\lcal_{\tt eetc} 
&= \inv{\Lambda^2} \left\{ \sum_{i,j=L,R}  V_{ij} \left({\bar e} \gamma_\mu P_i e \right)  \left( \bar t \gamma^\mu P_j c \right)  +S_{RR} \left( {\bar e} P_R e \right)  \left( \bar t P_R c \right) + T_{RR} \left( {\bar e} \sigma_{\mu \nu} P_R e \right) \left( \bar t \sigma_{\mu \nu} P_R c \right) + \text{H.c.} \right\} \,;\mcr
& V_{LL} = c_{lq}\up1 - c_{lq}\up3\,, \quad V_{RR} = c_{eu}\,, \quad V_{LR} = c_{lu}\,, \quad V_{RL} = c_{qe} \,,
\quad S_{RR} =  - c_{lequ}\up1\,, \quad T_{RR} = - c_{lequ}\up3\,.
\end{align}

\begin{figure}[htb]
$$\includegraphics[width=3in]{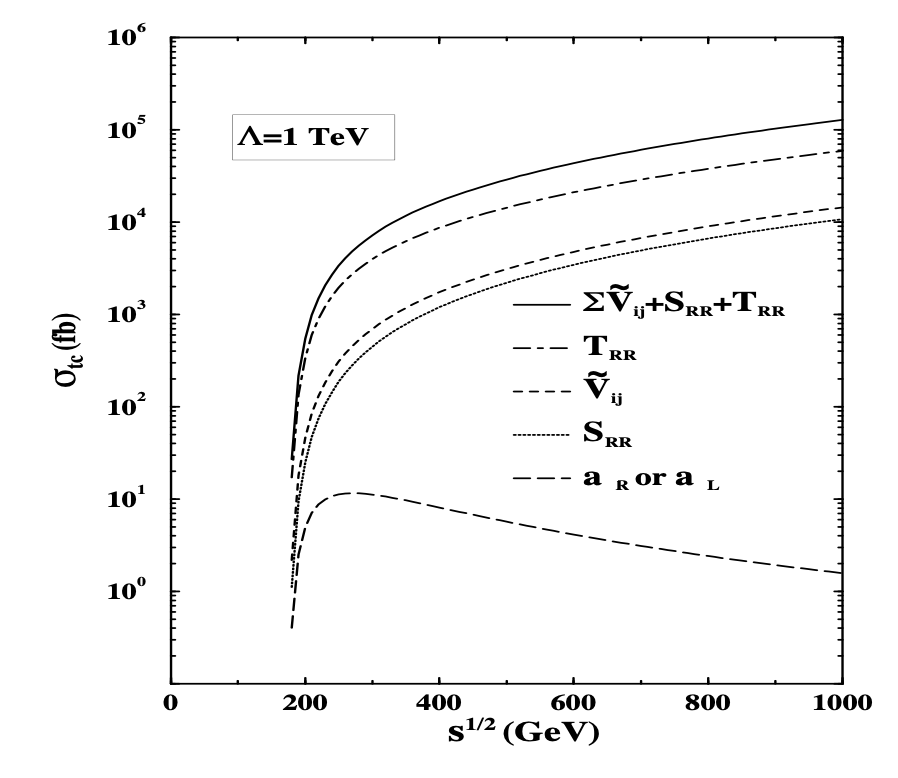} $$
\caption{ Plot of  $\sigma_{tc}$ (in $fb$) as a function of the c.m. energy $\sqrt s$ when $ \Lambda = 1$ TeV. The following cases are shown: all four-Fermi couplings are non-zero and equal 1, i.e., $ V_{ij}= S_{RR}= T_{RR}=1 = a_{L,R}$ (solid line), only $ T_{RR}=1$ (dot-dashed line), only one of the vector couplings $ V_{ij}$ equals 1 (dashed line), only $ S_{RR}=1$ (dotted line) and either $ a_L=1$ or $ a_R=1$ with the four-Fermi couplings set to zero (long-dashed line).}
\label{fig:tc}
\end{figure}

Assuming for simplicity that all coefficients are real, the total unpolarized cross section, plotted in Fig. \ref{fig:tc}, is given by
\beq
\sigma_{tc} = \sigma_0 \left[ (2 + m_t^2/s)\sum_{ij=L,R} \tilde V_{ij}^2  + \frac32 S_{RR}^2  + 8 (1+ 2 m_t^2/s) T_{RR}^2  \right]\,,
\eeq
where 
\beq
\sigma_0 =\frac{s}{16 \pi \Lambda^4} \left(1- \frac{m_t^2}s \right)^2 \,,\quad \tilde V_{ij} = V_{ij} + 4 b_i a_j \frac{\mw \mz}{s-\mz^2},
\eeq
where $ a_{L,R}$ were defined above, and $ b_L=-1/2+ \sin^2\theta_{\tt w},\, b_R = \sin^2\theta_{\tt w} $ are the couplings of a $Z$-boson to a left or a right-handed charged lepton, respectively. It is worth noting that, there being no significant SM contribution, the cross section scales as $ 1/\Lambda^4 $; also, $ \sigma \sim s/\Lambda^4$ for  $s\gg m^2_t $, but this expression is reliable only when $ \sqrt{s} $ is smaller than $ \Lambda $.

Though full phenomenological analysis of the observability of these effects can now be carried out, this lies beyond the scope of this review, and we refer the reader to the literature  \cite{BarShalom:1999iy,Han:1998yr} for details.

\section{EFT for Dark Matter physics}
\label{sec:DM}

\subsection{Dark Matter physics}
There are strong cosmological and astrophysical indications \cite{Zwicky:1933gu,Zwicky:1937zza,Rubin:1970zza} of the presence of a large amount of matter in the universe which has, apparently, very weak (if any) interactions with electromagnetic radiation; accordingly it is referred to as dark matter (DM). The current prevalent belief~\footnote{Though the possibility of black hole DM \cite{Bird:2016dcv} is under investigation, as are other more exotic alternatives \cite{Terazawa:2019ckf}.} is that DM has a particle-physics origin~\cite{Peebles:1994xt}, yet no evidence of DM has been found at either direct detection or collider experiments (although there are suggestive indications, most recently by the XENON1T collaboration \cite{Aprile:2018dbl}, and earlier by the DAMA-LIBRA collaboration \cite{Bernabei:2008yi}). The hypothesis that DM is composed by one or more fundamental particles not present in the SM is currently strongly favored because it can naturally explain the observed relic density \cite{Spergel:2006hy,Jarosik:2010iu,Hinshaw:2012aka,Ade:2013zuv,Aghanim:2018eyx} and lead to the observed properties in the large scale structure of the universe \cite{Hu:2001bc}. At the same time, the properties this type of DM candidates are consistent with the lack of signals in controlled experiments.

The main constraint on particle DM models comes from its relic density, dictated by the data obtained by satellite-borne experiments like WMAP \cite{Spergel:2006hy,Jarosik:2010iu,Hinshaw:2012aka} and PLANCK \cite{Ade:2013zuv}. These observations indicate that about 26\% of the energy density of the universe is made up of DM (when the universe is flat, as strongly favored by data \cite{Aghanim:2018eyx} and theoretical arguments). In addition, possible interactions of the dark sector with the SM are restricted by direct-detections experiments \cite{Aprile:2018dbl}, collider data \cite{Chala:2015ama,Aaboud:2016uro,Aaboud:2016tnv} and indirect detection experiments \cite{Conrad:2014tla,Bertone:2004pz,Feng:2010gw,Jungman:1995df}.

Several mechanisms have been used to ensure the expected DM relic density, and it proves useful to classify the models according to these. Broadly speaking there are two such mechanisms: a `freeze out' scenario, where the DM is assumed to have been in thermal equilibrium with the SM in the early universe, but later decoupled; and a  `freeze in' scenario where the dark and SM sectors have always been out-of equilibrium. 

\begin{figure}[htb!]
$$ \includegraphics[width=3in]{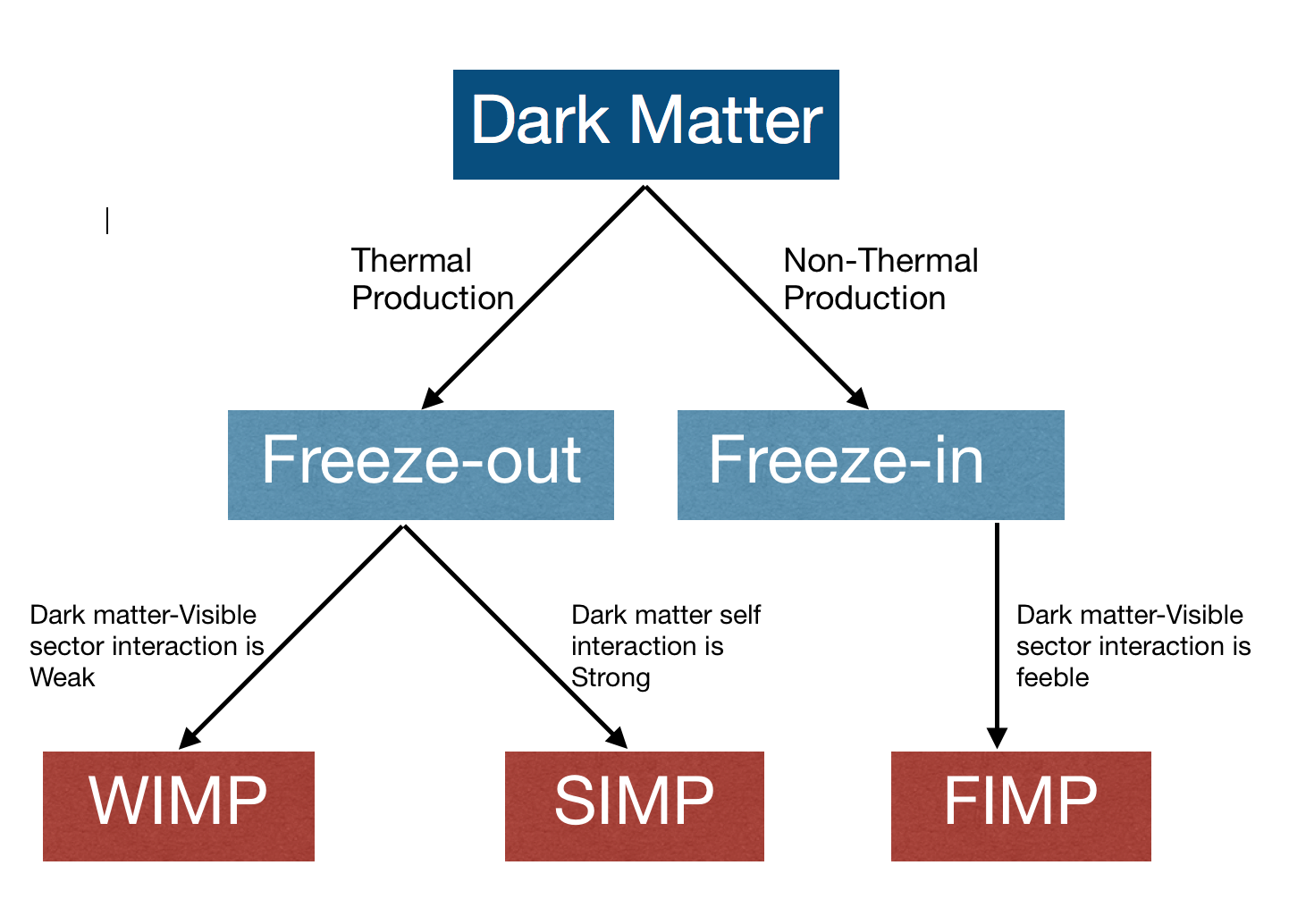} $$
\caption{Classification of possible particle DM candidates: WIMP, SIMP, FIMP (see text)}
\label{fig:classification}
\end{figure}

\begin{figure}[htb!]
$$ \includegraphics[width=3 in]{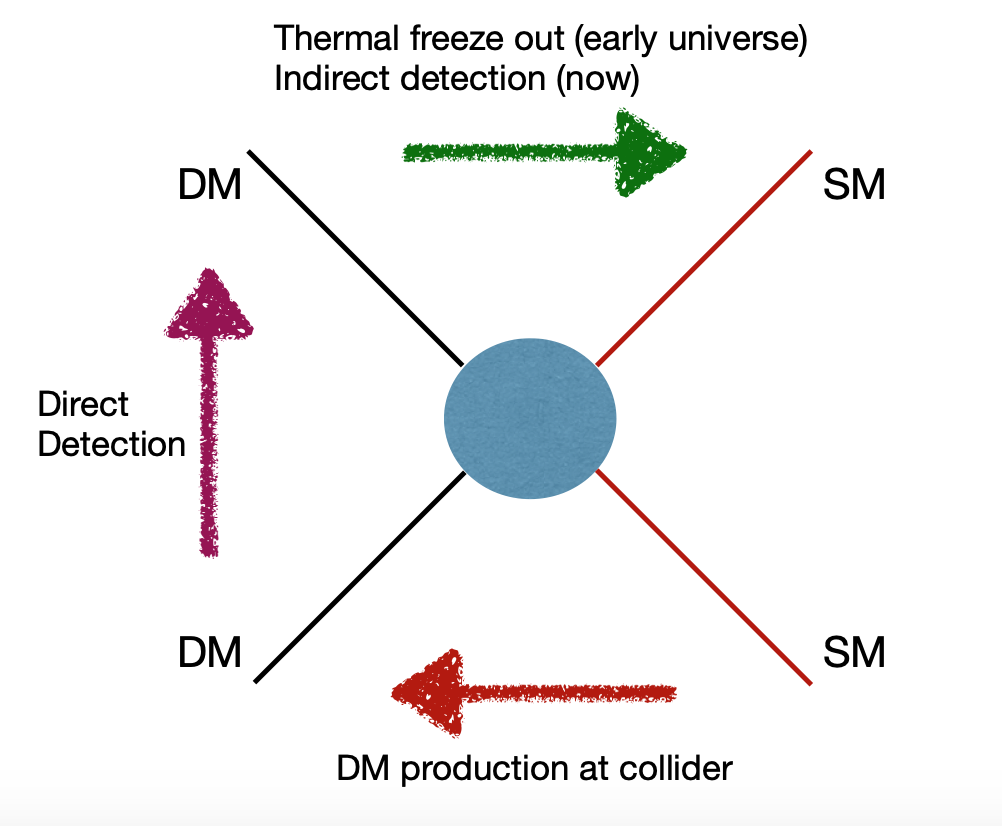} $$
\caption{Description of possible effects resulting from the interaction of a DM WIMP candidate with the SM.}
\label{fig:detection}
\end{figure}

Most of the DM candidates considered in the literature fall broadly in 3 categories: weakly interacting massive particle (WIMP) \cite{Arcadi:2017kky}; strongly interacting massive particles (SIMP) \cite{Hochberg:2014dra}; and feebly interacting massive particle (FIMP) \cite{Hall:2009bx}. Models with the first two usually rely on the freeze-out scenario to meet the relic density constraint, while FIMP-based models use non-thermal freeze-in processes. This DM taxonomy is summarized in Fig.~\ref{fig:classification}.  Historically the WIMP paradigm has been the most popular. Most such models exhibit DM interactions with the SM via $2\to2$ processes (see Fig.~\ref{fig:detection})~\footnote{Alternatives models where $3\to2$ processes dominate have also  been studied \cite{Carlson:1992fn, Hochberg:2014dra, Hochberg:2014kqa}.}, and are accessible at direct search and collider search experiments. We will mostly discuss WIMP-like DM in the context of EFT, though we will also touch upon prospects for a FIMP DM candidate.

The study of DM from an EFT viewpoint has been investigated by many authors and the literature is extensive. Space and time restrictions prevents us from presenting a comprehensive review of all these efforts; we will also fail to review all aspects of the few topics that we touch upon below. Our list of references will be tailored to the specific aspects of the topics discussed; this in no way should be viewed as an implied criticism on the papers not included, but only as a consequence of our working under the limitations just mentioned.

\subsection{Decoupling case}

In addition to having weak interactions with the SM (which is why DM is dark), any DM candidate must be stable (or very long lived). One way of ensuring stability is to assume the dark sector to have an exact, unbroken symmetry $ \gcal_{\tt DM} $ under which all SM fields are invariant, but not the dark ones. In this case the lightest dark particle with any given set of $ \gcal_{\tt DM} $ quantum numbers will be stable. Similarly, a simple way of ensuring that the dark sector is weakly coupled to the SM is to assume all dark fields are invariant under the SM gauge group, $ \gcal_{\tt SM}$. In the following we will adopt both these assumptions\footnote{There are many models which contain particles that transform non-trivially under both $ \gcal_{\tt DM} $ and  $ \gcal_{\tt SM} $, the most popular being supersymmetric models \cite{Jungman:1995df} where R-parity plays the role of $ \gcal_{\tt DM} $.}.

In both WIMP and FIMP cases it is the interactions between the dark and SM sectors that determine the DM abundance and possible production channels (see Fig. \ref{fig:detection}). Absent any clear indication as to the nature of these interactions we follow the simplest possibility and assume they are generated by the exchange of one or more particles that are invariant under both $ \gcal_{\tt DM} $ and  $ \gcal_{\tt SM} $; following common usage we refer to these particles as mediators. 

At energies below the mediator mass the SM-dark interactions then take the form\footnote{The same approach can be followed if the SM is extended by the addition of right-handed neutrinos; in this case the operators $ \ocal_{\tt SM} $ are constructed using the fields in this extension of the SM; as far as the authors are aware this scenario has not been fully explored in the literature.}
\beq
\lcal_{\tt int}=\inv{\Lambda^{n-4}} \ocal_{\tt DM} \ocal_{\tt SM}\,,
\label{eq:dm-sm.int}
\eeq
where $\Lambda = O(\text{mediator mass}) $, and $\ocal_{\tt SM} $ and $ \ocal_{\tt DM} $ are invariant under both $ \gcal_{\tt SM} $ and $ \gcal_{\tt DM} $; note that $ \ocal_{\tt DM} $ must have at least two dark-sector fields, since we assumed that all dark fields transform non-trivially under $ \gcal_{\tt DM} $. Following a common practice, will refer to these operators describing the interactions between the standard and dark sectors as {\em portal operators}.
 
Assuming that the dark sector can contain scalars ($\Phi$), Dirac fermions~\footnote{Extension to Majorana fields is straightforward.} ($\Psi$) and vector bosons ($X$),  one can construct DM-SM interaction operators up to any dimension; table \ref{tab:DM-SM} shows operators up to dimension six \cite{Duch:2014yma,Duch:2014xda,Macias:2015cna} where we defined
\bal
\ocal\up4_{\tt SM}  \in& \{|\phi|^4, \, \Box|\phi^2|,\, \bar{\psi}\phi \psi^{'},\, B_{\mu\nu}^2,\, (W_{\mu\nu}^I)^2,\, (G_{\mu\nu}^A)^2\}\,, \cr
\ocal\up4_{\tt dark}  \in&  \{|\Phi|^4, \, \Box|\Phi^2|,\,\Phi\bar{\Psi}P_{L,R} \Psi, \, X_{\mu\nu}^2\}\,, \cr
\ocal\up4_{\tt dark~\mu\nu} \in&  \{|\Phi|^\dagger X_{\mu\nu}\Phi, \, \Phi\bar{\Psi}\sigma_{\mu\nu}P_{L,R} \Psi, \, \bar{\Psi}(\gamma_\mu \dcal_\nu-\gamma_\nu \dcal_\mu)P_{L,R} \Psi\}\,;
\label{operadores}
 \end{align} 
 and
 \beq 
 \label{corrientes}
\begin{array}{ll} 
	\jcal_{\tt SM}^{(\psi) \mu} = \bar{\psi} \gamma^\mu \psi, & \qquad \jcal_{\tt SM}^{(\phi) \mu} = \frac{1}{2i} \phi^\dagger \overleftrightarrow{D}^\mu \phi,  \\
   \jcal_{\tt dark}^{(L,R) \mu} = \bar{\Psi} \gamma^\mu P_{L,R} \Psi, & \qquad \jcal_{\tt dark}^{(\Phi) \mu} = \frac{1}{2i} \Phi^\dagger \overleftrightarrow{\dcal}^\mu \Phi.
\end{array}
\eeq
In these expressions $ \psi,\,\psi' $ denote SM fermion fields~\footnote{No $P_{L,R} $ projectors are included because all SM fermion fields have definite chirality.}  
(such that all the above operators are gauge invariant) and $ \dcal $ the covariant derivative in the dark sector (replaced by an ordinary derivative if the sector is not gauged). 
We note that some of the operators in table \ref{tab:DM-SM} may be absent for some choices of $ \gcal_{\tt DM} $.

 \renewcommand{\arraystretch}{1.7}

\begin{table}[h!]
  \begin{center}
    \begin{tabular}{|c|c|c|} 
    \hline 
      \textbf{dim.} & \textbf{category} & \textbf{operators} \\
      \hline \hline
      4 & I & $ \vert \phi \vert^2 (\Phi^\dagger \Phi) $  \\
      \hline
      & II & $ \vert \phi \vert^2 \bar{\Psi} \Psi \hspace{1cm} \vert \phi \vert^2 \Phi^3 $\\
      5 & III & $ (\bar{\Psi} \Phi) (\phi^T \epsilon l) $ \\
      & IV & $ B_{\mu \nu} X^{\mu \nu} \Phi \hspace{1cm} B_{\mu \nu} \bar{\Psi} \sigma^{\mu \nu} \Psi $ \\
      \hline
      & V & $ \vert \phi \vert^2 \ocal\up4_{\tt dark} \hspace{1cm} \Phi^2 \ocal\up4_{\tt SM}  $ \\
      6 & VI & $ (\bar{\Psi} \Phi^2) (\phi^T \epsilon l) \hspace{1cm} (\bar{\Psi} \Phi) \slashed{\partial} (\phi^T \epsilon l) $ \\
      & VII & $ \jcal_{\tt SM}^\mu \jcal_{\tt dark\,\mu} $ \\
      & VIII & $ B_{\mu \nu} \ocal^{(4) \mu \nu}_{\tt dark} $ \\ \hline    
    \end{tabular}
  \end{center}
  \caption{Effective operators list up to dimension $6$ involving dark and SM fields; where $\phi$ stands for the SM scalar isodoublet, $B$ for the hypercharge gauge field, 
  and $l$ is a left-handed lepton isodoublet; also, $\epsilon = i \sigma_2$, where $\sigma_2$ is the corresponding Pauli matrix. Dark scalars, 
  Dirac dark fermions and vectors are denoted by $\Phi$, $\Psi$ and X respectively. The  operators $\ocal\up4$ in categories V and VIII are listed in \cref{operadores}, 
  and the vector currents in category VII in~\cref{corrientes}.} \label{tab:DM-SM}      
\end{table}

The phenomenology of all of these DM-SM interactions is not fully explored, but there have been significant efforts in several directions. The best studied are category I (Higgs portal coupling) \cite{Djouadi:2011aa, Arcadi:2017kky, Arcadi:2020jqf, Fedderke:2014wda}; category III (neutrino portal coupling) \cite{Macias:2015cna,Blennow:2019fhy}; category VII (vector portal coupling) \cite{Arcadi:2014lta, Fortuna:2020wwx}; and categories IV (spin-1 portal coupling) \cite{Barman:2020ifq}.
  
\begin{table}[h!]
\begin{tabular}{lcr}
    \begin{tabular}{|c|c|c|}
    \multicolumn{3}{c}{Scalar DM} \cr
    \hline
    Name & Operator & Coefficient \cr \hline
    C1 & $ (\chi^\dagger \chi) (\bar\qr  \qr )$ & $m_\qr /\Lambda^2$ \cr
    C2 & $ (\chi^\dagger \chi) (\bar \qr  \gamma_5 \qr )$ & $i m_\qr /\Lambda^2$ \cr \hline
    R1 & $ \chi^2 (\bar \qr  \qr )$ & $m_\qr /\Lambda^2$ \cr
    R2 & $ \chi^2 (\bar \qr  \gamma_5 \qr )$ & $i m_\qr /\Lambda^2$ \cr
    \hline
    \end{tabular}
\qquad
    \begin{tabular}{|c|c|c|}
    \multicolumn{3}{c}{Fermionic DM} \cr
    \hline
    Name & Operator & Coefficient \cr \hline
    D1 & $ (\bar\chi \chi) (\bar \qr  \qr ) $ & $m_\qr /\Lambda^3 $ \cr
    D2 & $ (\bar\chi \gamma_5 \chi) (\bar \qr  \qr ) $ & $i m_\qr /\Lambda^3 $ \cr
    D3 & $ (\bar\chi \chi) (\bar \qr  \gamma_5\qr ) $ & $i m_\qr /\Lambda^3 $ \cr
    D4 & $ (\bar\chi \gamma_5 \chi) (\bar \qr  \gamma_5 \qr ) $ & $m_\qr /\Lambda^3 $ \cr
    D5 & $ (\bar\chi \gamma^\mu \chi) (\bar \qr  \gamma_\mu  \qr ) $ & $1/\Lambda^2 $ \cr
    D6 & $ (\bar\chi \gamma^\mu \gamma_5 \chi) (\bar \qr  \gamma_\mu \qr ) $ & $1/\Lambda^2 $ \cr
    D7 & $ (\bar\chi \gamma^\mu \chi) (\bar \qr  \gamma_\mu \gamma_5 \qr ) $ & $1/\Lambda^2 $ \cr
    \hline
    \end{tabular}
\end{tabular}
\caption{Effective DM-SM operators assuming $\gcal_{SM} = U(1)_{EM}$ involving scalar DM (left), involving Dirac fermion as DM (right) \cite{Goodman:2010ku}. 
Generic quark fields are denoted by $\qr$ (to differentiate from $q$ that we use to denote a SM left-handed quark isodoublet); see the text for comments on the operator coefficients.}
\label{tab:DM-SM2}
\end{table}

The simplest choice of $\gcal_{\tt DM}$ is $\zBB_2$, under which all field in the dark sector are odd and all SM fields are even (and which would forbid $|\phi|^2 \Phi^3$ in category II). There are, however, other possibilities; for example, $ \gcal_{\tt DM} $ can be a non-Abelian gauge symmetry with $X$ the corresponding gauge boson and $ \Phi $ belonging to the adjoint representation  so that $ \Phi X_{\mu\nu} B^{\mu\nu} $ is invariant under $ \gcal_{\tt DM} \times \gcal_{\tt SM} $. The operators shown in table \ref{tab:DM-SM} may be classified further depending on the choice of mediators, and according to their LG and PTG character; for details, see \cite{Macias:2015cna}.

One may also consider the DM-SM operators when $\gcal_{\tt SM} $ is replaced by $ U(1)_{\tt EM}$, the gauge group for electromagnetism. This is adequate whenever the temperatures are low enough so that the SM local symmetry is broken down to $U(1)_{\tt EM} $, in which case the relevant SM particles (at temperatures above the QCD confinement transition) are the leptons, light quarks, the photon and, the gluon. In particular, the operators containing two SM fermions take the form
\bal
  \text{scalar DM}:& \quad  |\chi|^2 ( \overline{\psi} \Gamma \psi), ~\Gamma = \{ \mati, \gamma^5 \}\,, \cr
  \text{fermion DM}:&\quad  (\bar{\chi}\Gamma \chi)( \overline{\psi} \Gamma \psi),~\Gamma = \{ \mati, \gamma^5, \gamma^\mu, \gamma^{\mu} \gamma^5, \sigma^{\mu \nu} \}\,,
  \label{eq:ui.ops}
\end{align}
where $ \chi $  denotes the DM field and $ \psi $ a SM fermion. We list these dimension 5 and 6 operators in table~\ref{tab:DM-SM2} \cite{Goodman:2010ku}. 
The operators containing quarks are of special interest because the can contribute not only to relic density, 
but can also be probed in direct-detection experiments and collider searches at the LHC. 

Operators with coefficients $ \propto \Lambda^{-2} $ in table~\ref{tab:DM-SM2} correspond to linear combinations of those in categories V and VII in table \ref{tab:DM-SM}; those with coefficients $ \propto  \Lambda^{-3}$ to dimension 7 operators when written in terms of SM fermions and scalars. For example, D1 is generated from operators of the form $(\bar\chi \chi) (\bar q \phi d) $ where, as before, $q$ and $d$ denote, respectively, the left-handed quark isodoublet and right-handed down-type quark isosinglet fields; the factor of $m_\qr$ in the coefficient follows from replacing $\phi$ by its \vev\ and from assuming that the dimension-7 operator coefficient is of the same order as the corresponding Yukawa coupling.

Missing from table~\ref{tab:DM-SM2} are operators involving the photon field, and possible dark scalars or vectors. These correspond to categories III, IV, V, VI and VIII in table~\ref{tab:DM-SM}. For example, $(\bar{\Psi} \Phi) (\phi^T \epsilon l) $ in category III gives rise to $ (\bar{\Psi} \Phi P_L \nu ) $ of dimension 4,  $B_{\mu \nu} \bar{\Psi} \sigma^{\mu \nu} \Psi $ in category IV generates $F_{\mu \nu} \bar{\Psi} \sigma^{\mu \nu} \Psi $ of dimension 5, and $ \Phi^2 \tilde G_{\mu\nu}^A G^{A\,\mu\nu} $ of dimension 6 in category V is also of interest even below the QCD confinement transition. 

It is also worth noting that it is straightforward to extend the list of  DM-SM effective operators to include the presence of light right-handed neutrinos; this approach is of interest because it provides alternate avenues for understanding DM effects in conjunction with well-motivated explanations for the presence of neutrino masses, and the generation of the baryon asymmetry of the universe  via leptogenesis, within one framework; see for example, \cite{Barman:2021tgt}.

We now provide a few illustrations of the EFT approach when studying the properties and viability of a given DM scenario.

\subsection{Example 1: Vector mediators at temperatures below Electroweak Symmetry breaking }
\label{sec:dm.vect}

At temperatures below both the mediator mass and $\smvev$, the SM \vev\  ($\sim 246\,\gev$), the relevant quark-DM interactions are given in table \ref{tab:DM-SM2}; where it is useful to note that, for fermionic DM, only  $ \ocal_{\tt D5-D7}$ are suppressed only by $ 1/\Lambda^2 $. In this section we will consider the case where the DM-SM interaction is well described by (see \cite{Blennow:2015yca} for a detailed discussion)
\beq
\ocal_{\tt D5}= (\bar{\chi}\gamma^\mu\chi)(\bar{\qr}\gamma_\mu \qr).
\label{op-ex}
\eeq

This situation can be also described by a simple model containing the fermionic DM candidate $ \chi $, and vector mediator $X$ (not to be mistaken with a possible component of the dark sector listed in table \ref{tab:DM-SM}), that couples to both DM and quarks. A simple, consistent Lagrangian for this scenario is
\beq
\lcal_{\tt DM-SM} = - \inv4 X^{\mu \nu}X_{\mu \nu} + \half M_{\tt X}^2 (X_\mu - \partial_\mu s)(X^\mu - \partial^\mu s) + \bar\chi \left( i \slashed\dcal  - \mdm  \right) \chi + \sum_{\qr}  \bar\qr \left( i \slashed D - m_\qr \right) \qr\, ,
\label{eq:simplified}
\eeq
where $\dcal_\mu = \partial_\mu + i g_\chi X_\mu $ and~\footnote{Here $g_s$ denotes the strong coupling constant, $t^A$ the $\su3$ color generators in the fundamental representation, and $G_\mu^A$ the gluon fields.} 
$ D_\mu = \partial_\mu + i g_\qr X_\mu + i g_s t^A G_\mu^A $. 

We will be interested in this model only as a simple pedagogical realization of the effective DM-SM interactions in \cref{op-ex} (often called `simplified model'), and so we will only consider this aspect of the associated phenomenology. We will ignore other constraining aspects, namely, the fact that the mediator $X$ also generates 4-quark interactions that are severely restricted \cite{Zyla:2020zbs}: $\Lambda/g_\qr \gtrsim 5 \, \tev$. This simplified model is certianly not unique, see for example \cite{DeSimone:2016fbz}.

\begin{figure}[htb!]
$$
\includegraphics[width=.85\linewidth]{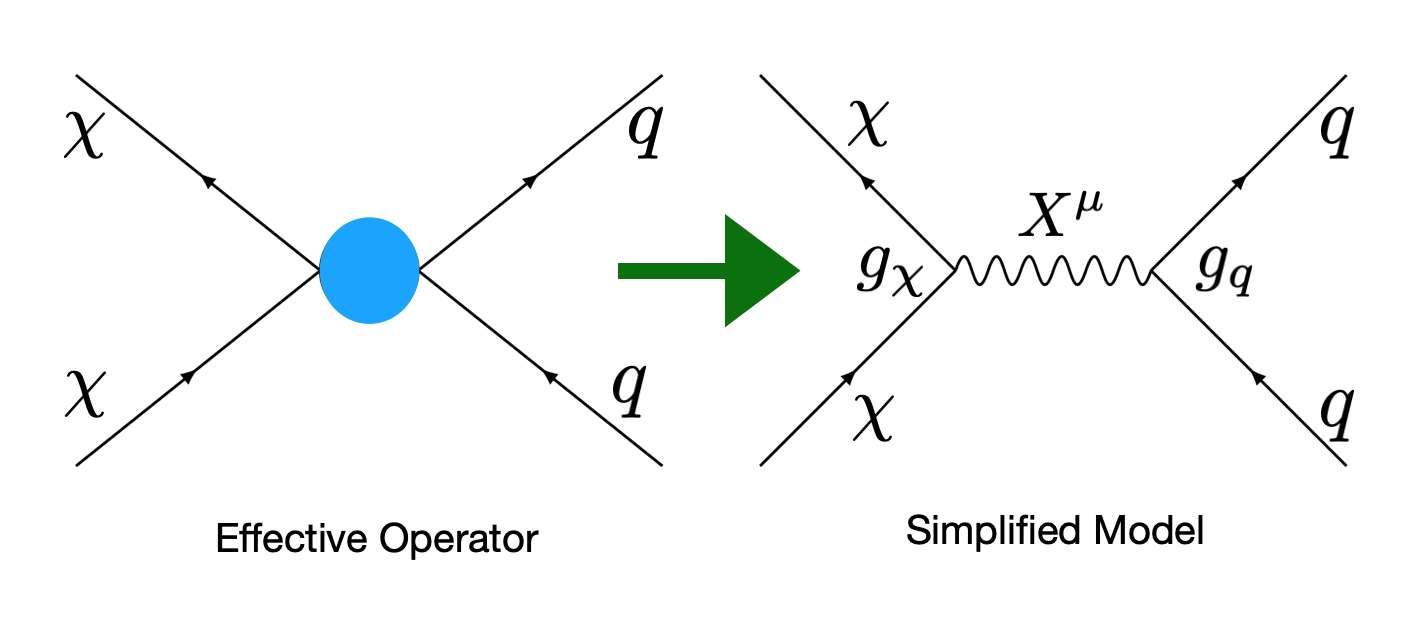}
$$
\caption{Diagram illustrating the generation of the effective operator portal $ \ocal_{\tt D5} $ in a simplified model (see text).}
\label{fig:EFT-simplified}
\end{figure}

The model is invariant under the following gauge transformation 
\beq
X_\mu \to X_\mu + \partial_\mu\omega\,,\quad s \to s + \omega\,,\quad \psi \to e^{- i g_\psi \omega} \psi\,~ (\psi=\chi,\ \qr)\,.
\eeq
In the unitary gauge, where $s=0$, the $X$ equation of motion is simply
\beq
\left( \Box \eta_{\mu\nu} - \partial_\mu \partial_\nu + M_{\tt X}^2 \eta_{\mu \nu} \right) X^\nu = j_\mu\,, \quad j_\mu = g_\chi \bar\chi \gamma_\mu \chi + \sum_\qr g_\qr  \bar\qr \gamma_\mu \qr\,.
\eeq
At energies below $ M_{\tt X}$ this has the simple solution $ X_\mu = j_\mu/M_{\tt X}^2$, which, when substituted into $\lcal_{\tt DM-SM}$ gives the effective Lagrangian (we omit the kinetic terms)
\beq
\leff = - \half \frac{j^2}{M_{\tt X}^2} \subset \inv{\Lambda^2} \ocal_{\tt D5}\,, \quad \inv{\Lambda^2} = - \frac{g_\chi g_\qr}{M_{\tt X}^2}\,.
\label{eq:lambda-mX}
\eeq
This process is illustrated in Fig. \ref{fig:EFT-simplified}. This argument shows that for energies below $ M_{\tt X} $ one can use $\lcal_{\tt DM-SM}$ interchangeably~\footnote{It is worth noting that  currently available calculation tools such as {\tt MicroOmegas} \cite{Belanger:2008sj} and {\tt MadGraph} \cite{Alwall:2011uj} can have difficulties in dealing with $\leff$, so using $\lcal_{\tt DM-SM}$ may be advantageous for this practical reason.}.

Now, for perturbation theory to be valid  we must have $g_{\chi,\,\qr}  \lesssim 4 \pi $; in addition, for the mediator $X$ to remain off-shell in $ \chi\chi \to \qr \qr $ we must also have $M_{\tt X} > 2m_\chi$. These two requirements then lead to 
\beq
\Lambda \gtrsim \frac{M_{\tt X}}{4\pi}> \frac{\mdm }{2\pi}\,.
    \label{EFT-valid}
\eeq

Next we calculate the relic density and determine its constraints on the parameter space of the model. As mentioned before, we assume $\chi$ to be a WIMP-like fermion DM candidate that was in thermal and chemical equilibrium with the hot soup of SM particles during the early universe, and which decoupled at some later epoch. The the relic density within the WIMP scenario (for a pedagogical discussion see Ref. \cite{Kolb:1990vq}) is based on the solution to the following Boltzmann equation:
\beq
\frac{d Y}{d x} = -0.264~ \sqrt{g_*}~M_{\tt pl}~ \frac{\mdm }{x^2}~\vevof{\sigma v}_{2_{\rm DM}\to 2_{\rm SM}} ~\left( Y^2-Y_{\tt eq}^{2} \right)\,, 
\label{eq:BEQ2to2}
\eeq
where $M_{\tt pl} $ denotes the Planck mass, $Y = n/s$ ($n$ is the DM density, $s$ is the total entropy density) and $x= \mdm /T$ ($T$ is the temperature). $Y_{\tt eq}$ denotes the value of $Y$ in thermal equilibrium 
\beq
 Y_{\tt eq}(x)=0.145~\frac{g_{\tt DM}}{g_*} x^{3/2}e^{-x}\,,
 \label{eq:yeq}
\eeq
and $g_{\tt DM}$ is the DM number of internal states and we assumed Maxwell-Boltzmann statistics; $g_*$ denotes the effective relativistic degrees of freedom
\beq
g_*=\sum_{i=\rm bosons}g_i \theta(T-m_i)+\frac{7}{8}\sum_{i=\rm fermions}g_i \theta(T-m_i)\,;
\label{eq:DOFeqn}
\eeq
here $ g_i $ are the internal degrees of freedom of particle $i$ with mass $ m_i $.

Finally, $\vevof{\sigma v}_{2_{\rm DM}\to 2_{\rm SM}} $ denotes the thermal average of the cross-section$\times$velocity for the process $ \chi\chi\to \qr\qr $ mediated by $ \ocal_{\tt D5} $ : 
\bal
\vevof{\sigma v}_{2_{\rm DM}\to 2_{\rm SM}} &=  \frac{3\mdm ^2}{2\pi\Lambda^4} \sum_\qr \left(g_\chi g_\qr \right)^2   \left\{ 2+\frac{m_\qr^2}{\mdm ^2} +\left[ \frac{8\mdm ^4 - 4 m_\qr^2 \mdm ^2 + 5 m_\qr^4 }{24 \mdm ^2 (\mdm ^2 - m_\qr^2)} \right]v^2  \right\} \sqrt{1- \frac{m_q^2}{\mdm ^2}}\,, \mcr
&= a + b v^2\,,
\label{ann-cross-sec}
\end{align}
where $m_\qr$ is the quark mass, $v$ is the M\"oller velocity $ v= \sqrt{(p_\chi.p_{\bar{\chi}})^2-m_{\tt DM}^4}/(E_\chi E_{\bar{\chi}})$. This interaction then gives rise to s-wave ($\propto a$) and p-wave ($\propto b$) contributions (see \cite{Kolb:1990vq} for details).

The relic abundance of DM after thermal freeze-out is given by (see \cite{Kolb:1990vq} for the derivation):
\beq
    \Omega_{\tt DM}{\sf h}^2 \simeq \frac{(1.04 \times 10^9 \text{GeV})x_f}{M_{\tt pl}\sqrt{g_*}(a + 3b/x_f)}\,; 
    \label{theo.relic}
\end{equation}
where ${\sf h}\sim 0.674$ is the Hubble parameter in units of 100 km/s/Mpc, $\Omega_{\tt DM} = \rho_{\tt DM}/\rho_c$, and $\rho_c, \rho_{\tt DM} $ denote the critical and DM densities, respectively;  $x_f=\mdm /{T_f}$, with $ T_f$ the freeze-out temperature, corresponding to the time where $Y - Y_{\tt eq} \simeq Y_{\tt eq}$ (in practice $ x_f \sim 20-25$). This expression for the relic density should be compared to the current PLANCK result~\cite{Aghanim:2018eyx}
\beq
\Omega_{\tt DM} {\sf h}^2 = 0.11933\pm 0.00091 \,; 
\label{relic-value}
\eeq
that corresponds to $\vevof{\sigma v}_{2_{\rm DM} \to 2_{\rm SM}} \sim 1.5\times10^{-9}~\rm{GeV}^{-2}$, which is also a typical cross-section for the weak interactions; hence the generic name WIMP associated with this type of model. We can now use \cref{ann-cross-sec,theo.relic,relic-value} to constrain the parameters of the model $\{\mdm ,\,\Lambda,\, g_\chi,\, g_{\qr} \}$

The operator $ \ocal_{\tt D5}$  can also contribute to coherent DM-nucleon scattering probed in direct search experiments \cite{Aprile:2018dbl}. Again citing only the final expressions (for a derivation see \cite{Feng:2010gw}) the corresponding amplitude $ \mcal $ is given by
\bal
    |\mathcal{M}|^2_{\tt DM-nucleus} &= (4\mdm  \, m_N)^2 [Zf_p + (A-Z)f_n]^2, \mcr
    \frac{f_{n,p}}{m_N} &= \sum_{\qr=u,d,s} f_{T_q}^{(n,p)} \frac{g_\qr }{m_\qr \, \Lambda^2} + \frac{2}{27}\left[ 1- \sum_{\qr=u,d,s} f_{T_q}^{(n,p)} \right] \sum_{\qr=c,b} \frac{g_\qr }{m_\qr \, \Lambda^2}\,,    
    \label{eq:dir}
    \end{align}
where $m_N$ denotes the nucleon mass, and $Z$ and $A$, respectively, the proton and nucleon numbers of the target nucleus. The numerical parameters $   f_{T_q}^{(n,p)} $ represent the values of appropriate nuclear form factors that embody the nuclear physics effects of the scattering process under consideration; details of the calculation can be found in \cite{Walecka:1977vc, Anand:2014kea}, some typical values: $f_{T_u}\up p=0.0153$, $f_{T_d}\up p=0.0191$, $f_{T_s}\up p=0.0447$ \cite{Belanger:2008sj}.
The latest experimental XENON1T limits \cite{Aprile:2018dbl} require the DM-Nucleon spin-independent cross-section derived from \cref{eq:dir} to be less than $\sim 10^{-47} \rm{cm}^2$ for a DM mass $\sim 100$ GeV, which leads to a further restriction on the model parameters.

\begin{figure}[htb!]
$$
\includegraphics[width=.50\linewidth]{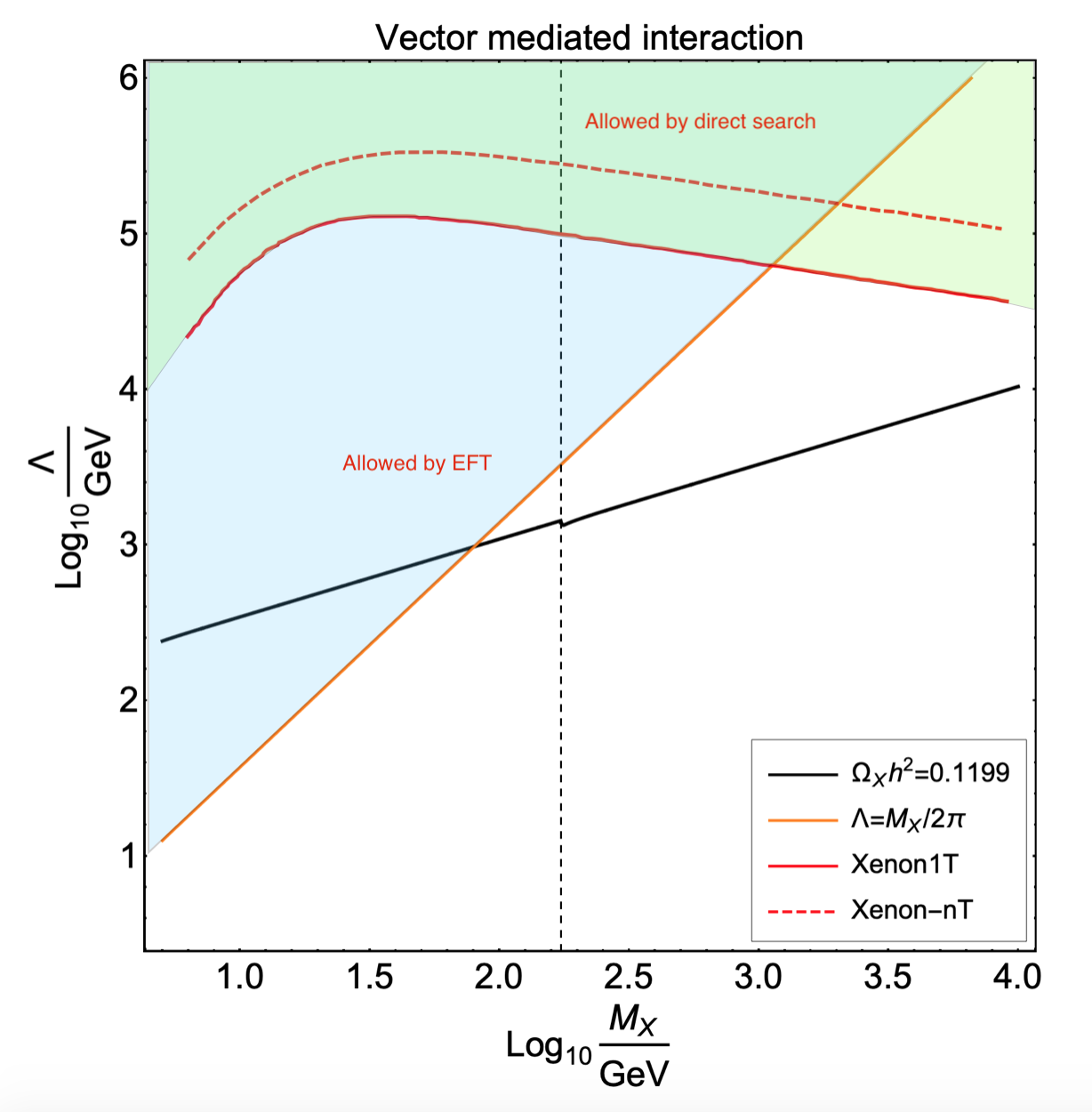}
$$
\caption{Available parameter space for the DM portal operator  $\overline{\chi} \gamma^\mu \chi \overline{\qr} \gamma_\mu \qr$. Green region: allowed by the direct detection constraint from Xenon1T (region below the dashed line gives the expected exclusion region from Xenon-nT); the EFT approached used is applicable within the blue region; and the relic abundance constraint restricts parameters to the black line.}
\label{fig:allowed}
\end{figure}

In Fig.~\ref{fig:allowed}, we show the constraints on this EFT operator model from relic density (black thick line), direct search bound from XENON1T (red thick line and above), future direct search sensitivity in XENON-nT (red dashed line and above), and the validity of EFT \cref{EFT-valid} (orange thick line), assuming $g_\chi=g_\qr=1$. We see that, in fact, there are no allowed values of $ \{ \Lambda,\, \mdm  \} $, 
and one can check that is continues to hold for all values of $ g_{\qr,\,\chi} $: this model is ruled out.

This situation is common to most of the operators which contribute to spin-independent~\footnote{The nomenclature corresponds to the nucleon interactions: a vector quark current generates a  coupling independent of the nucleon spin, while the coupling generated by an axial-vector quark current is proportional to the nucleon spin.}  direct search cross-sections in single-portal models: the direct detection constraints allow only relatively small couplings, wile the relic abundance data demands relatively large ones. In contrast, operators like D6 and D7 in Table \ref{tab:DM-SM2} contribute only to the DM-SM spin-dependent cross sections that are much smaller (the amplitude is not $ \propto A,\,Z$ as in~\cref{eq:dir} ), and for which existing constraints are weaker (see for example latest PANDA bound \cite{Fu:2016ega}); models with these portal operators are allowed \cite{Belyaev:2018pqr}. 

Alternatively, we can imagine that the mediator $X$ also couples to the leptons~\footnote{These are the fields after \ssb, to be distinguished from the left-handed lepton isodoublets $l$.} $\ell$ with gauge coupling $ g_\ell $, so that at energies below $ M_{\tt X} $ the  effective interaction $ (\bar{\chi}\gamma^\mu\chi)(\bar{\ell}\gamma_\mu \ell) $ is also generated. If $ g_\qr = g_\ell $ the model is again disallowed, but if $ g_\ell \gg g_\qr $ the relic abundance constraint can be met due to the relatively large leptonic cross section, while the small quark coupling allows meeting the direct detection limit. 

We conclude this section noting that more general analyses of the EFT approach to SM-DM interactions have appeared in the literature: a limited study of DM EFT operators involving leptons was carried out in \cite{Rawat:2017fak}; Ref. \cite{Fortuna:2020wwx} provides a study of the viability of the operator portals in Table \ref{tab:DM-SM} when the DM candidate mass is below $ \mz/2$; while the case where DM is a singlet Majorana fermion have been considered in \cite{Barman:2020plp,Matsumoto:2014rxa}.

\subsection{Example 2: Freeze-in scenario in DM-EFT}
\label{sec:DM-freeze-in}

DM can also reach the correct relic density via the so-called freeze-in scenario, which occurs when the DM-SM interactions are extremely weak. In this case, DM density is assumed to be zero in the very early universe $Y(x\sim 0)=0$ (unlike the WIMP case, where $Y(x\sim0)=Y_{\tt eq}$; see \cref{eq:BEQ2to2,eq:yeq}) and increases from the annihilation or decay products of other particles that are in thermal equilibrium. The relic density can again be obtained from the solution of the  appropriate Boltzmann equation, with the difference that now the initial DM density is zero.

As an illustration of this scenario we consider the portal described by the following operator in category IV ({\it cf.} table \ref{tab:DM-SM}; for other portal operators and a general discussion see \cite{Blennow:2013jba})
\bea
\ocal_{\tt IV}= B_{\mu \nu} X^{\mu \nu} \Phi\,.
\eea
It is straightforward to show that this effective operator cannot be generated at tree level in a theory with scalar, vector and fermions, so its coefficient will be naturally small (if the theory is also weakly-coupled). For simplicity we assume that the dark symmetry that stabilizes the DM candidate against decay is simply a `dark parity' (a $Z_2$ symmetry) under which both $\Phi$ and $X$ are odd.

\begin{figure}[htb!]
$$
\includegraphics[scale=0.20]{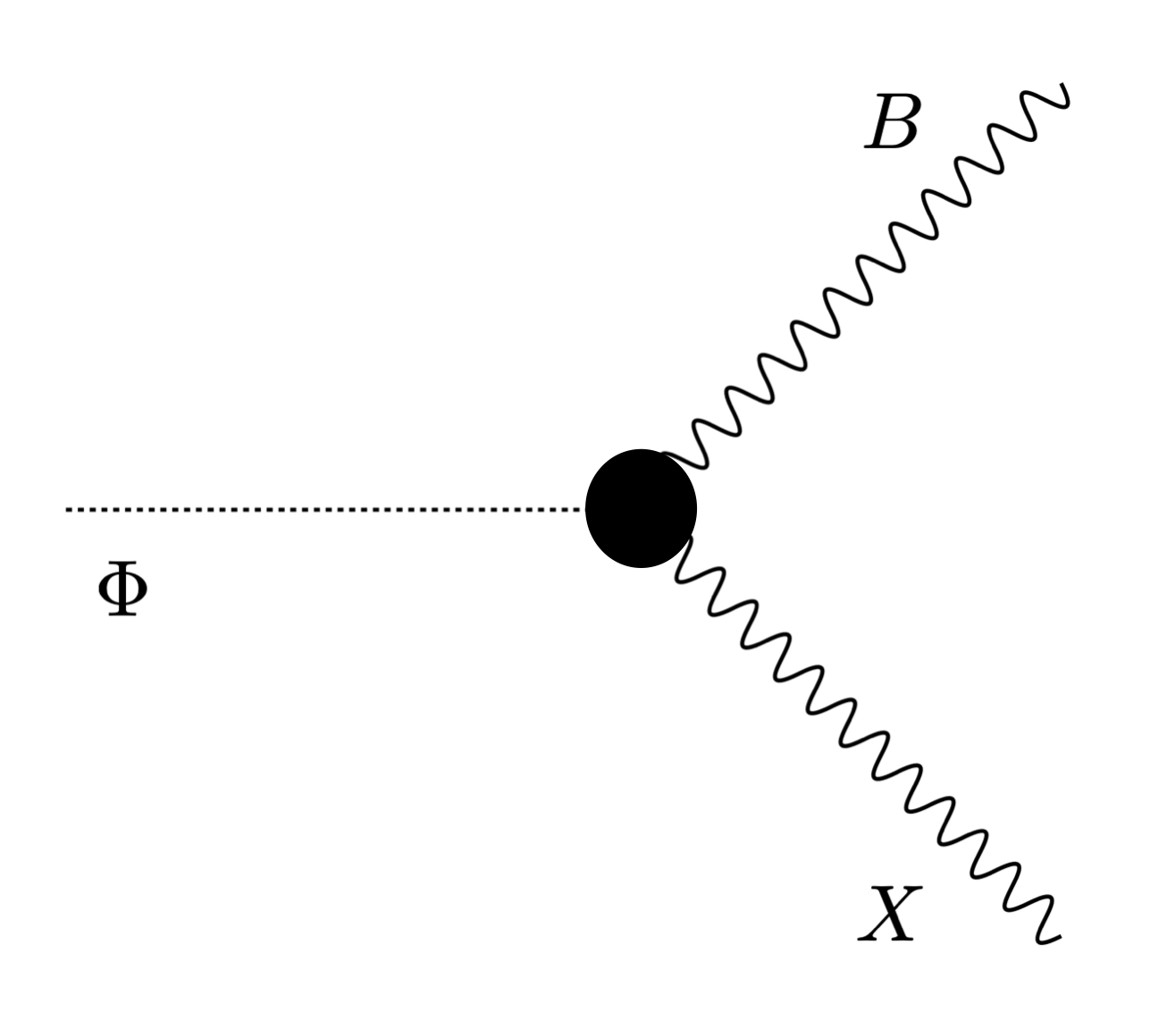}
$$
\caption{Decay of $\Phi \to XB$ before EWSB and $\Phi \to X \gamma(Z)$ after EWSB, which contributes to the freeze-in production of $X$.}
\label{fig:phi-decay}
\end{figure}

\begin{figure}[htb!]
$$
\includegraphics[scale=0.45]{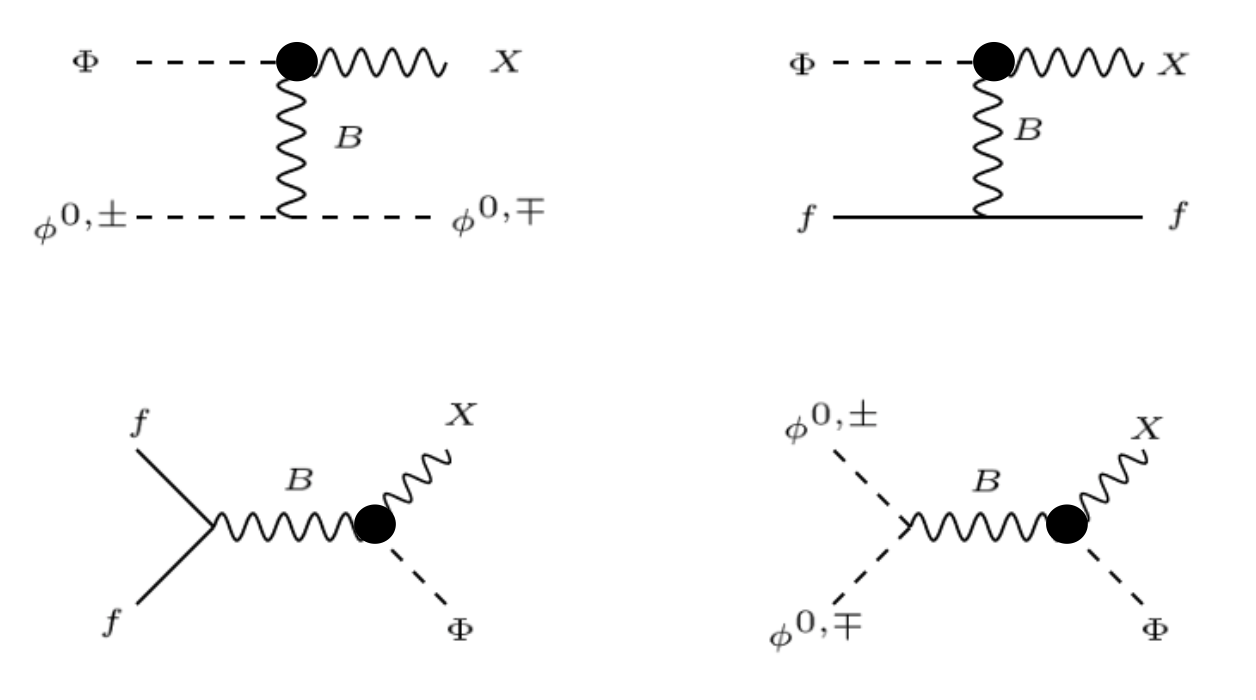}
$$
\caption{$X$ production from scattering processes with $t$ and $s$-channel $B$ exchange. The diagrams with Goldstone bosons $\phi^{0,\pm}$ vanish identically, leaving only two diagrams with SM fermions ($f$). 
}\vspace{0.2cm}
\label{fig:bewsb}
\end{figure}

We now assume that the dark sector in this model contains only the scalar and the vector in $ \ocal_{\tt IV} $; in this case the lighter of these particles will be the DM candidate. Here we will consider the case of vector DM, so that $m_\Phi > m_X $, and assume the $ \Phi $ are in thermal equilibrium with the SM because of a sufficiently strong $| \Phi|^2 |\phi|^2$ interaction, and produce the $X$ through the decay $ \Phi \to X B $ (or, at temperatures below that of electroweak symmetry breaking, $ \Phi \to XZ,\,X\gamma $), see Fig.~\ref{fig:phi-decay}; in addition, there are scattering processes that also produce the $X$ as shown in Fig.~\ref{fig:bewsb}. We emphasize that we assume these processes are rare because of the weak $\Phi BX$ coupling, in particular, the $X$ never equilibrates with the SM.

Then, assuming the initial abundance for DM ($X$) to be zero, neglecting Pauli blocking and stimulated emission effects, with dominant DM production coming from $\Phi \to BX$,
the Boltzmann equation (BEQ) for DM yield ($Y_x=n_X/s$) as a ratio of DM number density $n_X$ and the comoving entropy density in the visible sector $s$ can be written as: 
\bea
\begin{split}
- \frac{dY_X^\text{D}}{dT} &= \frac{g_\Phi m_\Phi^2 \Gamma_{\Phi\to X,B}}{2\pi^2\, s\, {\sf H}} K_1\left(m_\Phi/T\right),
\end{split}
\label{eq:yld-temp-decay}
\eea
where {\sf H} is the Hubble parameter, $K_1(x)$ the Bessel function of first kind, and the decay width $\Gamma_{\Phi\to X,B}$  the decay width defined as:
\beq
\Gamma_{\Phi\to X,B}=\int d\Pi_X d\Pi_B \frac{\left|\mathcal{M}\right|_{\Phi\to X,B}^2}{2 g_\Phi\, m_\Phi} \left(2\pi\right)^4\delta^4\left(p_X+p_B-p_\Phi\right)\, ,
\eeq
where $\left|\mathcal{M}\right|_{\Phi\to X,B}^2$ is the matrix element for the decay and $d\Pi= d^3 p/[2E\left(2\pi\right)^3]$ the Lorentz-invariant phase-space elements.

The total DM density per entropy $Y^{\tt tot}_X$ is obtained by adding the contributions from the annihilation  (Fig.~\ref{fig:bewsb}) and decay processes:
\bal
Y_X^\text{\tt tot} &= Y_X^\text{D}+Y_X^\text{ann}\,,\mcr
&= \int_{T_{\tt min}}^{T_{\tt max}} dT\frac{m_\Phi^2 \Gamma_{\Phi\to X,B}}{2\pi^2}\frac{K_1\left(m_\Phi/T\right)}{s(T) {\sf H}(T)} \mcr
& \quad + \inv{512\pi^6}\sum_{i,j,k}\int_{T_{\tt min}}^{T_{\tt max}}\frac{dT}{s(T) {\sf H}(T)}\int_{\bar s=0}^\infty d\bar s d\Omega \left(\frac{\sqrt{\bar s}}{2}\right)^2 \left|\mathcal{M}\right|_{i,j\to X,k}^2 \inv{\sqrt{\bar s}} K_1\left(\frac{\sqrt{\bar s}}{T}\right). 
\label{eq:totyld1}
\end{align}
$i,j,k$ denote particles in the initial and final state (Fig.~\ref{fig:bewsb}) $s(T)$ the entropy density and $ \bar s$ the CM energy squared; $T_{\tt min}=2.7K$ represents present temperature of Universe, while maximum temperature available can be assumed to be the temperature characteristic to the reheating phase expected at the end of the inflationary epoch \cite{Kolb:1990vq}, $T_{\tt max}=T_{\tt RH}$; this is essentially a free parameter  as is very loosely bounded from Big Bang Nucleosynthesis, which requires $T_{\tt RH}\gtrsim 4.7~\rm MeV$~\cite{deSalas:2015glj}, while simple inflationary scenarios require $T_{\tt RH}\sim 10^{16}~\rm GeV$~\cite{Linde:2005ht} for a successful inflation. This parameter can therefore be chosen large or small, with significant effects on the  freeze-in process for DM; here we choose the following hierarchy:
\beq
\Lambda \gtrsim T_{\tt RH} > m_\Phi > m_X.
\label{eq:fi-hierarchy}
\eeq

\begin{figure}[htb!]
$$
\includegraphics[width=.55\linewidth]{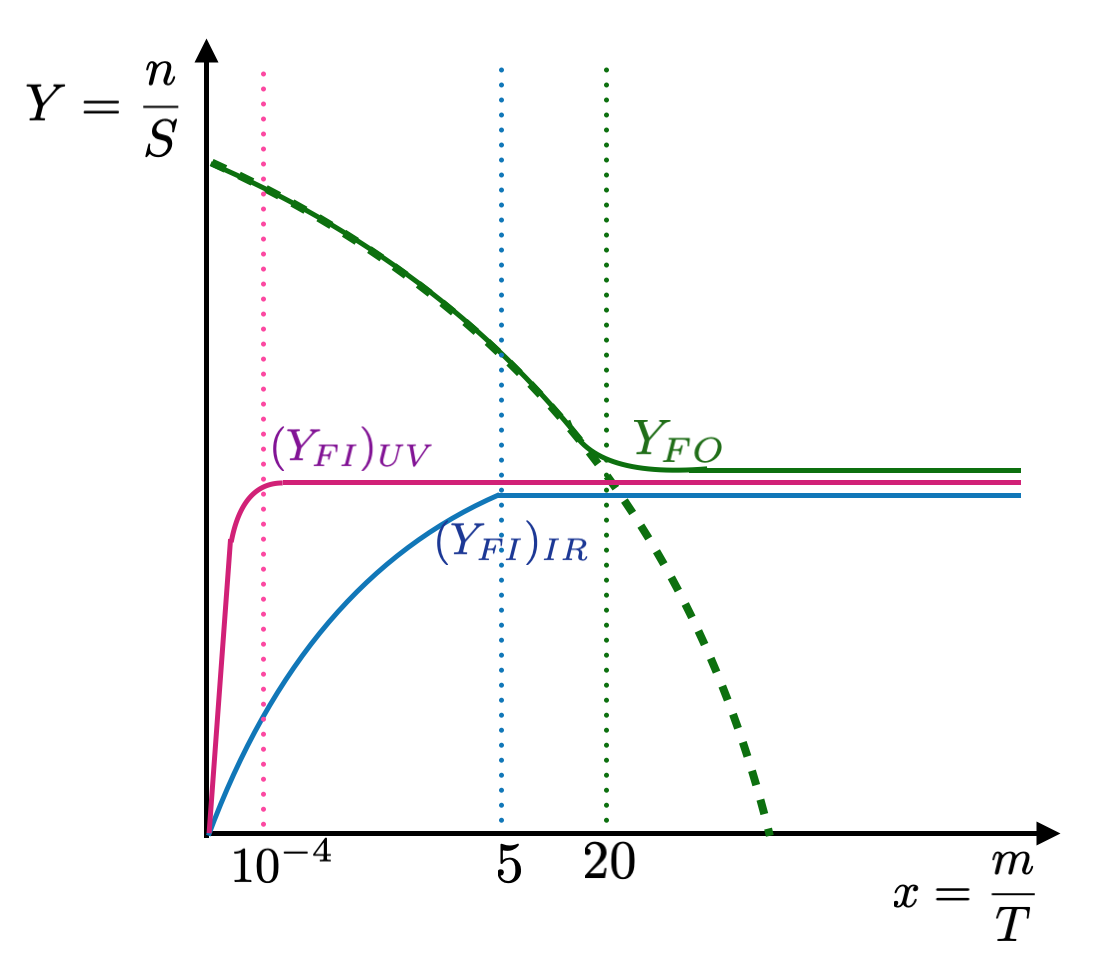}
$$
\caption{A cartoon illustration of the evolution of the DM abundance $Y$ in the ultraviolet and infrared freeze-in scenarios (purple and blue lines, respectively) compared to the freeze-out scenario (green line).}
\label{fig:UV}
\end{figure}

There are two interesting special cases of \cref{eq:fi-hierarchy}: {\it(i)} when $T_{\tt RH} \gg m_\Phi$ (``ultraviolet'' freeze in \cite{Elahi:2014fsa}),  significant DM production occurs at very high temperatures and the freeze-in temperature $T_{\tt FI}\sim T_{\tt RH} $, ($x_{\tt FI} \lesssim 10^{-4}$); and {\it(ii)} when $T_{\tt RH} \gtrsim m_\Phi$  (``infrared'' freeze-in),  DM production is slow and freeze-in occurs at a low temperature $T_{\tt FI} \sim \mdm = m_X$ (typically $x_{\tt FI}\sim 1-5$), and where renormalizable (dimension $\le4$)  operators  are the main interactions responsible for DM production. Choosing, for example \cite{Barman:2020ifq}, $m_\Phi=500\,\gev, m_X=100\,\gev$, the ultraviolet scenario corresponds to $T_{\tt RH}=10^8\,\gev$ and  $\Lambda\simeq10^{16}\,\gev$; while in the infrared scenario $T_{\tt RH}=1\,\tev,~\Lambda\simeq10^{13}\,\gev$. An illustration of these two cases and the comparison to the freeze-out scenario is presented in Fig.~\ref{fig:UV}.

We close this section by noting that the above model illustrates but one of the realization of the freeze-in paradigm; this approach (still within the EFT context) has also been studied for various other portal operators, see for example \cite{Barman:2020plp, Biswas:2019iqm}.
 
\subsection{Example 3: Collider searches for DM}

\begin{figure}[htb!]
$$
\includegraphics[width=.50\linewidth]{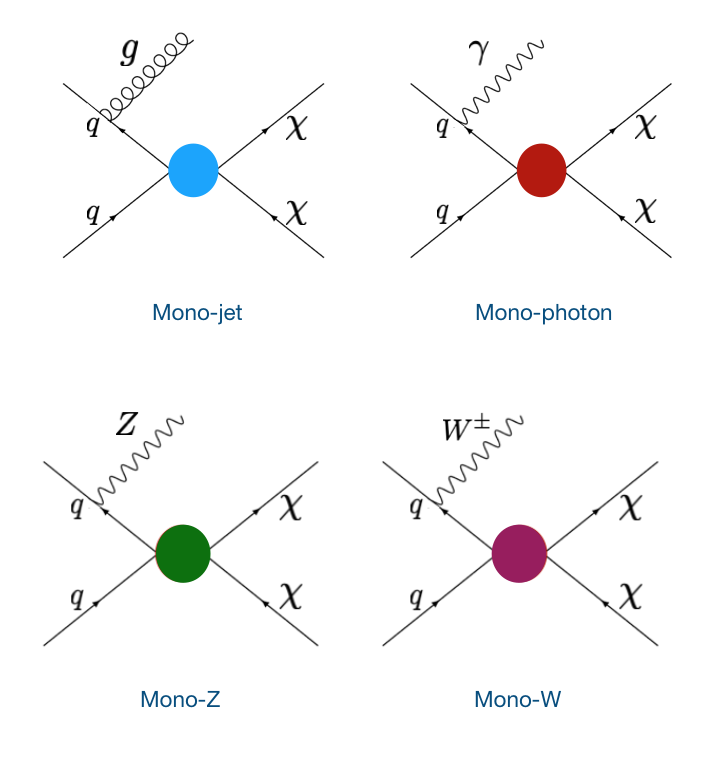}
$$
\caption{Production of dark particles in association with a gluon jet, photon, $Z$ or $W^\pm$ to give rise mono-X signatures.}
\label{fig:mono-gamma.jet}
\end{figure}

DM collider searches provide an independent tool for probing the dark sector, though due to their assumed weak interactions with the SM, dark particles would not be seen directly, and their presence must be inferred using other signatures. The standard approach is based on the observation that the total momentum transverse to the colliding beams is very small, therefore, if dark particles are produced at a collision, the transverse momentum they carry as they leave the detector must be balanced by the momentum of another particle which may be detected. The simplest signal is then the production of a single SM particle with large transverse momentum or missing energy. This is illustrated in Fig. \ref{fig:mono-gamma.jet} for DM production at the LHC; the details of this DM search strategy at the LHC is elaborated in many articles (see for example, \cite{Goodman:2010ku,Fox:2011pm}).

In this section we will assume, as an illustrative example of this approach to DM detection, that the leading SM-DM interactions are well described by the operator $\ocal_{\tt D5}=(\bar{\chi}\gamma^\mu\chi)(\bar{\qr}\gamma_\mu \qr)$  in Table \ref{tab:DM-SM2}, and will study some of its effects at the LHC. This operator generates final states of the form [missing (transverse) energy] + X (where X can be a jet or a photon). For the calculations described below it is important to note that this has an important irreducible SM background process: neutrino production in association with a photon or gluon via an intermediate $Z$ boson: $ \qr \bar\qr \to Z + \gamma/g \to \nu \bar\nu + \gamma/g $. In addition, an important reducible background is $\qr \bar\qr^{'} \to W^{\pm} + \gamma/g \to \nu \ell^{\pm} + \gamma/g $, when the charged lepton ($\ell^\pm$) is soft and missed at the detector, leading to the same signal.

The calculation is straightforward: knowing the interaction Lagrangian ($ \propto \ocal_{\tt D5}$), standard field-theory technology can be used to obtain the number of DM pairs produced in association with a jet or photon, which can then be compared to the SM background. For a given luminosity one can then determine the region in the $ \Lambda - \mdm $ plane where the LHC would be able to experimentally detect the production of DM using this signature. Though in principle straightforward, a realistic calculation is involved and we refer the reader to the literature for the details \cite{Aad:2013oja, Carpenter:2012rg, Berlin:2014cfa, Petrov:2013nia, Belyaev:2018pqr}. The final result is relatively simple: the absence of a signal in this channel at the LHC with 13 TeV C.M. energy \cite{Sirunyan:2017jix}, implies $ \Lambda > 1\,\tev $ for $ \mdm \lesssim 250 \, \gev $ (beyond this value the cross section drops significantly), whenever the EFT parameterization remains valid. 

In closing this section we note that it is important to note that the parameter space is further restricted by the relic density and direct search constraints; the interplay of DM direct search versus collider search using an EFT parameterization has been studied in several publications, see, for example, \cite{Buchmueller:2014yoa,Belyaev:2018pqr,Chang:2013oia}. In addition, the types of mediators that generate $ \ocal_{\tt D5} $ will also generate $ \qr^4 $ effective interactions, but the implications are model dependent (see below and section \ref{sec:dm.vect}).

\subsubsection{Simplified Model Approach}

The use of EFT in practical calculations for DM production at hadron colliders presents a practical obstacle. As we have repeatedly noted, the effective parameterization is valid only if the typical energy associated with an effective operator lies well below the NP scale $ \Lambda $; for the process in Fig. \ref{fig:mono-gamma.jet} this means that the $ \chi\bar\chi $ and/or the $q \bar\chi $ invariant masses must lie below $ \Lambda$. Unfortunately, neither of these quantities can be measured: the quark energies are known only as a distribution determined by their distribution functions inside the proton, and only the transverse momentum of the DM pair can be measured, not their total momentum (or energy). 
As a result, the applicability of the EFT approach is difficult to guarantee.

Faced with this, a natural alternative is to adopt a specific model containing a mediator. Calculations can then be carried out using standard field-theory technology and simulation packages; if desired, one can then translate the results to the EFT language. For the above example the natural model is the one already discussed in section \ref{sec:dm.vect}: at energies well below $ M_{\tt X} $ (see \cref{eq:simplified}) the model generates $ \ocal_{\tt D5} $ with $ \Lambda^2 = - M_{\tt X}^2/(g_\chi g_\qr) $ as noted in \cref{eq:lambda-mX}. Calculations are significantly simplified by implementing the Lagrangian in \cref{eq:simplified} within {\tt Feynrules} \cite{Christensen:2008py}, or {\tt CalcHep} \cite{Belyaev:2012qa}, and then using simulation packages like {\tt Madgraph} \cite{Alwall:2011uj} and {\tt Pythia} \cite{Sjostrand:2006za} to generate mono-X plus missing energy events.

\begin{figure}[htb!]
$$
\includegraphics[scale=0.38]{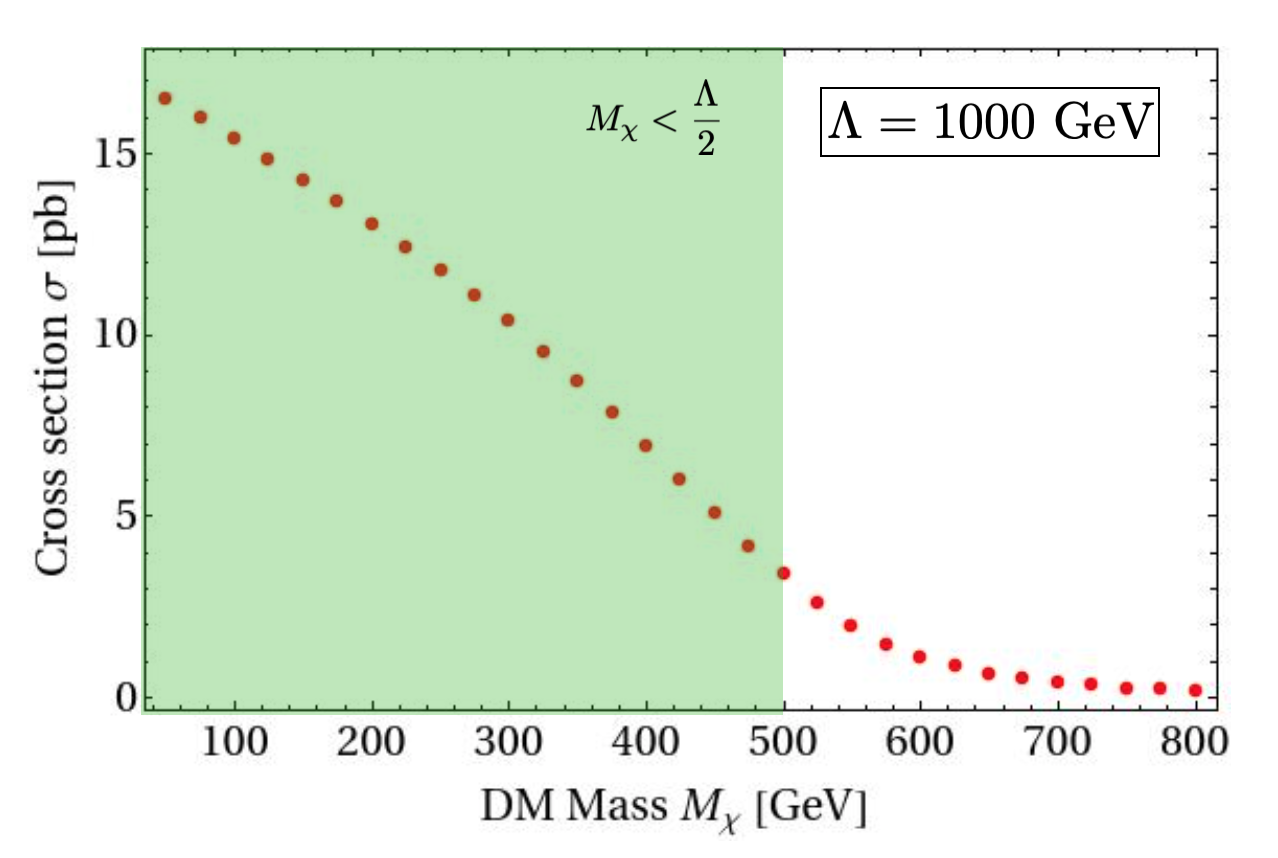}
$$
\caption{ 
DM production cross-section as a function of DM mass ($M_\chi = \mdm$) at LHC for $\sqrt{s}=14$ TeV for the operator $\ocal_{\tt D5}$ of Table \ref{tab:DM-SM2} generated by mediator of mass $M_{\tt X}=\Lambda$= 1000 GeV; the shaded region corresponds to that where the DM invariant mass is below $\Lambda $ (see text).
}
\label{fig:MET}	
\end{figure}

As an introduction to this approach we present here the results of a simpler calculation: DM pair production the the LHC, without requiring the additional photon or gluon (processes containing one or more SM particles with high transverse momentum plus ``missing'' transverse energy have been studied extensively in the literature, see for example \cite{Aad:2013oja, Carpenter:2012rg, Berlin:2014cfa, Petrov:2013nia})). Taking~\footnote{Different values of the couplings can be obtained by appropriately rescaling $ \Lambda $ in the results below. As in section \ref{sec:dm.vect} we shall ignore the limits form the reaciton $ \qr \qr \to \qr\qr$ mediated by the $X$.} $ g_\chi = g_\qr = 1 $, so that $ \Lambda = M_{\tt X} $ one can then calculate the total cross section for $ pp \to \chi\chi $ at the LHC (Fig.~\ref{fig:MET}); the shaded area labeled $ M_\chi = \mdm < \Lambda/2 = M_{\tt X} /2 $ corresponds to the region where EFT is applicable.

The region where this model is equivalent to its EFT counterpart corresponds to that where the invariant mass $ M_{\tt inv} $ of the final state DM particles is below $ \Lambda $. Unfortunately, as we noted above, $ M_{\tt inv} $ is not measurable; still one can estimate the region in parameter space where the EFT is applicable by the following procedure: let $ \sigma_{\tt exp} $ is the measured cross section for the process of interest, and $\sigma_{\tt theo}( M_{\tt inv} <\Lambda) $ the corresponding theoretical cross section, obtained using the same cuts as for $ \sigma_{\tt exp} $ and, in addition, restricting $ M_{\tt inv} $ as noted. Using these define
\bea
R_\Lambda=\frac{\sigma_{\tt theo}( M_{\tt inv} <\Lambda)}{\sigma_{\tt exp}}\,;
\eea
then the EFT approximation will be valid for those values of  $\Lambda $ (and other parameters that enter the expression of $ \sigma_{\tt theo}$) for which $ R_\Lambda \sim 1 $.

A more detailed discussion on the validity of DM EFT at the LHC can be found in
\cite{Busoni:2013lha,Busoni:2014sya,Busoni:2014haa}. It is worth noting, however, that this problem does not arise at $e^-e^+$ colliders since in this case the CM energy of the hard cross section is known.

\subsection{Example 4: Neutrino portal Dark Matter}

Of the dimension 5 operators in table \ref{tab:DM-SM} the one in category III,
\beq
\ocal_{\tt III}=(\bar{\Psi} \Phi) (\phi^T \epsilon \ell) \quad \stackrel{\tt SSB}\longrightarrow \quad \frac{\smvev + h}{\sqrt{2}} \left( \bar\Psi \Phi \right) \nu,
\label{eq:nu.port}
\eeq
describes a type of DM that interacts with visible sector mainly through the neutrinos. The dark sector contains, at the least, a fermion $ \Psi $ and a scalar $ \Phi $ that, for consistency (and naturality), one must also have a Higgs portal coupling $ \ocal_{\tt I} = |\phi|^2 |\Phi|^2 $. The operator $ \ocal_{\tt III} $ can be generated at tree level, which we assume (see below for a specific model).

\begin{figure}[htb!]
$$
\includegraphics[scale=0.4]{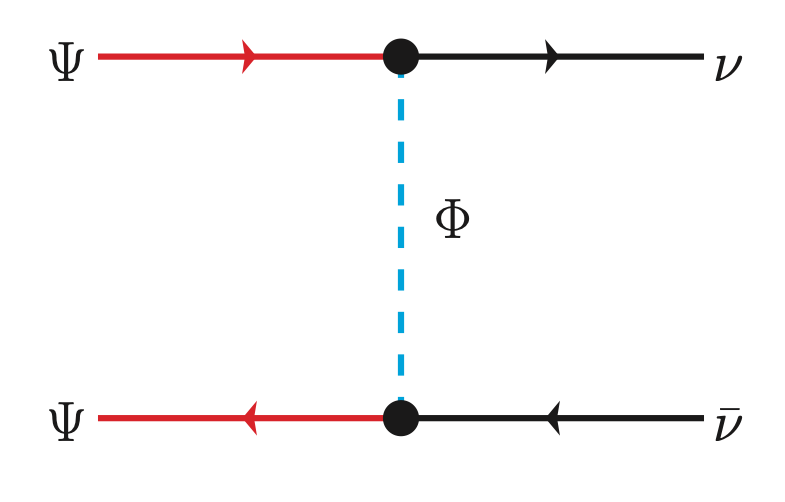}
$$	
\caption{Annihilation channels for DM ($\Psi$) to neutrinos providing the leading DM-SM interaction in the neutrino-portal scenario; the heavy dots denote vertices generated by $ \ocal_{\tt III}$ in \cref{eq:nu.port}.}
\label{fig:nu.port-f1}
\end{figure}

\begin{figure}[htb!]
$$
\includegraphics[scale=0.4]{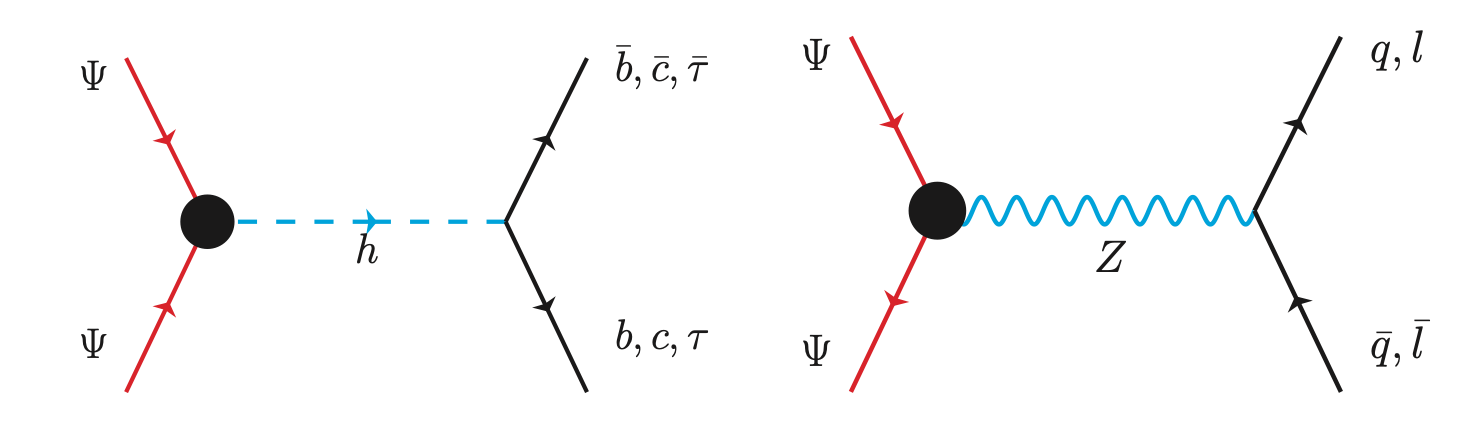}
$$	
\caption{Leading couplings of the DM to the quarks ($q,\,b,\,c)$ charged leptons ($l,\,\tau$) within the neutrino portal scenario. The heavy dots refer to vertices derived from operators in categories II and VII of table \ref{tab:DM-SM}, assumed to be loop-generated (see text). }
\label{fig:nu.port-f2}
\end{figure}

If the $ \Phi $ is heavier than the $ \Psi $ it will decay promptly, $ \Phi \to \Psi \nu $ though the interaction provided by \cref{eq:nu.port}; in this case the dark fermion is the DM candidate~\footnote{The opposite holds if the dark scalars are lighter; we will not consider this scenario here.}. In this scenario the $\Psi\Phi\nu$ DM-neutrino vertex generates the leading DM-SM interaction~\footnote{The $ \Psi \Phi\nu h$ in $ \ocal_{\tt III}$ does not have a significant phenomenological impact.} and determines the DM relic abundance ({\it cf.} Fig. \ref{fig:nu.port-f1}). In contrast, we assume that the DM couplings to the quarks occurs through the exchange of vector bosons and the Higgs (see Fig. \ref{fig:nu.port-f2}), and that the  $ \Psi\Psi Z,\,\Psi\Psi h $ couplings, described by the operators in categories II and VII, occur only at one loop~\footnote{Operators $ \ocal_{\tt II,\,VII}$  in table \ref{tab:DM-SM} are PTG; we  justify below our assumption that they are loop generated within the present scenario.}, which ensures that the coupling of the DM to the nucleons is naturally suppressed.

With these ingredients one can apply the (by now) standard machinery to determine the restrictions on the operator coefficients derived from relic abundance, direct and indirect detection and collider constraints, the procedure is similar to the one described in more detail in section \ref{sec:dm.vect} and will not be further pursued here; full details can be found in \cite{Macias:2015cna}.

These features allow the model to meet all experimental and observational constraints without fine tuning and without severe restrictions on the DM mass, hence the interest in \cref{eq:nu.port}. In contrast, models where operators in categories I and II of table \ref{tab:DM-SM} provide the leading DM-SM couplings require large DM masses or  $ \mdm  \sim m_{\tt h}/2 $ (for recent discussions see {\it eg.} \cite{Athron:2018hpc,Arcadi:2019lka}). Note also that category IV operators are loop generated and have difficulties satisfying the relic abundance requirement in a freeze-out scenario (see Sect. \ref{sec:DM-freeze-in}). The thermal-averaged cross section is given by
\beq
\vevof{\sigma v}_{\Psi\Psi \to \nu\nu} = \frac{(c_{\tt III}\smvev/\Lambda)^4}{256\pi \mdm^2} \left(\frac{\mdm^2} {m_\Phi^2 + \mdm^2}\right)^2 \,,
\label{eq:av.cs}
\eeq
where $ \mdm = m_\Psi$, $ c_{\tt III} $ is the coefficient of the neutrino portal operator in \cref{eq:nu.port}, and $ \Lambda $ is the mass scale of the mediator(s) that generate this operator. As in  previous sections, this expression can be used in \cref{eq:BEQ2to2} to calculate the DM relic abundance \cite{Kolb:1990vq}. Because of the small error on this quantity ({\it cf.} \cref{relic-value}) the effect of this constraint is to impose a relation between the model parameters contributing to \cref{eq:av.cs}.

The terms in the effective Lagrangian that are relevant for direct detection are given by
\beq
\lcal_{\tt DM-Z,h} = \frac{\smvev c_{\tt II}}{16\pi^2 \Lambda} h \bar\Psi \Psi - \frac g{2 \cw} \frac{\smvev^2 }{16\pi^2 \Lambda^2} \bar\Psi \slashed{Z} \left ( c_{\tt VII}\up{\tt L} P_L + c_{\tt VII}\up{\tt R} P_R \right) \Psi + \cdots \,,
 \label{ZHex}
\eeq 
where we assumed that the type II and VII operators are loop generated, so we wrote the operator coefficients as $ c_{\tt II,\,VII}/(16\pi^2)$ with $ c_{\tt II,\,VII} = O(1) $. This can then be used to obtain the DM-nucleon cross section, determine the values of $ \Lambda $ and $ c_{\tt II,\,VII} $ allowed by the current limits. 

It is worth noting that from the effective-theory point of view the  operator coefficients contributing to the relic abundance and direct detection are independent, so meeting the corresponding constraints is straightforward. This will not necessarily be the case for specific models that realize the neutrino portal scenario. We now turn to this `ultraviolet completion' to illustrate the interplay of the effective theory and model building approaches.

\subsubsection{Ultraviolet completion}

Constructing a model that leads to a tree-generated $ \ocal_{\tt III}$ in \cref{eq:nu.port} is straightforward \cite{Gonzalez-Macias:2016vxy}. It is only necessary to note that the two factors in the operator can be generated by the exchange of a mediator fermion $ \chi $ with couplings $ \bar\Psi\Phi\chi $ and $ \bar\chi \phi^T\epsilon\ell$. Specifically, the Lagrangian is given by
\bal
\lcal =& \lcal_{\tt SM} + \bar\chi(i \slashed\partial - \mdm) \chi + \bar\fm(i \slashed\partial - M) \fm + |\partial\Phi|^2 - m_\Phi^2 |\Phi|^2 \cr
& \quad - \left(  \bar l Y\up\nu \fm \tilde\phi + \bar\chi  y_{\tt DM}^\dagger \fm \Phi +  {\rm H.c.} \right)-\lambda_x|\Phi|^2|\phi|^2\,,
\label{eq:nu.port.lag}
\end{align}
where $\fm$ denote the Dirac fermion mediators, assumed to be 3 in number, and $ \chi $ the fermionic DM field; $ Y\up\nu $ denote the mediator-SM Yukawa couplings and $y_{\tt DM}$ the mediator-DM ones. As noted above we also assume $ \mdm < m_\Phi $; generation numbers are not displayed. 

At this point it is worth pausing to compare and contrasting the EFT and model approaches. In the first one we have a large number of unknown parameters (the operator coefficients) which reduces predictability, but facilitates accommodating experimental constraints. In a model the number of parameters is reduced, so there are, in general, more observables that can be predicted; however, meeting all experimental constraints may prove more challenging. We now illustrate this using the comparing the above model to the neutrino portal EFT.

\begin{figure}[htb!]
$$
\includegraphics[scale=1]{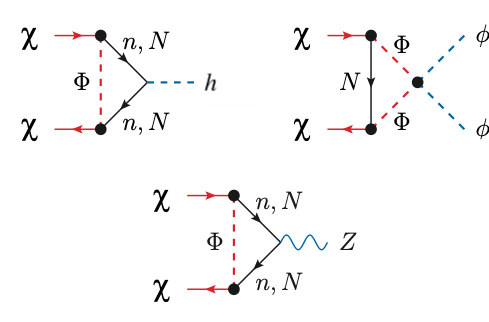}
$$	
\caption{One loop graphs generating the leading DM coupling to the $Z$ and $h$ bosons for the model of \cref{eq:nu.port.lag}.}
\label{fig:nu.port-f3}
\end{figure}

The first thing to notice is that upon \ssb\ the mediators $\fm$ will mix with the SM neutrinos:
\beq
\bar l Y\up\nu \fm \tilde\phi \stackrel{\tt SSB}\longrightarrow \smvev \bar\nu_L Y\up\nu F + \cdots\,,
\eeq
so that the mass eigenstates will be linear combinations of $F$ and $ \nu $, one, which we denote by $N$ will be heavy (mass $ \sim M$), the other, $n_L$ will be massless and corresponds to the physical neutrinos~\footnote{Generating a small mass for the $n_L$ can be achieved by giving the $F$ a small Majorana mass.}; because of the mixing, the couplings of the $n_L$ to the $W$ and $Z$ bosons will be different from those of the $\nu_L$ . The second thing to notice is that this mixing also generates a $n\chi\Phi$ coupling which in its turn will generate $\chi\chi Z$ and $\chi\chi h$ vertices at one loop, see Fig. \ref{fig:nu.port-f3}; these realizes the assumption made in the EFT approach \cref{ZHex}.

In contrast, the modification to the neutrino couplings to the $W$ and $Z$ were not included in the EFT discussion. These are described by the effective operators $ \ocal_{\phi l}\up{1,3} $ in table \ref{tab:dim-6-1}; these can certainly be added to the effective theory and the corresponding coefficients constrained by current electroweak data. The difference is that within the context of this model, the coefficients of these purely SM operators are related to those describing DM-SM interactions; specifically, we have
\beq
\Delta \Gamma(Z \to nn) \sim \eta^2 \,,  \qquad \vevof{\sigma v}_{\Psi\Psi \to \nu\nu} \sim \eta^2 \times y_{\tt DM}^2\,; \qquad \eta \sim  \frac\smvev\Lambda Y\up\nu\,.
\eeq
In order to obey the constraints on the invisible $Z$ width, $\eta$ must be small, but then the relic abundance requirement demands $y_{\tt DM}$ to be relatively large, and this imposes significant restrictions on parameter space as $y_{\tt DM}$ must also be small enough for the theory to remain perturbative (a tacit assumption throughout).

There are additional constraints to be included: the $W$-boson couplings to the leptons are also modified, so restrictions follow from $ \tau \to \nu\nu\mu,\,\nu\nu e $ and $ \pi \to \nu\mu$ decays; for sufficiently light $N$, the model allows the decays $ h \to Nn,\,NN$ and, at 1-loop, $ h \to \Psi \Psi $ when $\mh > 2 \mdm$, all of which are constrained by the limits on the Higgs invisible width. The process followed in the EFT approach can now be repeated, and the allowed regions in parameter space identified. One finds that, $ m_\Phi > \mdm + 10\,\gev$, the electroweak and relic-abundance restrictions allow only $ \mdm \lesssim 35\,\gev $, or $ \mdm \sim\mh/2 $, providing another illustration of the  restrictions that often occur when a model realization of the EFT is used. Details can be found in \cite{Gonzalez-Macias:2016vxy}.

\section{Summary and Conclusions}
\label{sec:summary}

The effective theory approach is an important tool in studying physics beyond the SM; recently it has received additional attention because of the absence of a specific hint as to the  nature of that new physics, aside from the very strong indications that it is present. In this review we have touched upon several aspects of effective theories, describing several important theoretical and phenomenological aspects such as the decoupling theorem, the role of gauge invariance, the equivalence theorem and the effects and characterization of loop and tree level generated operators. We listed all effective operators relevant for the SM up to dimension 7 (assuming weakly coupled and decoupling heavy physics) allowing for the possibility that light right-handed neutrinos are present; though we argued that in most cases PTG (potentially tree-generated) operators of dimension $ \le6 $ are sufficient for studying most types of new physics effects, though we also noted that higher-dimensional or loop-generated operators must sometimes be considered.

There is a monumental body of literature devoted to the phenomenology of effective theories, studying all aspects of possible deviations from the SM, and recently global analyses of the constraints on the Wilson coefficients have become available (see, {\it e.g.} \cite{Marzocca:2020jze}; for a review see \cite{Brivio:2017vri}). Areas of particular interest include Higgs, top-quark and vector-boson physics, and flavor-changing processes; this review contains examples of such applications. To make this review manageable, we covered by a minute aspect of these investigations, and even these in a simplified manner. We refer the reader to the literature for wider and deeper discussions.

As noted above, the EFT approach is readily extended to the study of DM-DM and DM-SM interactions. Given our current ignorance of the nature of DM these studies must allow for a variety of DM candidates. We provided the relevant effective operators assuming that the DM-SM interactions are generated by the exchange of neutral mediators, and briefly discussed some applications intended as illustrations of the possible effects that can be described using the effective theory approach, including model realizations of the effective theory. We may also note here that we are unaware of a comprehensive review of DM physics in an EFT context, so we hope that, despite the various gaps we have noted, the discussion here presented will be of use and interest.

We concentrated on two general aspects of effective theory (weakly-coupled and decoupling NP effects in the SM, and mediator driven DM-SM interactions), but there are many other paradigms where effective theory is useful. These include, among others, low energy QCD \cite{Scherer:2012xha,Ecker:1989yg,Gasser:1984gg,Gasser:1983yg}, strongly coupled new physics beyond the SM \cite{Georgi:1992dw,Giudice:2007fh,Agashe:2004rs,Redi:2011zi,Contino:2010rs,Cacciapaglia:2014uja,Bruggisser:2018mrt,Marzocca:2012zn,Anastasiou:2009rv}, 
baryogenesis \cite{deVries:2017ncy,Murphy:2017omb,Huang:2015bta,Bruggisser:2018mrt}, leptogenesis \cite{Biondini:2013xua,Barman:2021tgt}, and holographic models  \cite{Erdmenger:2020lvq,Erdmenger:2020flu} (to mention a few). The effective theory parameterization is also becoming a standard way of presenting constraints on new physics at the LHC, another aspect of the field that we have not reviewed in any detail. Despite these limitations, this review (hopefully) serves the purpose of sketching a broad picture of the effective theory approach, and will attract researchers to contribute to this field in this exciting era where high luminosity (high statistics) allows unprecedented probes of fundamental physics. 

{\bf Acknowledgments}

SB would like to acknowledge the funding from DST SERB grant CRG/2019/004078, Govt. of India. SB also acknowledges discussion and technical help from Dr. Basabendu Barman, Mr. Sudhakantha Girmohanta, and Mr. Soumen Kumar Manna.

\appendix
\section{SMEFT dimension 7 operators}
\label{sec:ap-A}
In this appendix we provide the list of dimension 7 operators for the case of a single family (flavor diagonal case); more details can be found in \cite{Bhattacharya:2015vja}. The extension to multiple families and some aspects of the renormalization group evolution of these operators can be found in \cite{Liao:2020zyx,Liao:2016hru}.

To simplify the notation we find it useful to define the composite operators
\beq
N = \phi ^{\dagger}\eps\ell \,; \qquad E= \phi^\dagger\ell \,,
\eeq
The nomenclature is motivated by the fact that, in the unitary gauge $N,\,E$ have terms  proportional to the left-handed neutrino and electron fields, respectively. As before we denote by PTG operators that are potentially tree-generated, and by LG those that are necessarily loop-generated.

with these preliminaries one can classify the dimension 7 operators in the following categories:

\subsubsection{Operators with 2 fermions}

These operators  are of the form~\footnote{Field strength tensors correspond to $[D,D]$ commutators contained in terms with $s=2$ in Eq.~(\ref{eq:2fops}). }
\beq
\overline{\psi^c} \Gamma \psi' \vp^r D^s\,, \quad r+s=4,~ r,s \ge0\,,
\label{eq:2fops}
\eeq
where $ \vp $ denotes $\phi $ or $ \phit=\epsilon \phi^*$, and $ \psi $ a fermion in the \nsm:
\beq
\psi \in \{ q,\,u,\,d,\,\ell,\,e,\,\nu,~q^c,\,u^c,\,d^c,\,\ell^c,\,e^c,\,\nu^c \}\,,
\eeq
where the charge conjugate fields are defined as $\psi^c = C \bar\psi^T$ ($C$ is Dirac charge conjugation matrix), and $\Gamma=\{ 1,\gamma^\mu, \sigma^\mn\}$, where $\sigma^\mn=\frac{i}{2}[\gamma^\mu, \gamma^\nu]$. All these operators conserve baryon number but violate lepton number by two units: $|\Delta L|=2,\, \Delta B=0$. 

\bit
\item{$r=4,~s=0$: 2 PTG operators:}
\beq
(\Ncb N) |\phi|^2 ,\quad \ncb\nu |\phi|^4\,.
\label{eq:r4s0}
\eeq

\item{$r=3,~s=1$: 4 PTG operators:}
\beq
(\ecb \gamma^\mu N )(\ptd \stackrel \leftrightarrow D_\mu \phi), \quad
(\ncb \gamma^\mu N )(i \phi^\dagger \stackrel \leftrightarrow D_\mu \phi), \quad
(\ncb \gamma^\mu E )(\ptd \stackrel \leftrightarrow D_\mu \phi), \quad 
(\ncb \gamma^\mu N)( \partial_\mu |\phi|^2 )\,.
\label{eq:r3s1}
\eeq
where $\phi^\dagger \stackrel \leftrightarrow D_\mu \phi =\phi^\dagger D_\mu\phi - ( D_\mu\phi)^\dagger \phi$. 
\item{$r=s=2$: 9 PTG operators.} 
\beq
\begin{array}{lll}
 ( \lcb D_\mu \ell)( \ptd D^\mu \phi), &\quad \Ncb(D_\mu\ptd  D^\mu \ell),  &\quad (\lcb D\phi) (\ell \eps D\phi), \cr
  [\Ncb \sigma^\mn (\ptd \WW_\mn \ell)], &\quad (\ncb D_\mu e)(\ptd D^\mu \phi), &\quad (\ncb \nu) |D\phi|^2, \cr
    (\ncb \sigma^\mn e)(\ptd \WW_\mn \phi) , & \quad(\Ncb\sigma^\mn N) B_\mn, &\quad  |\phi|^2 (\ncb \sigma^\mn \nu) B_\mn  ;
\end{array}
\label{eq:r2s2}
\eeq
where $\WW_\mn = \tau^I  W_\mn^I$.

\item{$r=1,~s=3$: 8 LG operators:}
\beq
\begin{array}{llll}
(\partial^\mu \ncb) \gamma^\nu N B_\mn, & \quad\ncb\gamma^\mu (\ptd D^\nu \ell) B_\mn, & \quad
(\partial^\mu \ncb) \gamma^\nu (\ptd  \WW_\mn \ell)  , &\quad \ncb\gamma^\mu (\ptd  \WW_\mn D^\nu \ell),  \cr
(\partial^\mu \ncb) \gamma^\mu N \tilde B_\mn,  & \quad\ncb\gamma^\mu (\ptd  D^\nu \ell)\tilde  B_\mn,  &\quad
(\partial^\mu \ncb) \gamma^\mu (\ptd \WW_\mn \ell) , &\quad \ncb\gamma^\mu (\ptd  \tilde\WW_\mn D^\nu \ell);
\end{array}
\eeq
where $\tilde X_\mn=\frac{1}{2}\epsilon_{\mu\nu\rho\sigma}X^{\rho\sigma}$ denote the dual tensors.

\item{$r=0,~s=4$: 6 LG operators:}
\bea
&& \ncb\nu \times \{ ( G^A_\mn)^2,\, (W^I_\mn)^2,\, (B_\mn)^2,\,( \tilde G^A_\mn G^A_\mn),\, (\tilde W^I_\mn W^I_\mn),\, (\tilde B_\mn B_\mn) \}.
\eea
\eit

\subsubsection{Operators with 4 fermions}

These operators are of the form $ \psi^4 D$ (operators with 4 fermions and one covariant derivative) or $ \psi^4 \varphi $ 
(operators with 4 fermions and one scalar); they all violate $|B-L|$ by two units with $ |\Delta B|=0,\,1$.

\bit
\item{$\psi^4 D $: 21 LG operators.}
Using Fierz rearrangements these can be cast in either of two forms:
\beq
(L_1 \sigma^\mn L_2)(L_3 \gamma_\nu \stackrel \leftrightarrow{D_\mu} R)\,, \qquad (L_1 \sigma^\mn L_2) D_\mu (L_3 \gamma_\nu R)\,;
\label{eq:4f.D}
\eeq
where $L$ and $R$ denote, respectively, left and right-handed fermion fields. 

\item{$\psi^4 \phi $: 33 PTG operators.}
Using Fierz transformations one can readily see that these take one of the two forms:
\beq
(L_1^T C L_2)(L_3^T C L_4) \vp\,, \qquad (L_1^T C L_2)(R_1^T CR_2) \vp\,;
\label{eq:4f.phi}
\eeq
where $ \varphi = \phi,\, \phi^c$. 

The allowed field combinations are listed in table \ref{tab:4f} for a single family. In certain cases, however, the operators vanish when some of the fields are in the same family; those operators can be found in \cite{Liao:2020zyx}.

\eit

\begin{table}[ht]
{\footnotesize
$$
\renewcommand{\arraystretch}{1.1}
\begin{array}{|c|cccc|cc|}
\hline
 \multicolumn{6}{|c}{\ocal=(L_1 \sigma^\mn L_2)(L_3 \gamma_\nu \stackrel \leftrightarrow{D_\mu} R),}		&			  \cr
 \multicolumn{6}{|c}{\qquad (L_1 \sigma^\mn L_2) D_\mu (L_3 \gamma_\nu R)}		&			  \cr
 \hline
	&	L_1		&	L_2		&	L_3		&	R	&	\Delta L	&	\Delta B			\cr
\hline\hline
1	&	d^c		&	d^c		&	d^c		&	e	&	 1	&	 -1					\cr
2	&	d^c		&	\ell		&	\ell		&	u	&	 2	&	 0					\cr
3	&	d^c		&	\ell		&	d^c		&	q^c	&	 1	&	-1					\cr
\hline
4	&	q		&	d^c		&	\ell		&	\nu	&	 2	&	 0					\cr
5	&	q		&	\ell		&	d^c		&	\nu	&	 2	&	 0					\cr
6	&	d^c		&	\ell		&	q		&	\nu	&	 2	&	 0					\cr
7	&	\ell		&	e^c		&	\ell		&	\nu	&	 2	&	 0					\cr
8	&	q		&	u^c		&	\nu^c	&	\ell^c	&	-2	&	 0					\cr
9	&	q		&	\nu^c	&	u^c		&	\ell^c	&	-2	&	 0					\cr
10	&	u^c		&	\nu^c	&	q		&	\ell^c	&	-2	&	 0					\cr
11	&	u^c		&	\nu^c	&	e^c		&	d	&	-2	&	 0					\cr
12	&	u^c		&	e^c		&	\nu^c	&	d	&	-2	&	 0					\cr
13	&	\nu^c	&	e^c		&	u^c		&	d	&	-2	&	 0					\cr
14	&	u^c		&	d^c		&	d^c		&	\nu	&	 1	&	-1					\cr
15	&	q		&	\nu^c	&	q		&	d	&	-1	&	 1					\cr
\hline
16	&	q		&	\nu^c	&	\nu^c	&	q^c	&	-2	&	 0					\cr
17	&	u^c		&	\nu^c	&	\nu^c	&	u	&	-2	&	 0					\cr
18	&	d^c		&	\nu^c	&	\nu^c	&	d	&	-2	&	 0					\cr
19	&	\ell		&	\nu^c	&	\nu^c	&	\ell^c	&	-2	&	 0					\cr
20	&	\nu^c	&	e^c		&	\nu^c	&	e	&	-2	&	 0					\cr
\hline
21	&	\nu^c	&	\nu^c	&	\nu^c	&	\nu	&	-2	&	 0					\cr
\hline
\end{array}
\qquad
\begin{array}{|c|cccc|cc|c|}
\hline
	 \multicolumn{7}{|c}{\ocal=(L_1^T C L_2)(L_3^T C L_4) \vp }		&			  \cr
\hline
	&	L_1	&	L_2	&	L_3	&	L_4	&	\Delta L	&	\Delta B	  & \vp \cr
\hline
\hline
1	&	\ell	&	\ell	&	\ell	&	e^c	&	 2	&	 0	&  \phi  \cr
2	&	q	&	d^c	&	\ell	&	\ell	&	 2	&	 0	&  \phi  \cr 
3^{**}	&	q	&	\ell	&	\ell	&	d^c	&	 2	&	 0	&  \phi \cr 
4	&	u^c	&	d^c	&	d^c	&	\ell	&	 1	&	-1	&  \phi  \cr 
5	&	d^c	&	d^c	&	d^c	&	\ell	&	 1	&	-1	&  \phit \cr 
6	&	u^c	&	\ell	&	d^c	&	d^c	&	 1	&	-1	&  \phi  \cr 
\hline
7	&	q	&	u^c	&\nu^c	&	e^c	&	-2	&	 0	&  \phit \cr 
8	&	q	&	e^c	&\nu^c	&	u^c	&	-2	&	 0	&  \phit \cr 
9^*	&	q	&	q	&	q	&\nu^c	&	-1	&	 1	&  \phit \cr 
\hline
10	&	q	&	u^c	&\nu^c	&\nu^c	&	-2	&	 0	&  \phi  \cr 
11	&	q	&	d^c	&\nu^c	&\nu^c	&	-2	&	 0	&  \phit \cr 
12	&	q	&\nu^c	&\nu^c	&	u^c	&	-2	&	 0	&  \phi  \cr 
13	&	q	&\nu^c	&\nu^c	&	d^c	&	-2	&	 0	&  \phit \cr 
14	&	\ell	&	e^c	&\nu^c	&\nu^c	&	-2	&	 0	&  \phit \cr 
15	&	\ell	&\nu^c	&\nu^c	&	e^c	&	-2	&	 0	&  \phit \cr 
\hline
16	&	\ell	&\nu^c	&\nu^c	&\nu^c	&	-2	&	 0	&  \phi  \cr 
\cline{1-8}
\end{array}
\qquad
\begin{array}{|c|cccc|cc|c|}
\hline
	 \multicolumn{7}{|c}{\ocal=(L_1^T C L_2)(R^T_1 CR_2) \vp }		&				  \cr
\hline
	&    L_1	&	L_2	&	R_1	&	R_2	&	\Delta L	&	\Delta B	& \vp \cr
\hline
\hline
1	&	d^c	&	\ell	&	u	&	e	&	 2	&	 0	 & \phi  \cr
2	&	\ell	&	\ell	&	q^c	&	u	&	 2	&	 0				 & \phi  \cr
3^*	&	q	&	q	&	d	&	\ell^c	&	-1	&	 1				 & \phit \cr
4	&	q	&	e^c	&	d	&	d	&	-1	&	 1				 & \phit \cr
\hline
5	&	q	&	d^c	&	\nu	&	e	&	 2	&	 0			&	  \phi  \cr
6	&	u^c	&	\ell	&	u	&	\nu	&	 2	&	 0			&	  \phi  \cr
7	&	d^c	&	\ell	&	u	&	\nu	&	 2	&	 0			&	  \phit \cr
8	&	d^c	&	\ell	&	d	&	\nu	&	 2	&	 0			&	  \phi  \cr
9	&	\ell	&	e^c	&	\nu	&	e	&	 2	&	 0			&	  \phi  \cr
10^{**}	&	q	&	\ell	&	q^c	&	\nu	&	 2	&	 0			&	 \phi \cr
11	&	\ell	&	\ell	&	\ell^c	&	\nu	&	 2	&	 0			&	  \phi  \cr
12	&	q	&\nu^c	&	u	&	d	&	-1	&	 1			&	  \phit \cr
13	&	q	&\nu^c	&	d	&	d	&	-1	&	 1			&	  \phi  \cr
\hline
14	&	q	&	u^c	&	\nu	&	\nu	&	 2	&	 0			&	  \phi  \cr
15	&	q	&	d^c	&	\nu	&	\nu	&	 2	&	 0			&	  \phit \cr
16	&	\ell	&	e^c	&	\nu	&	\nu	&	 2	&	 0			&	  \phit \cr
\hline
17	&	\ell	&\nu^c	&	\nu	&	\nu	&	 2	&	 0			&	  \phi  \cr
\hline
\end{array}
$$
} 
\caption{
Field combinations that can contribute to the operators containing 4 fermions, with one derivative and no scalar fields (left column) -- \cref{eq:4f.D}, and with one scalar and no derivatives (center and right columns) -- \cref{eq:4f.phi}. The entries with one (two) asterisks have 2 (3) possible $\su2$ contractions (assuming only family-diagonal couplings, see text).}
\label{tab:4f}
\end{table}

Amongst operators those do not contain right handed neutrinos 20 are PTG operators:
\beq
{\small
\begin{array}{llll}
\ocal_1 = (\lcb \eps D^\mu \phi) (\ell \eps D_\mu \phi),\quad & 
\ocal_2 = (\ecb \gamma^\mu N )(\phi \epsilon  D_\mu \phi),\quad &   
\ocal_3 = ( \lcb \eps D_\mu \ell)( \phi \eps D^\mu \phi),\quad &  
\ocal_4 = \Ncb (D_\mu \phi \eps D^\mu \ell)\cr 
\ocal_5 = (\Ncb \ell )\eps(	\bar e \ell	),\quad &  
\ocal_6 = (\Ncb N) |\phi|^2,\quad &  
\ocal_7 = [\Ncb \sigma^\mn (\phi\eps\WW_\mn\ell)],\quad &   
\ocal_8 = (\Ncb \sigma^\mn N) B_\mn \cr
\ocal_9 = (\bar d q	)\eps(	\Ncb \ell	),\quad &  
\ocal_{10} = [(\overline{q^c} \phi)	\eps \ell )( \bar d \ell ),\quad &  
\ocal_{11} = (\Ncb q )\eps(	\bar d \ell	),\quad &   
\ocal_{12} = (\lcb	\eps q )( \bar d N )\cr   
\ocal_{13} = (\bar d N )(	u^T C e ),\quad &  
\ocal_{14} = (\Ncb \ell	)( \bar q u ),\quad &  
\ocal_{15} = (\bar u d^c )(	\bar d N ),\quad & 
\ocal_{16} = [\overline{q^c}  (\phi^\dagger q) ]\eps( \bar\ell d )\cr 
\ocal_{17} = (\overline{q^c} \eps q	)( \bar N d	),\quad &  
\ocal_{18} = (\bar d d^c )(	\bar d E ),\quad & 
\ocal_{19} = (\bar e  \phi^\dagger q)(	\overline{d^c}	d	) ,\quad  &  
\ocal_{20} = (\bar u N	)( \bar d d^c	)  
\end{array}
\label{eq:PTG-nonu}
} 
\eeq

\bibliography{Bibliography}

\end{document}